\documentclass[varenna,cite,overfull]{cimento}

\usepackage{amssymb}
\usepackage{graphicx}
\usepackage{amsmath}
\usepackage{calc}
\usepackage{dsfont}
\usepackage{mathrsfs}
\usepackage{color}
\usepackage{geometry}
\newcommand{\tp}{{\it T}}
\newcommand{\op}[1]{{\rm #1}}

\newcommand{\dd}{\partial}
\newcommand{\D}{\mbox{d}}
\newcommand{\diff}[2]{\frac{\D#1}{\D#2}}
\newcommand{\pdiff}[2]{\frac{\dd#1}{\dd#2}}

\newcommand{\basind}[1]{{\scriptscriptstyle(}#1{\scriptscriptstyle)}}
\newcommand{\symm}[2]{\underset{(#1)}{\op{Symm}}\left[\,#2\,\right]}

\newcommand{\A}{\overset{(0)}{C}}
\newcommand{\B}{\overset{(2)}{C}}
\newcommand{\C}{\overset{(3)}{C}}

\title{Rotation in relativity and the propagation of light}
\author{E. Kajari, M. Buser, C. Feiler \atque W.~P. Schleich}
\institute{Institut f\"ur Quantenphysik, Albert-Einstein-Allee 11, Universit\"at Ulm, D-89069 Ulm, Germany}

\begin{document}

\maketitle

\vspace*{-0.4cm}
{\it``Whatever nature has in store for mankind, unpleasant as it may be, men must accept,}\\
{\hspace*{6.4cm} {\it for ignorance is never better than knowledge.'' }}\\
{\hspace*{11.4cm} Enrico Fermi}
\vspace*{0.4cm}


\begin{abstract}
We compare and contrast the different points of view of rotation in general relativity, put forward by Mach, Thirring and Lense, and G\"odel. Our analysis relies on two tools: (i) the Sagnac effect which allows us to measure rotations of a coordinate system or induced by the curvature of spacetime, and (ii) computer visualizations which
bring out the alien features of the G\"odel Universe. In order to keep the paper self-contained, we summarize in several appendices crucial ingredients of the mathematical tools used in general relativity. In this way, our lecture notes should be accessible to researchers familiar with the basic elements of tensor calculus and general relativity.
\end{abstract}


\section{Introduction\label{SecIntroduction}}

For centuries, physicists have been intrigued by the concepts of rotation. Galileo Galilei, Vincenzo Viviani, Isaac Newton, Jean Bernard L\'eon Foucault, Ernst Mach, Albert Einstein, George Sagnac, Hans Thirring, Josef Lense, Hermann Weyl and Kurt G\"odel form an impressive line of researchers trying to gain insight into this problem. Key ideas such as inertial forces, Mach's principle, frame dragging, gravito-magnetism, compass of inertia and rotating universes are intimately connected with these pioneers.
Even today, there are still many open questions. One of them is related to the meaning of Mach's principle and best described by the following quote from James A. Isenberg and John A. Wheeler \cite{Isenberg79}:
\begin{quotation}
{\it ``Mystic and murky is the measure many make of the meaning of Mach''}.
\end{quotation}
The present lecture notes start from a brief historical development of the notion of rotation and then focus on the Sagnac effect as a tool to measure rotations. Moreover, we bring out some of the alien features of the G\"odel Universe by presenting computer visualizations of two scenarios.

\subsection{Concepts of rotation from Foucault to G\"odel}

During a service in the great Cathedral of Pisa Galilei made a remarkable discovery watching the chandeliers swinging in the wind. The period of a pendulum is independent of the amplitude of its oscillation. As a clock, he used his pulse. Had the church service lasted longer, he might have noticed what his student Viviani mentioned later \cite{Aczel}: 
\begin{quotation}
{\it ``... all pendulums hanging on one thread deviate from the initial vertical plane, and always in the same direction.''}
\end{quotation} 
Contained in this cryptic remark is the by now familiar fact that the Coriolis force causes the plane of oscillation of a pendulum to slowly drift. Viviani's observation was rediscovered by Foucault in 1851. He was the first to interpret this phenomenon as a consequence of the rotation of the Earth relative to the absolute space of Newton.

A new era in the investigation of rotation was ushered in by the development of general relativity. Stimulated by Mach's criticism of Newton's rotating bucket argument \cite{Mach1893}, the question whether rotations should be considered as absolute or relative motions was one of the principles guiding Einstein in his formulation of general relativity~\cite{Einstein16}. However, he was not the first to be inspired by Mach's ideas. For the contributions of the brothers Benedikt and Immanuel Friedl\"ander we refer to \cite{Pfister07}. 

Einstein's theory of general relativity made it possible for the first time to analyze Mach's principle in a mathematically rigorous way. Thirring and his coworker \mbox{Lense~\cite{Thirring18,Lense18}} performed this task in 1918 and addressed the questions: Are there Coriolis- and centrifugal-like forces in the center of a rotating hollow sphere? And, does the rotation of a solid sphere influence the motion of a nearby body by ``dragging forces'' not present in Newton's theory? According to general relativity the answer to both questions is yes. However, the expected effects were experimentally not accessible at that time. Nevertheless, there were early experiments aiming at the measurement of ``dragging forces'' in Arnold Sommerfeld's institute in Munich. According to Otto Scherzer \cite{Scherzer}, a former assistant of Sommerfeld, an experiment was set up to measure the influence of a heavy, rotating sphere on a Foucault pendulum. Unfortunately, the mass, which weighed a few tons, fell off the device driving the rotation, and started running around in the lab. The experiment was aborted.

We emphasize that the results of Thirring and Lense did not imply that Mach's principle is fully incorporated in the theory of general relativity, as~e.g.~discussed by Hermann Weyl \cite{Weyl24}. Due to the intriguing, but vague formulation of Mach's principle, the controversy is still ongoing \cite{Pfister95,Ciufolini95}.

In 1949 a new issue in the study of rotation within general relativity emerged from a remarkable discovery by the ingenious mathematician G\"odel. Trying to explain~\footnote{This statement is unsourced but most likely correct, see \cite{Jung06,Wheeler98} and references therein.} 
the observed statistical distribution of the angular momenta of galaxies \cite{Gamov46}
\footnote{Rotation curves of galaxies depict the velocity of stars or gas orbiting the center of a galaxy against the distance of the stars from the center. It is interesting to note, that investigations of rotation curves of galaxies made in the seventies of the last century, are nowadays widely accepted as evidence for the existence of dark matter.}, 
G\"odel found an exact cosmological solution of Einstein's field equations \cite{Goedel49} whose energy-momentum tensor corresponds to a homogeneous and rotating ideal fluid. Today, we refer to this solution as G\"odel's Universe. G\"odel soon recognized that his solution exhibits a very puzzling property, namely the existence of closed timelike world lines. This feature which is incompatible with the common notion of causality forced him and many others to a ``in-depth reconsideration of the nature of time and causality in general relativity'' \cite{Ellis00}. Until this fateful discovery it had been tacitly assumed, that Einstein's field equations might automatically prevent themselves from pathologies of this kind. Moreover, G\"odel noticed that his solution was not in agreement with cosmological observations. Indeed, it could not explain the expansion of the universe, which was commonly accepted after the discovery of Hubble's law \cite{Hubble29} in 1929. This disagreement caused him to search for a cosmological solution which exhibits both features: rotation and expansion. Only one year later, G\"odel published a new solution \cite{Goedel50} which satisfied both requirements. George~F.~R.~Ellis \cite{Ellis00} described the influence of G\"odel's contributions to the progress of general relativity as follows:
\begin{quotation}
{\it 
``These papers stimulated many investigations leading to fruitful developments.
This may partly have been due to the enigmatic style in which
they were written: for decades after, much effort was invested in giving
proofs for results stated without proof by G\"odel.''
}
\end{quotation}
This is neither the place nor the time to mention further developments stimulated by G\"odel's articles \cite{Delgado02,Ciufolini03}. Instead, we will focus in this paper on G\"odel's first solution \cite{Goedel49} -- mainly because it still allows an analytical approach to most questions and therefore provides a suitable first encounter with the subject of rotating universes.

\subsection{Space-based research}

Today, we are in the unique position of being able to observe the Lense-Thirring field \cite{Laemmerzahl01,Mashoon01,Ruffini03,Schaefer04,Tourrenc04}. The present proceedings represent a living testimony to the outstanding experimental progress in tests of general relativity. Three domains of physics come to mind: (i) solid state devices are at the very heart of the Gravity Probe B experiment \cite{GravityProbeBFinal}, (ii) laser technology is put to use in the laser ranging to the LAGEOS satellites \cite{Ciufolini04,Ciufolini07}, and (iii) modern astronomy made the recent observation of the relativistic spin precession in a double pulsar possible \cite{Breton08}. Indeed, Gravity Probe~B utilizes for the readout of the rotation axes of the mechanical gyroscopes the London moment of the spinning superconducting layer on the balls. The LAGEOS experiment relies on the precession of the nodal lines of the trajectories of the two satellites. In this case, the readout takes advantage of modern laser physics. Finally, the impressive confirmation \cite{Breton08} of the prediction of relativistic spin precession \cite{Connell75,Barker77} in a double pulsar falls into this category of outstanding experiments on this subject. 

Now is the time to dream of future projects in space \cite{Jentsch04,CosmicVision,Kleppner06,Tino07,Turyshev07}. 
The development of optical clocks \cite{Laemmerzahl04,Schiller07}, frequency combs \cite{Hall06,Haensch06}, and atom interferometers \cite{Borde89,Cronin07} has opened new avenues to measure rotation. Matter wave gyroscopes and accelerometers \cite{Canuel06,Antoine07} have reached an unprecedented accuracy, which might allow us to perform new tests of gravito-magnetic forces. Such ideas have a long tradition. Indeed, a few decades ago there were already proposals to search for preferred frames \cite{Scully79,Haugan80,Scully81,SchleichLesHouches} and probe the Lense-Thirring effect using ring laser gyroscopes \cite{Chow85}. Today's discussions of matter wave gyroscopes can build on and take advantage of these results obtained for light waves. Here the Sagnac effect \cite{Post67} plays a central role. With its help we can measure the rotation of a coordinate system and catch a glimpse of the curvature of spacetime.

\subsection{Goal of the paper}

In these lecture notes we study the influence of rotation on the propagation of light. Here we focus on two main themes: \mbox{(i) the} Sagnac effect, and (ii) visualizations of scenarios in the G\"odel Universe.

Motivated by the important role of the Sagnac effect in today's discussions of tests of general relativity using light or matter waves, we dedicate the first part of our lecture notes to a thorough analysis of rotation in general relativity. In this context we introduce an operational approach towards Sagnac interferometry based on light. Our approach is stimulated by the following passage from the introduction of the seminal article \cite{Scully81} by Scully \etal
\begin{quotation}
{\it ``The proposed experiment provides but one example of the possible applications of quantum optics to the study of gravitation physics. In order to make the analysis clear and understandable to readers with backgrounds in quantum optics, we have avoided the use of esoteric techniques (e.g., Fermi--Walker transport) and have instead carried out an explicit general relativistic analysis in order to obtain the necessary representation of the metric in the ring interferometer laboratory.''}
\end{quotation}
Our intention is to make these ``esoteric techniques'' accessible to the quantum optics community and to point out their advantages in the description of local satellite experiments. Indeed, local and conceptually simple experiments provide a necessary counterpart to the sophisticated tests of general relativity based on cosmological observations and their intricate theoretical foundations. Therefore, our first aim is to develop a measurement procedure relying solely on local measurements which properly separates the different sources of rotation.  Throughout our analysis, we concentrate on a description which is geared towards a one-satellite-experiment. We assume that all the necessary Sagnac interferometers are contained within a single satellite, so that we can locally expand the metric around its world line \footnote{
The authors of \cite{Scully81} derive their expressions for the Sagnac effect using the weak field approximation $g_{\mu \nu}(x^\sigma)=\eta_{\mu \nu}+h_{\mu \nu}(x^\sigma)$ with $h_{\mu \nu}$ being a small, global perturbation of the flat spacetime metric $\eta_{\mu\nu}$. However, we can apply this approximation only when the underlying spacetime is nearly flat everywhere. Otherwise, some peculiarities concerning the coordinates might occur. In order to illustrate this point, we take the two-dimensional surface of a sphere with the usual metric induced by its embedding in the three-dimensional Euclidean space. In this case no coordinates can cover the whole surface and allow for a decomposition of the metric into a dominant flat Euclidean part and a small correction to it. Such a decomposition only exists, when we restrict the coordinates to a small region on the surface using e.g. Riemann normal coordinates. Thus, in order to keep our considerations of the Sagnac effect as general as possible, we pursue in sect.~\ref{ProperRefSagnac} an approach which does not rely on a background metric.}.
Moreover, we will restrict ourselves to the proper time delay between two counter-propagating light rays, since it provides the fundamental quantity necessary for the description of the Sagnac effect within the framework of general relativity. Other measurement outcomes such as the interference fringes, which are due to the phase shift between the two counter-propagating light rays, are intimately related to this proper time delay.

\subsection{Outline}
Our lecture notes are organized as follows:
In sect.~\ref{section_SagnacEffect} we introduce an elementary version of the Sagnac effect within the framework of general relativity. For this purpose we first briefly recall in \ref{SecSagnacOriginal} the original experiment of Sagnac and then proceed in \ref{SecDerivationTimeDelay} with the derivation of the Sagnac time delay for a stationary metric.

The aim of sect.~\ref{SecProperRefCoord} is to provide suitable coordinates for local measurements in a single satellite. After a short motivation in \ref{AppMotiviationProper}, we define in \ref{ProperRef} the local coordinates of a proper reference frame attached to an accelerating and rotating observer. They provide us with a valuable power-series expansion of the metric coefficients around the world line of the observer. With the help of this expansion, we establish in sect.~\ref{ProperRefSagnac} the two dominant terms in the Sagnac time delay. Moreover, we briefly discuss a measurement strategy for separating the rotation of the observer from the influence of the curvature of spacetime. 

We conclude our discussion of the Sagnac effect by applying these results to two very different situations. In sect.~\ref{SecRotatingFrame} we introduce the metric coefficients of a rotating reference frame in Minkowski spacetime, study the light cone diagram and then analyze the resulting Sagnac time delay. In sect.~\ref{PropertiesGoedel} we address the G\"odel Universe and follow a similar path. We start with a brief historic motivation and then present the metric. We continue with the exploration of some of its peculiar features such as time travel by considering the corresponding light cone diagram. Finally, we investigate the Sagnac time delay and compare it to the results obtained for the rotating reference frame.

The second part of these lectures is mainly devoted to the question: How does the inherent rotation of the ideal fluid in G\"odel's Universe affect the visual perception of an observer? In sect.~\ref{SecHowThings} we answer this question by visualizing two scenarios in G\"odel's Universe. In order to keep the notes self-contained, we give in \ref{SecFundRaytracing} a short introduction into the world of computer graphics. Here we discuss a fundamental version of ray tracing, which is a simple yet powerful technique to render realistic images of a given scenario. Moreover, we also point out the changes necessary for the successful application of those techniques to obtain visualizations of relativistic models such as the G\"odel Universe. We then devote \ref{LightPropagationGoe} to a thorough analysis of these visualizations. In the first scenario, discussed in \ref{SecViewinner}, we consider an observer located inside a hollow sphere. Its inner checkered surface appears warped due to the peculiar propagation of light in G\"odel's Universe. Moreover, we point out the existence of an optical horizon which restricts the view of any observer to a limited spatial region. The second scenario, highlighted in \ref{SecViewsmallObj}, visualizes the view of an observer on a small terrestrial globe. We emphasize that in general small objects in G\"odel's Universe enjoy two images.

In order to lay the foundations of the individual sections, we summarize several important concepts of general relativity in the appendices~\ref{AppendixBasicConcept}, \ref{TetradsAndTransport} and~\ref{AppendixSpecialCoordinates}. For example, in appendix~\ref{AppendixBasicConcept} 
we analyze the transformation of a metric to its Minkowski form at a fixed point in spacetime. We discuss symmetries and Killing vectors as well as world lines and geodesics of test particles and of light. A comparison between parallel transport and Fermi-Walker transport concludes this introduction into some important aspects of general relativity. Appendix~\ref{TetradsAndTransport} provides insight into the concept of orthonormal tetrads and their orthonormal transport. Here, special attention is devoted to the Fermi-Walker transport and to a natural generalization of it, the so-called proper transport. Appendix~\ref{AppendixSpecialCoordinates} establishes Riemann normal coordinates and proper reference frame coordinates together with the leading-order contributions in the corresponding metric expansions.
In appendix \ref{CalcInPropRef} we provide a series expansion of the Sagnac time delay in proper reference frame coordinates and derive the explicit expressions for the first two leading-order contributions. Finally, we briefly sketch the analytical solution of the geodesic equation for light rays which emanate from the origin in G\"odel's Universe in appendix \ref{Intnullgeo}.

\subsection{Notation and conventions}

In this article we use $(1,-1,-1,-1)$ as signature for any metric. Greek indices denote both space and time components of tensors and will run from $0$ to $3$, whereas Latin indices indicate only the spatial components and therefore just take on the values $1$,$2$ and $3$. Throughout the paper, we retain the speed of light $c$ in all our calculations. 
In table \ref{Definition} we summarize several fundamental equations of tensor calculus and general relativity.
\begin{table}[h]
\caption{Definitions and basic equations of differential geometry and general relativity. The curve parameters of massive particles and of light
are the proper time $\tau$ and the evolution parameter $\lambda$, respectively. Newton's gravitational constant $G$ and the speed of light $c$ define the constant ${\kappa= {8\pi G}/{c^4}}$. Moreover, $\Lambda$, $\rho$ and $p$ denote the cosmological constant, the mass density and the pressure of the ideal fluid, respectively.}
  \label{Definition}
  \begin{tabular}{ll}
    \hline
      Metric of Minkowski spacetime & $(\eta_{\mu\nu})\equiv \op{diag}(1,-1,-1,-1)$     \\
      Line element and proper time & $\D s^2=g_{\mu\nu}\,\D x^\mu\,\D x^\nu\equiv c^2\,\D \tau^2$  \\
    
    \hline
      Christoffel symbols     & $\Gamma^\mu_{\;\alpha\beta}\equiv
\frac{1}{2}\,g^{\mu\nu}\left(g_{\nu\alpha,\beta}+
g_{\nu\beta,\alpha}-g_{\alpha\beta,\nu}\right)$   \\
     Curvature tensor &  $R^\mu_{\;\,\alpha\beta\gamma}\equiv
\Gamma^\mu_{\;\alpha\gamma,\beta}-\Gamma^\mu_{\;\alpha\beta,\gamma}
+\Gamma^\mu_{\;\rho\beta}\Gamma^\rho_{\;\alpha\gamma}-
\Gamma^\mu_{\;\rho\gamma}\Gamma^\rho_{\;\alpha\beta}$     \\
     Ricci tensor and scalar curvature &  $R_{\alpha\beta}\equiv R^\mu_{\;\,\alpha\mu\beta}$ \quad\text{and}
     \quad  $R\equiv R^\mu_{\;\,\mu}$ \\
     \hline
      Covariant derivative of a contravariant vector & 
$V^\alpha_{\,\;;\beta}\equiv V^\alpha_{\;\,,\beta}+\Gamma^\alpha_{\;\mu\beta}V^\mu$\\
      Covariant derivative of a covariant vector & 
$V_{\alpha;\beta}\equiv V_{\alpha,\beta}-\Gamma^\mu_{\;\alpha\beta}V_\mu$\\
    \hline
     Four-velocity of massive particles and light & $u^\mu(\tau)\equiv\frac{\D x^\mu}{\D \tau}$ and
     $u^\mu(\lambda)\equiv\frac{\D x^\mu}{\D \lambda}$\\
     Geodesic equation &
     $\frac{\D^2x^\mu}{\D\lambda^2}+\Gamma^\mu_{\;\alpha\beta}\frac{\D x^\alpha}{\D \lambda}
     \frac{\D x^\beta}{\D\lambda}\equiv u^\mu_{\;\,;\nu}u^\nu=0$ \\
     Constraint for particles and light &  $g_{\mu\nu}u^\mu u^\nu=c^2$ and $g_{\mu\nu}u^\mu
      u^\nu=0$ \\
     \hline
      Einstein's field equations &  $R_{\mu\nu}-\frac{1}{2}g_{\mu\nu}R=
\kappa T_{\mu\nu}+\Lambda g_{\mu\nu}$ \\
      Energy-momentum tensor for an ideal fluid &  $T_{\mu\nu}\equiv\left(\rho+\frac{p}{c^2}\right)
u_\mu u_\nu-p g_{\mu\nu}$ \\
    \hline
  \end{tabular}
\end{table}



\section{Formulation of the general relativistic Sagnac effect\label{section_SagnacEffect}}

The goal of the present section is to derive an exact expression for the Sagnac time delay measured in a reference frame corresponding to a time independent metric. We start with a brief discussion of Sagnac's original experiment and then continue with the derivation of the Sagnac time delay within the framework of general relativity.

\subsection{Sagnac's original experiment\label{SecSagnacOriginal}}

\begin{figure}
\hspace{0.07\textwidth}
\includegraphics[width=0.42\textwidth]{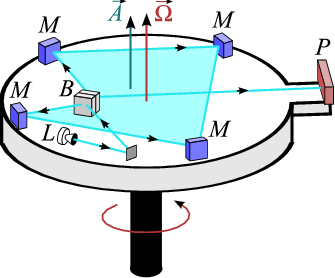}
\hfill
\includegraphics[width=0.31\textwidth]{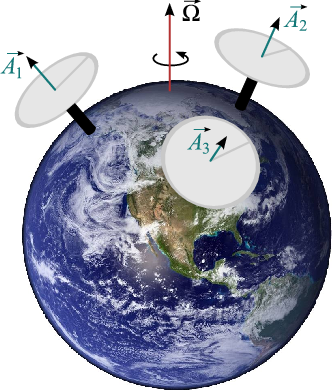}
\hspace{0.07\textwidth}
\caption{Sagnac's original experiment (left) and the measurement of the angular velocity vector $\boldsymbol{\Omega}$ of the Earth (right) with the help of three Sagnac interferometers with area vectors $\mathbf{A}_i$.}
\label{SagnacOriginal}
\end{figure}

In 1913 George Sagnac performed the experiment \cite{Sagnac,Hazelett79} summarized by the left picture of fig.~\ref{SagnacOriginal}: On a horizontal platform which carries all optical components, including a mercury arc lamp $L$ and a fine-grained photographic plate $P$, a light ray is split at the separator $B$ into a clockwise and a counterclockwise-propagating beam. Both beams are then reflected successively by four mirrors $M$ and travel around a circuit with enclosed area $\mathcal{A}=|\mathbf{A}|$. They recombine again at the beam splitter which superimposes them on the photographic plate $P$, leading to interference fringes. 

Once the platform is in rotation, a difference $\Delta t$ in the arrival times of the clockwise and counterclockwise-propagating beams arises, which translates into a shift of the fringes at the photographic plate. By comparing the fringe positions corresponding to rotations in clockwise or counter-clockwise direction with approximately the same rate, Sagnac observed that $\Delta t$ is proportional to the area $\mathbf{A}$ enclosed by the light beams and to the angular velocity $\boldsymbol{\Omega}$ of the rotating platform. The classical expression 
\begin{equation}
\Delta t=\frac{4}{c^2} \mathbf{A}\cdot\boldsymbol{\Omega}
\label{famsagnac}
\end{equation}
for the Sagnac time delay constitutes a very good approximation of the relativistic expression derived in~\ref{SagRotatingReferenceFrame} in the limit of small rotation rates. Furthermore Sagnac established, that eq.~\eqref{famsagnac} is independent of the location of the center of rotation and of the shape of the enclosed area.

From today's perspective it is interesting to note that Sagnac's interpretation of his results points towards the existence of the luminiferous either \cite{Hazelett79}:
\begin{quotation}
{\it ``The observed interference effect is clearly the optical whirling effect due to the movement of the system in relation to the ether and directly manifests the existence of the ether, supporting necessarily the light waves of Huygens and of Fresnel.''}
\end{quotation}

When lasers found their way into Sagnac interferometry in form of ring-laser gyros, they provided such an enormous increase in sensitivity \cite{Chow85,Post67} that the Sagnac effect is nowadays a backbone of modern navigation systems. Moreover, it can be used for measurements of geophysical interest \cite{Stedman97}, e.g. when one is looking for the time dependence of magnitude and direction of the angular velocity vector of the Earth \cite{Schreiber04}. Equation~\eqref{famsagnac} suggests that one needs at least three Sagnac interferometers with linearly independent area vectors $\mathbf{A}_i$ to recover all three components of the angular velocity vector $\boldsymbol{\Omega}$ of the Earth as illustrated in the right picture of fig.~\ref{SagnacOriginal}. Finally, further improvements of earth-bound Sagnac interferometers may allow a direct measurement of the Lense-Thirring effect in a not too far future \cite{Stedman03}.

\subsection{Sagnac time delay for a stationary metric \label{SecDerivationTimeDelay}}

In this subsection, we present an elementary derivation of the Sagnac time delay within the framework of general relativity for the case of a stationary spacetime. Since many roads lead to Rome, we also want to draw attention to several other approaches. In \cite{Scully81,SchleichLesHouches} the authors analyze the Sagnac effect in the limit of weak gravitational fields, whereas \cite{Ashtekar75} provides a general derivation of the Sagnac time delay for stationary spacetimes. Investigations based on arbitrary spacetimes without any restriction to certain symmetry properties of the spacetime can be found in \cite{Bazanski98,Bazanski99}.\\

\subsubsection{Mathematical description of the arrangement}

\begin{figure}
\hspace{0.07\textwidth}
\includegraphics[width=0.37\textwidth]{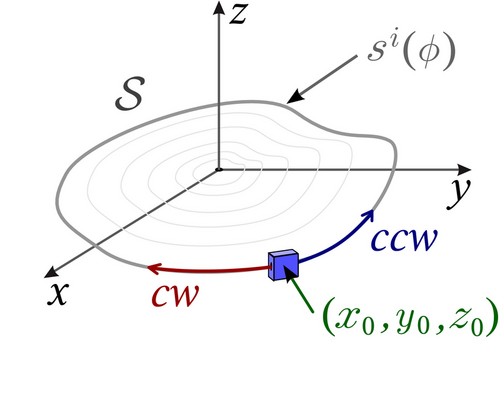}
\hfill
\includegraphics[width=0.35\textwidth]{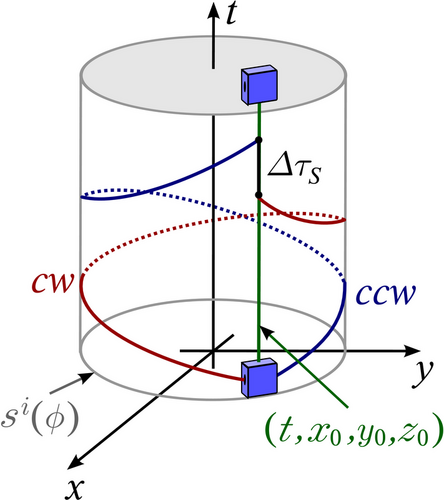}
\hspace{0.07\textwidth}
\caption{Propagation of light in a Sagnac interferometer depicted in space (left) and in spacetime (right). In the latter we suppressed the $z$-coordinate. In the spatial diagram (left), light propagates from the observer located at the fixed position $q^i=(x_0,y_0,z_0)$ in clockwise (cw) and counter-clockwise (ccw) direction along the closed spatial curve $\mathcal{S}$ described by $s^i(\phi)$. The spacetime diagram (right) shows the definition of the Sagnac time delay $\Delta\tau_S$ in terms of the proper time difference between the arrivals of the two light rays measured along the world line of the observer at rest.}
\label{SagnacTubeRestFrame}
\end{figure}

Our derivation of the Sagnac time delay requires a reference frame for our observer and his experimental setup in which the components $g_{\mu\nu}$ of our stationary metric do not depend on time. We denote the coordinates of this reference frame by $x^\mu=(t,x,y,z)$ and suppose that the observer is located at the fixed spatial point $q^i=(x_0,y_0,z_0)$, as shown in the left picture of fig.~\ref{SagnacTubeRestFrame}. From there, he sends out two light rays in opposite directions which, forced by an appropriately arranged set of mirrors,  travel along the null curves that correspond to the closed spatial curve $\mathcal{S}$. For simplicity, we assume that $\mathcal{S}$ is spacelike and that we can parameterize the curve $\mathcal{S}$ uniquely by the angle $\phi\in[0,2\pi)$, thereby using the notation $s^\mu(\phi)=(0,s^i(\phi))=(0,x(\phi),y(\phi),z(\phi))$. We denote the position of the observer at rest by $q^i=s^i(\phi_0)$ with the corresponding curve parameter $\phi_0$.

\subsubsection{Null curves of the counter-propagating beams}

As indicated by the spacetime diagram on the right of fig.~\ref{SagnacTubeRestFrame}, the light rays arrive after one circulation at different coordinate times at the observer, thus giving rise to the Sagnac proper time delay $\Delta\tau_S$ along the observer's world line. 
In order to derive an explicit formula for this proper time delay, we parameterize the counter-propagating light beams on $\mathcal{S}$ by the null curve $x^\mu(\phi)=(t(\phi),s^i(\phi))$, which have to satisfy the condition
\begin{equation}
g_{\mu\nu}\big|_\mathcal{S}\,\diff{x^\mu}{\phi}\,\diff{x^\nu}{\phi}
=g_{00}\big|_\mathcal{S}\left(\diff{t}{\phi}\right)^2+2\, g_{0i}\big|_\mathcal{S}\diff{s^i}{\phi}\,\diff{t}{\phi}
+g_{ik}\big|_\mathcal{S}\diff{s^i}{\phi}\,\diff{s^k}{\phi}=0\,.
\label{quadrEq}
\end{equation}
Since the metric does not depend on time in our chosen reference frame, we have introduced the abbreviation $g_{\mu\nu}(x^\mu(\phi))\equiv g_{\mu\nu}(s^i(\phi))\equiv g_{\mu\nu}\big|_\mathcal{S}$ to indicate that the metric coefficients have to be taken along the spacelike curve $\mathcal{S}$.

The two solutions of the quadratic equation~\eqref{quadrEq} for $(\D t/\D \phi)$ read
\begin{equation}
\left(\diff{t}{\phi}\right)_{\pm}=-\frac{g_{0i}}{g_{00}}\bigg|_\mathcal{S}\diff{s^i}{\phi}
\pm\sqrt{\frac{\gamma_{ik}}{g_{00}}\bigg|_\mathcal{S}\diff{s^i}{\phi}\,\diff{s^k}{\phi}}
\label{LightSolutions}
\end{equation}
with
\begin{equation*}
\gamma_{ik}\equiv \frac{g_{0i}g_{0k}}{g_{00}}-g_{ik}\,.
\end{equation*}
At this point we have to impose a further restriction: In order to guarantee the existence of two solutions $(\D t/\D \phi)_{\pm}$ the spacelike curve $\mathcal{S}$ must be contained in a region $\mathcal{R}$ of spacetime where the conditions
\begin{equation}
g_{00}\big|_\mathcal{S}>0 \,, \quad \mathcal{S}\subset\mathcal{R}
\label{ConditionsOnMetric}
\end{equation}
and
\begin{equation}
\gamma_{ik}\bigg|_\mathcal{S}\diff{s^i}{\phi}\,
\diff{s^k}{\phi}>0 \,,\quad \mathcal{S}\subset\mathcal{R}
\label{ConditionsOnMetric2}
\end{equation}
are satisfied for all points in $\mathcal{S}$.

The first condition, given by eq.~\eqref{ConditionsOnMetric}, implies that the spacetime curve ${x^\mu(\lambda)=(\lambda,s^i_0)}$ is timelike for any fixed spatial point $s^i_0=\text{const}$ on $\mathcal{S}$. Only in this case it is possible to relate the coordinate time $t$ with the physically measured proper time $\tau$ of a fixed observer at $s^i_0\in\mathcal{S}$. Since this requirement means physically that all mirrors defining $\mathcal{S}$ have to move on timelike curves, this condition is a priori fulfilled.

Concerning the second condition, eq.~\eqref{ConditionsOnMetric2}, we would like to mention that the quantities $\gamma_{ik}$ constitute the components of the local spatial metric, as specified in~\cite{Landau}. In case the chosen reference frame is realized by material objects, the coefficients $\gamma_{ik}$ represent a positive definite matrix and condition~(\ref{ConditionsOnMetric2}) is automatically fulfilled. 

Since we have presumed that $\mathcal{S}$ is a spacelike curve which satisfies
\begin{equation*}
g_{\mu\nu}\big|_\mathcal{S}\diff{s^\mu}{\phi}\,\diff{s^\nu}{\phi}=
g_{ik}\big|_\mathcal{S}\diff{s^i}{\phi}\,\diff{s^k}{\phi}<0\,,
\end{equation*}
it directly follows from the eqs. (\ref{LightSolutions}), (\ref{ConditionsOnMetric}) and \eqref{ConditionsOnMetric2} that the two solutions possess opposite signs, where ${(\D t/\D \phi)_{+}>0}$ and ${(\D t/\D \phi)_{-}<0}$. Being only interested in solutions which are located on the future light cone and for which the coordinate time $t(\phi)$ increases with increasing angle $\phi$, we conclude that the solution ${(\D t/\D \phi)_{+}>0}$ corresponds to the counterclockwise (ccw)-propagating beam. Since we have to reverse the direction of rotation for ${(\D t/\D \phi)_{-}<0}$, we can identifying the second solution with the clockwise (cw)-propagating beam. 

\subsubsection{Final expression for the time delay}

When we integrate the time coordinate $t(\phi)$ along the opposite paths of the beams, we find the expression
\begin{equation}
t_{\pm}=\int_{\phi_0}^{\phi_0\pm 2\pi}\left(\diff{t}{\phi}\right)_{\pm}\,\D\phi
=\pm \int_{0}^{2\pi}\left(\diff{t}{\phi}\right)_{\pm}\,\D\phi
\label{IntLightSolutions}
\end{equation}
for the arrival coordinate times $t_\pm$ after one circulation. Here we have used the time independence of the metric coefficients, as well as their periodicity in the angular coordinate~$\phi$.

Hence, the difference $\Delta t=t_{+}-t_{-}$ between the arrival times of the ccw- and the cw-propagating 
beams reads
\begin{equation*}
\Delta t= \int_{0}^{2\pi}\left(\diff{t}{\phi}\right)_{+}+\left(\diff{t}{\phi}\right)_{-}\,\D\phi\,.
\end{equation*}
When we insert eq.~(\ref{LightSolutions}), we arrive at
\begin{equation*}
\Delta t=-2 \displaystyle{\int_0^{2\pi}}\frac{g_{0i}}{g_{00}}\bigg|_\mathcal{S}\diff{s^i}{\phi}\,\D\phi\,.
\end{equation*} 
The connection 
\begin{equation*}
\Delta \tau =\frac{1}{c}\,\sqrt{g_{00}(q^r)}\,\Delta t\,,
\end{equation*}
between the coordinate time difference $\Delta t$ and the corresponding proper time difference $\Delta \tau$ measured by the observer along his world line $q^\sigma(\tau)=(c\tau,q^i)$ allows us to cast the Sagnac time delay $\Delta\tau_S$ into the form
\begin{equation*}
\Delta\tau_S=-\frac{2}{c}\,\sqrt{g_{00}(q^r)}\int_0^{2\pi}\frac{g_{0i}}{g_{00}}
\bigg|_\mathcal{S}\diff{s^i}{\phi}\,\D\phi\,.
\end{equation*}
Thus, the spatial line integral
\begin{equation}
\Delta\tau_S=-\frac{2}{c}\,\sqrt{g_{00}(q^r)}\oint\limits_\mathcal{S}\frac{g_{0i}}{g_{00}}\,\D s^i 
\label{GenSagnacTimeDelay}
\end{equation}
relates the Sagnac proper time delay $\Delta\tau_S$ of two counter-propagating light rays to the metric coefficients $g_{00}$ and $g_{0i}$ evaluated along the spacelike curve $\mathcal{S}$. We note that for $\Delta\tau_S>0$ the cw-beam arrives before the ccw-beam. The opposite situation occurs for negative Sagnac time delays $\Delta\tau_S<0$. 

\subsubsection{Form invariance\label{SecFormInvariance}}

It is not difficult to show that the Sagnac time delay, given by eq.~(\ref{GenSagnacTimeDelay}), is form invariant\footnote{We can understand this form invariance on a deeper level by making use of the geometrical derivation of the Sagnac time delay provided by Ashtekar and Magnon in \cite{Ashtekar75} together with the three-dimensional formalism of Geroch \cite{Geroch71} for spacetimes endowed with a Killing vector field $\xi^{\mu}(x^\sigma)$. In our derivation of the Sagnac time delay, we started from a stationary metric and utilized adapted coordinates in which this metric is time independent. In this case, the corresponding Killing vector field reads $\xi^{\mu}(x^\sigma)=(1,0,0,0)$, see appendix \ref{AppSymmetries}.} under the special class of coordinate transformations
\begin{equation}
x'^0=x'^0(x^0,x^k)\,,\quad x'^i=x'^i(x^k)\,,
\label{SpecialTrafo}
\end{equation}
which also satisfy the additional condition
\begin{equation}
\pdiff{x^0}{x'^0}\bigg|_\mathcal{S}=\text{const}>0 \quad\forall\quad x^i(\phi)\in \mathcal{S}\,.
\label{SpecialTrafo2}
\end{equation}  
These coordinate transformations neither change the frame of reference nor the direction of the arrow of time. In particular, purely spatial coordinate transformations belong to this class.

\section{Coordinates appropriate for local satellite experiments\label{SecProperRefCoord}}

In the preceding section we derived an expression for the Sagnac time delay $\Delta\tau_S$ in terms of the metric coefficients. Two different physical effects contribute to $\Delta\tau_S$: inertia and gravitation.
Purely inertial effects depend only on the acceleration and rotation of the chosen reference frame of the observer and can in principle be completely eliminated by performing the measurement in an appropriately adapted reference frame. Gravitational terms on the other hand originate from the curvature of spacetime itself and cannot be globally removed by choosing a different frame of reference. In order to identify the origin of these different effects, we choose a certain class of local coordinates that define the so-called {\it proper reference frame} of the observer. In the present section we lay the foundations for the subsequent analysis of the Sagnac effect by establishing proper reference frame coordinates and the corresponding metric expansion.

\subsection{Motivation\label{AppMotiviationProper}}

The Earth has approximately the shape of an oblate ellipsoid and despite its curvature, Euclidean geometry works quite well for distance measurements on its surface as long as they are restricted to sufficiently small regions. The same property holds true also for curved spacetime. Indeed, in a sufficiently small region around a fixed point $P$ in spacetime the metric appears to be flat and all laws of nature can be reduced to their special-relativistic form. Riemann developed the adequate mathematical formalism \cite{Riemann1854} and thereby established the so-called Riemann normal coordinates \cite{Eisenhart64,Synge,AlvarezGaume81,Hatzinikitas00}. They constitute a first step towards the definition of the proper reference frame. For the sake of completeness, we provide an introduction to Riemann normal coordinates in appendix~\ref{AppendixSpecialCoordinates}.

The second important step was initiated by the development of general relativity. According to the equivalence principle all physical experiments performed by a freely falling and non-rotating observer in his local spatial neighborhood should lead to the same outcome as if they would have been performed in flat Minkowski spacetime. Thus, physical intuition suggests that it should always be possible to introduce coordinates, such that the transformed metric reduces to a flat spacetime metric for all points on the geodesic of the freely falling observer.
However, Riemann normal coordinates just guarantee a flat spacetime metric in a sufficiently small region around a {\it single} spacetime point $P$ and not along the {\it whole geodesic}. For this reason it was not obvious in the early days of general relativity, whether the intuitive notion of the equivalence principle mentioned above could be put on a rigorous mathematical footing.

It was the young Enrico Fermi \cite{Fermi22,Jantzen02} who made the next important contribution by showing that it is always possible to introduce local coordinates around any given spacetime curve in such a way, that the Christoffel symbols vanish along this curve and the metric takes its Minkowski form there. Inspired by his work, many investigations followed, in particular the seminal article by Manasse and Misner \cite{Manasse63,WannQuan79b}. In order to deal with the freely falling observer they specialized earlier ideas of Fermi and Synge \cite{Synge} to what they called ``Fermi normal coordinates''. These coordinates can be regarded as a natural generalization of Riemann normal coordinates. However, they also correspond to a limiting case of proper reference frame coordinates, as will be seen later.

Since our ultimate goal is the theoretical description of the Sagnac time delay measured in a satellite based experiment it is necessary to extend these considerations to a non-geodesic motion and to allow for a possible rotation of the observer. The coordinates most suitable for such a situation have been established by Ehlers \cite{Ehlers73}, and Misner, Thorne and Wheeler~\cite{MTW}. They are called the local coordinates of the ``proper reference frame''. With their help it is possible to identify the different contributions which arise  in the Sagnac time delay, eq.~\eqref{GenSagnacTimeDelay}. We now define these coordinates and present the corresponding expansion of the metric around the world line of the observer in these coordinates \cite{Nesterov76,Ni78,Li79a,Marzlin94,Nesterov99}.

\subsection{Construction of coordinates \label{ProperRef}}

Following Ehlers \cite{Ehlers73}, and Misner, Thorne and Wheeler \cite{MTW}, we now introduce the local coordinates of the proper reference frame\footnote{Unfortunately the name of these coordinates varies in the literature. For example, in \cite{Marzlin94,Nesterov99} they are called ``Fermi coordinates''.} for an observer moving along an arbitrary world line and carrying with him ``spatial coordinate axes'' which rotate. 

\subsubsection{Building blocks}

We denote the world line of the observer by $p^\mu(\tau)$ and use his proper time $\tau$ as curve parameter, giving rise to the four-velocity 
\begin{equation*}
u^\mu(\tau)=\diff{p^\mu}{\tau}
\end{equation*} 
and to the four-acceleration
\begin{equation*}
a^\mu(\tau)=u^\mu_{\;\,;\nu}\,u^\nu\,.
\end{equation*} 

In order to identify spatial directions, the observer carries with him three spacelike vectors $e^\mu_{\,\basind{i}}(\tau)$, where ${(i)\in\{1,2,3\}}$ labels the individual basis vector. It is reasonable to attach the timelike tangent vector
\begin{equation}
 e^\mu_{\,\basind{0}}(\tau)=\frac{1}{c}\,u^\mu(\tau)
\label{e0}
\end{equation} 
to the latter, which completes the four-tetrad $e^\mu_{\,\basind{\alpha}}(\tau)$ with ${(\alpha)\in\{0,1,2,3\}}$. For a brief introduction to the tetrad formalism we refer to appendix~\ref{AppTetrad}. 

In order to ensure the uniqueness of the construction of the coordinates, we need to add (i) the relativistic orthogonality condition 
\begin{equation}
 e^\mu_{\,\basind{\alpha}}(\tau)\, e^\nu_{\,\basind{\beta}}(\tau)\,g_{\mu\nu}(p^\sigma(\tau))
=\eta_{\basind{\alpha\beta}}
\label{RelOrthogonality}
\end{equation}
for all $p^\mu(\tau)$, and (ii) the transport law 
\begin{equation}
e^\mu_{\;\basind{\alpha};\nu}\, u^\nu=-\Omega^{\mu}_{\;\;\nu}\,e^{\nu}_{\;\basind{\alpha}}
\label{ProperTransport}
\end{equation} 
introduced in appendix~\ref{AppTetradTransport}. 
The diagonal matrix $\eta_{\basind{\alpha\beta}}=\op{diag}(1,-1,-1,-1)$ in eq.~(\ref{RelOrthogonality}) resembles the Minkowski metric with invariant tetrad indices. The first term in the antisymmetric transport matrix
\begin{equation}
\Omega^{\mu\nu}=-\frac{1}{c^2}\left(a^\mu u^\nu- a^\nu u^\mu \right)+\frac{1}{c}\,u_\rho\, \omega_\sigma \,
\varepsilon^{\rho\sigma\mu\nu}
\label{TransportMatrix}
\end{equation}
entering eq.~(\ref{ProperTransport}) contains the \mbox{four-velocity} ${u^\mu(\tau)}$ and the \mbox{four-acceleration} ${a^\mu(\tau)}$ of the observer and represents the Fermi-Walker transport of the tetrad along $p^\mu(\tau)$. The second expression $u_\rho\, \omega_\sigma \,\varepsilon^{\rho\sigma\mu\nu}/c$ characterizes the rotation of the spatial tetrad vectors $e^\mu_{\,\basind{i}}(\tau)$ in the subspace orthogonal to the four-velocity $u^\mu(\tau)$. Therefore, the identity~\eqref{e0} is automatically preserved by the transport eq.~(\ref{ProperTransport}) for all points on the world line. 

The transport law~(\ref{ProperTransport}) constitutes a natural generalization of the Fermi-Walker transport. In order to highlight its significance in the definition of proper reference frame coordinates, we call eq.~(\ref{ProperTransport}) the {\it proper transport law}.

\subsubsection{Exploration of the spatial neighborhood with spacelike geodesics}

\begin{figure}[h]
\centering
\includegraphics[height=0.35\textheight]{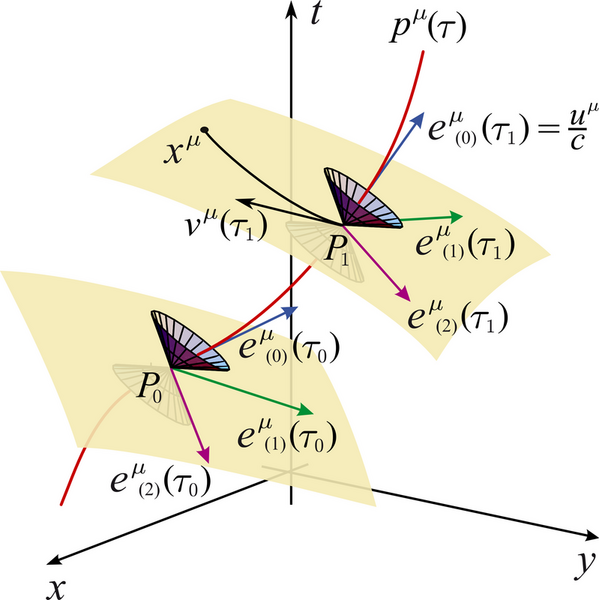}
\caption{Construction of proper reference frame coordinates $x^{\basind{\sigma}}$ for the neighborhood of the world line $p^\mu(\tau)$ of an accelerating and rotating observer with proper time $\tau$. The initial tetrad $e^\mu_{\,\basind{\alpha}}(\tau_0)$ at the point $P_0$ is properly transported according to eq.~(\ref{ProperTransport}) along $p^\mu(\tau)$ to $P_1$, resulting in the tetrad $e^\mu_{\,\basind{\alpha}}(\tau_1)$. The ``infinitesimal'' light cones at $P_0$ and $P_1$ illustrate the timelike and spacelike regions of the tangent space attached to these points. We now define proper reference frame coordinates geometrically in two steps: (i) we follow the world line from $p^\mu(\tau_0)$ to $p^\mu(\tau_1)$ and (ii) draw a unique spacelike geodesic 
${x^\mu=x^\mu(\tau,v^{\basind{i}} s)}$ from the point $p^\mu(\tau_1)$ to $x^\mu$ such that the initial tangent vector $v^\mu(\tau_1)$ is orthogonal to the four-velocity $u^\mu(\tau_1)$. For $\tau_0=0$, the proper reference frame coordinates $x^{\basind{\alpha}}$ which correspond to the spacetime point $x^\mu$ are given by the proper time $x^{\basind{0}}=c\,\tau_1$ and by the scaled tetrad components of the initial tangent vector $x^{\basind{i}}=v^{\basind{i}}(\tau_1)\,s$.}
\label{FigProperReferenceFrame}
\end{figure}

We now proceed with the explicit construction of proper reference frame coordinates $x^{\basind{\alpha}}$ shown in fig.~\ref{FigProperReferenceFrame}. We define the time coordinate $x^{\basind{0}}=c\,\tau$ as in terms of the proper time $\tau$ measured by the clock of the accelerated observer along $p^\mu(\tau)$. 
In order to define the spatial coordinates $x^{\basind{i}}$ we introduce the tangent vector 
\begin{equation*}
v^\mu(\tau)=v^{\basind{i}}(\tau)\,e^\mu_{\,\basind{i}}(\tau)
\end{equation*} 
with the additional normalization condition
\begin{equation*}
(v^{\basind{1}})^2+(v^{\basind{2}})^2+(v^{\basind{3}})^2=1 \,.
\end{equation*}
By construction, this tangent vector is orthogonal to the four-velocity ${u^\mu(\tau)=c\,e^\mu_{\,\basind{0}}(\tau)}$ of the observer at $p^\mu(\tau)$. 

When we now draw {\it spacelike geodesics} ${x^\mu=x^\mu(\tau,v^{\basind{i}} s)}$ from the initial point $p^\mu(\tau)$ in all spacelike directions $v^\mu(\tau)$ orthogonal to $u^\mu(\tau)$, we are able to explore the spatial neighborhood of the point $p^\mu(\tau)$. We employ the arclength $s$ as curve parameter of the spacelike geodesics. For a sufficiently small spatial neighborhood around $p^\mu(\tau)$, there exists a one-to-one correspondence between the tetrad components 
${v^{\basind{i}}(\tau)\, s}$ of the scaled initial tangent vector $v^\mu(\tau)\, s$ and the spacetime point $x^\mu$, which we would like to express in proper reference frame coordinates. Hence, the simplest idea is to identify the spatial coordinates $x^{\basind{i}}$ of the proper reference frame with the tetrad components ${v^{\basind{i}}\, s}$ of the initial tangent vector. According to this construction, the connection between proper reference frame coordinates $x^{\basind{\alpha}}$ and the original coordinates $x^\mu$ is established by inserting ${x^{\basind{0}}=c\,\tau}$ and ${x^{\basind{i}}=v^{\basind{i}} s}$ into the spacelike geodesics
\begin{equation}
x^\mu=x^\mu(\tau,v^{\basind{i}} s)=x^{\mu}(x^{\basind{0}}/c\,,\,x^{\basind{i}})\,.
\label{ProperRefGeodesicDef}
\end{equation}
Appendix \ref{AppProperRef} explores this coordinate transformation in more detail by making use of a formal expression for the spacelike geodesic given by eq.~(\ref{ProperRefGeodesicDef}).

We conclude by briefly recapitulating this geometrical construction using fig.~\ref{FigProperReferenceFrame}. Suppose, we want to assign local coordinates to the point $x^\mu$ in the spatial neighborhood of $P_1$. For simplicity, we take the origin of our proper reference frame coordinates to be the initial point $P_0$, which implies $\tau_0=0$. From $P_0$ we follow the world line $p^\mu(\tau)$ until we are able to draw a unique, spacelike geodesic ${x^\mu=x^\mu(\tau,v^{\basind{i}} s)}$ from a point on the world line to $x^\mu$. In fig.~\ref{FigProperReferenceFrame} this point is represented by $P_1$ with coordinates $p^\mu(\tau_1)$ and the initial tangent vector $v^\mu(\tau_1)$ of the spacelike geodesic is assumed to be orthogonal to the four-velocity $u^\mu(\tau_1)$. Then, the local coordinate time corresponding to the point $x^\mu$ reads ${x^{\basind{0}}=c\,\tau_1}$. On the other hand, the spatial coordinates $x^{\basind{i}}$ correspond to the tetrad components $v^{\basind{i}}(\tau_1)\,s$ of the scaled initial tangent vector ${v^\mu(\tau_1)\,s=(v^{\basind{i}}(\tau_1)\,s)\,e^\mu_{\,\basind{i}}(\tau_1)}$. 

\subsubsection{Caveat emptor}

We note, that these local coordinates are only valid within a sufficiently small region $\mathcal{D}$ around the world line $p^\mu(\tau)$ of the observer. This region ensures the one-to-one correspondence between the coordinates of the spacetime point ${x^\mu\in\mathcal{D}}$ and the tetrad components of the scaled, initial tangent vector $v^{\basind{i}}(\tau_1)\,s$ at $p^\mu(\tau)$. However, the curvature of spacetime can cause two spacelike geodesics with different initial conditions to coincide in a spacetime point $y^\mu$. In this case, the one-to-one correspondence breaks down and ${y^\mu\notin\mathcal{D}}$. For the sake of simplicity, we restrict ourselves for the remainder of these notes to spacetime points $x^\mu\in\mathcal{D}$.

Moreover, as already pointed out by L.~Synge~\cite{Synge}, a spacelike geodesic is a somewhat artificial object when considered from the operational point of view. Indeed, spacelike geodesics are not immediately linked to physically accessible objects such as light rays or timelike world lines of massive particles -- disregarding the conceptional difficulties which arise due to the idealization one usually makes in the descriptions of light rays and world lines. However, the analysis of lightlike or timelike geodesics in proper reference frame coordinates offers a possibility to establish such a relation between spacelike geodesics and physically accessible objects. Appendix~\ref{AppRadarDistance} therefore provides an approximate solution to the geodesic equation in the spatial neighborhood of the observer's world line.

\subsection{Metric expansion}

We now return to the discussion on the suitability of proper reference frame coordinates for local satellite experiments, alluded to already at the end of the motivation in~\ref{AppMotiviationProper}. In particular, we provide the power-series expansion of the metric coefficients around the world line of the observer which in proper reference frame coordinates reads ${p^{\basind{\sigma}}(\tau)=(c\,\tau,0,0,0)}$.

\subsubsection{Leading-order contributions}

The power-series expansion of the metric coefficients for the spatial neighborhood around the world line  $p^{\basind{\sigma}}(\tau)$ is then carried out in terms of the spatial coordinates $x^{\basind{i}}$, and takes the form
\begin{equation}
g_{\basind{\mu \nu}}(x^{\basind{\sigma}})=g_{\basind{\mu\nu}}(p^{{\basind{\sigma}}}) +\sum_{n=1}^\infty \frac{1}{n!}\, g_{\basind{\mu \nu},\basind{i_1},\ldots,\basind{i_n}}(p^{{\basind{\sigma}}})\, x^{\basind{i_1}}\cdot\ldots\cdot x^{\basind{i_n}} \,.
\label{ProperMetricExpans}
\end{equation}
We emphasize, that the expansion coefficients 
${g_{\basind{\mu \nu},\basind{i_1},\ldots,\basind{i_n}}(p^{{\basind{\sigma}}})}$ still depend on the coordinate time $x^{\basind{0}}$ through ${p^{\basind{\sigma}}=p^{\basind{\sigma}}(\tau)}$. 

The acceleration and the rotation of the observer crucially affect the outcome of local experiments within the satellite. In Newtonian mechanics these effect arise from fictitious, inertial forces. However, in general relativity inertial forces are treated on the same footing as gravitation -- they are both absorbed in the metric of spacetime. But since we are dealing with a metric expansion in the rest frame of our accelerating and rotating observer using proper reference frame coordinates, we expect the four-acceleration $a^{\basind{\mu}}$ and the tetrad rotation vector $\omega^{\basind{\mu}}$ to enter the expansion coefficients ${g_{\basind{\mu \nu},\basind{i_1}}(p^{{\basind{\sigma}}})}$ 
and ${g_{\basind{\mu \nu},\basind{i_1},\basind{i_2}}(p^{{\basind{\sigma}}})}$. As discussed in appendix~\ref{AppProperRef}, the zeroth components of the four-acceleration $a^{\basind{\mu}}$ and of the four-vector $\omega^{\basind{\mu}}$ vanish, that is $a^{\basind{0}}=0$ and $\omega^{\basind{0}}=0$.
As a consequence, the spatial components $a^{\basind{i}}$ and $\omega^{\basind{i}}$ are the only parameters which characterize the acceleration and rotation of the observer in the metric expansion.

For the purpose of Sagnac interferometry within a satellite, it suffices to focus on the first two leading terms in eq.~(\ref{ProperMetricExpans}). In appendix~\ref{AppProperRef} we derive the expressions
\begin{align}
g_{\basind{0 0}}(x^{\basind{\sigma}})=&1
-\frac{2}{c^2}\,a_{\basind{i_1}} x^{\basind{i_1}}
+R_{\basind{0}\basind{i_1}\basind{i_2}\basind{0}}(p^{\basind{\sigma}})\,
x^{\basind{i_1}} x^{\basind{i_2}}
\label{ProperMetricExpansion}\\
&\phantom{1}+\frac{1}{c^2}\left(\frac{1}{c^2}\, a_{\basind{i_1}} a_{\basind{i_2}}
 +\omega_{\basind{i_1}} \omega_{\basind{i_2}}
 -\omega^{\basind{l}}\omega_{\basind{l}}\,\eta_{\basind{i_1 i_2}}\right)\,
x^{\basind{i_1}} x^{\basind{i_2}}
+\mathcal{O}(x^3)\,,
\nonumber\\
g_{\basind{0 k}}(x^{\basind{\sigma}})=&
\frac{1}{c}\,\varepsilon_{\basind{0  k l i_1}}\,
\omega^{\basind{l}} x^{\basind{i_1}} + \frac{2}{3}\,
R_{\basind{0}\basind{i_1}\basind{i_2}\basind{k}}(p^{\basind{\sigma}})\, x^{\basind{i_1}} x^{\basind{i_2}}
+\mathcal{O}(x^3)\,,
\nonumber\\
g_{\basind{j k}}(x^{\basind{\sigma}})=&-\delta_{\basind{j k}}
+\frac{1}{3}\,R_{\basind{j}\basind{i_1}\basind{i_2}\basind{k}}(p^{\basind{\sigma}})\, x^{\basind{i_1}} x^{\basind{i_2}}
+\mathcal{O}(x^3)\,.
\nonumber
\end{align}
We briefly illustrate the notation by two examples. For this purpose, we first note that the spatial co- and contravariant components of $\omega_{\basind{i}}(\tau)$ and $\omega^{\basind{i}}(\tau)$ differ by a minus sign, since the raising and lowering of the indices is carried out by the Minkowski metric $\eta_{\basind{\alpha\beta}}$ along the observer's world line. As a consequence, we find the relations
\begin{equation*}
\omega^{\basind{l}}\omega_{\basind{l}}\,\eta_{\basind{i_1 i_2}}
=-\omega^{\basind{l}}\omega_{\basind{l}}\,\delta_{\basind{i_1 i_2}}
=\boldsymbol{\omega}^2\,\delta_{\basind{i_1 i_2}}\quad\text{and}\quad
\varepsilon_{\basind{0  k l i_1}}\,
\omega^{\basind{l}} x^{\basind{i_1}}=-(\boldsymbol{\omega}\times \mathbf{x})_{\basind{k}}
\end{equation*}
by identifying the spatial coordinates $x^{\basind{i}}$ and the non-vanishing vector components $\omega^{\basind{i}}$ with $\mathbf{x}$ and $\boldsymbol{\omega}$.
In the second expression, we have made use of the correspondence~(\ref{EpsilonTetradForm2}) between the covariant components of the antisymmetric tensor $\varepsilon_{\basind{\alpha \beta \gamma \delta}}$ in proper reference frame coordinates and the Levi-Civita symbol $\Delta^{\alpha \beta \gamma \delta}$.

\subsubsection{Special cases of proper reference frame coordinates\label{SpecialCases}}

Before we proceed, we briefly review three examples of proper reference frame coordinates: 

(i) An observer moving along a geodesic and carrying with him a the Fermi-Walker-transported tetrad $e^\mu_{\;\basind{\alpha}}(\tau)$ represents the most elementary case. Indeed, we know from the geodesic equation and from the definition of the Fermi-Walker transport, discussed in appendix~(\ref{AppTetradTransport}), that the four-acceleration and the spatial rotation of the tetrad vanish, that is $a^{\basind{i}}=0$ and $\omega^{\basind{i}}=0$.
Hence, all first-order contributions in the expansion of the metric, eq.~(\ref{ProperMetricExpansion}), disappear, which implies that proper reference frame coordinates constitute the local coordinates of a freely falling, inertial observer in this case. The only correction to the metric of flat spacetime originates from the components of the curvature tensor in the second-order. In particular, these terms represent the gravitational field gradients acting in the neighborhood of the inertial observer. This special case corresponds to the so-called Fermi normal coordinates of \cite{Manasse63}.

(ii) We now consider an accelerated observer whose tetrad is still Fermi-Walker transported giving rise to $a^{\basind{i}}\neq 0$ and $\omega^{\basind{i}}=0$. In this case we obtain a first-order contribution~${-2\,a_{\basind{i_1}} x^{\basind{i_1}}}/c^2$ to the metric coefficient ${g_{\basind{0 0}}(x^{\basind{\sigma}})}$, as well a second-order term. The first-order expression is the major contribution to the gravitational frequency shift measured between a light source and an accelerated observer along his world line. Here we refer to the familiar red shift experiments \cite{Pound59} using the M\"ossbauer effect in the accelerated frame of an earth-bound laboratory.

(iii) Finally, for an observer who is accelerating as well as rotating, that is $a^{\basind{i}}\neq 0$ and $\omega^{\basind{i}}\neq 0$, we also encounter the first-order terms ${\varepsilon_{\basind{0  k l i_1}}\,\omega^{\basind{l}} x^{\basind{i_1}}}/c$ in the metric coefficients ${g_{\basind{0 k}}(x^{\basind{\sigma}})}$. As shown in the next section, these terms account for the leading-order contribution of the Sagnac time delay between two counter-propagating light rays. For this reason, the standard literature calls an observer {\it non-rotating} if his tetrad vectors are Fermi-Walker transported along the world line such that $\omega^{\basind{i}}=0$. 

We emphasize, that rotation as well as acceleration are in this way {\it absolute} quantities \cite{Ehlers73} and not relative ones. Hence, they provide a coordinate independent characterization of the observer's state of motion, or equivalently, of the gravitational field acting in its immediate local neighborhood. This fact expresses itself in coordinate independent values of the tetrad rotation $\omega^{\basind{i}}$ and of the four-acceleration $a^{\basind{i}}$. The question, how to use a Sagnac interferometer to decide whether an observer is rotating or not, will be addressed in the next section.

\section{Sagnac time delay in a proper reference frame \label{ProperRefSagnac}}

The expression for the Sagnac time delay, eq.~(\ref{GenSagnacTimeDelay}), suffers from a dilemma frequently encountered in general relativity. It contains an implicit dependence on the coordinates used for the description of the experiment. Since coordinates have no immediate physical meaning in general relativity -- unless they are operationally defined -- our formula for the Sagnac time delay~(\ref{GenSagnacTimeDelay}) does {\it not} provide a direct relationship between measurable quantities on both sides of the equation. For this reason, it is necessary to perform additional measurements which define the underlying coordinate system. Only under this condition, a measurement of the Sagnac time delay is capable of determining unknown parameters in the metric under consideration\footnote{This point is illustrated for the Sagnac time delay in G\"odel's Universe in \cite{Kajari04}.}. 

These arguments suggest the use of proper reference frame coordinates as a tool to circumvent the problem of coordinate dependence. As discussed in the previous section, they are defined in terms of (i) the proper time measured by the observer, and (ii)~spacelike geodesics which emerge from his world line. Due to their unique geometric construction, these coordinates constitute invariants under general coordinate transformations. Consequently, the Sagnac time delay and the unknown parameters entering the metric are connected in an invariant way. This invariant formulation stands out most clearly when we restrict ourselves to a sufficiently small spatial region around the world line of the observer. Here we can take advantage of the power-series expansion of the metric coefficients~(\ref{ProperMetricExpansion}) in proper reference frame coordinates.

In this section, we first setup the machinery for the description of the Sagnac time delay $\Delta\tau_S$, eq.~(\ref{GenSagnacTimeDelay}), within a proper reference frame. We then discuss the first two leading orders in the expansion of $\Delta\tau_S$ and analyze the influence of inertial and gravitational effects. We conclude with a comparison of some measurement schemes which allow for the determination of the tetrad rotation vector $\omega^{\basind{i}}$ and several coefficients of the curvature tensor.

\subsection{Framework for the Sagnac time delay measurement}

\begin{figure}
\hspace{0.07\textwidth}
\includegraphics[width=0.37\textwidth]{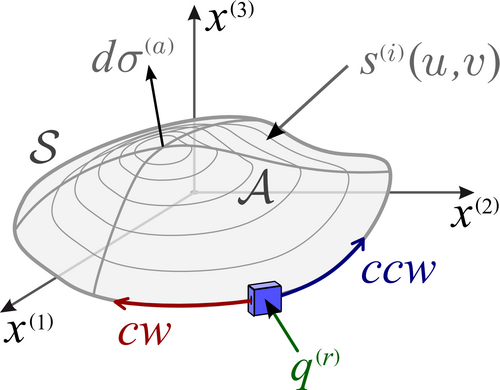}
\hfill
\includegraphics[width=0.35\textwidth]{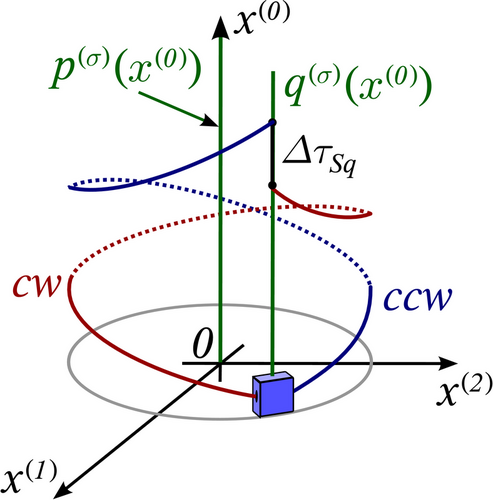}
\hspace{0.07\textwidth}
\caption{Measurement of the Sagnac time delay within a proper reference frame depicted in space (left) and spacetime (right). In the spatial diagram (left), an observer located at the fixed position $q^{\basind{r}}$ emits two counter-propagating light rays along the spatial curve $\mathcal{S}$, which enclose an arbitrarily shaped surface~$\mathcal{A}$. The parameters $u,v\in\mathds{R}$ span $\mathcal{A}$ by $s^{\basind{i}}(u,v)$ giving rise to the infinitesimal surface normal $d\sigma^{\basind{a}}$. 
In the spacetime diagram (right), the first observer moves along the world line $p^{\basind{\sigma}}(x^{\basind{0}})$ and defines the proper reference frame coordinates according to the procedure discussed in \ref{ProperRef}. The proper time delay $\Delta\tau_{Sq}$ between the arrivals of the two light rays is measured by the second observer along his world line~$q^{\basind{\sigma}}(x^{\basind{0}})=(x^{\basind{0}},q^{\basind{r}})$. Since both observers are spatially separated, they have to take into account the redshift factor $\sqrt{g_{\basind{00}}(q^{\basind{r}})}$ when comparing their proper times $\tau_p$ and $\tau_q$ for a time independent metric.}
\label{FigSagnacProper}
\end{figure}

As illustrated in the left picture of fig.~\ref{FigSagnacProper}, we suppose that the counter-propagating light rays, which emerge from the fixed position $q^{\basind{r}}$, travel along the positively oriented, closed spatial curve $\mathcal{S}$. We parameterize $\mathcal{S}$ in proper reference frame coordinates by $s^{\basind{i}}(\phi)$. The surface $\mathcal{A}$ of arbitrary shape is bounded by $\mathcal{S}$. We parameterize $\mathcal{A}$ in terms of the variables $u,v\in \mathds{R}$ via $s^{\basind{i}}(u,v)$. In particular, the infinitesimal surface normal $d\sigma^{\basind{a}}$, defined by the covariant components 
\begin{equation*}
\D \sigma_{\basind{a}}=\varepsilon_{\basind{0 a m n}}\pdiff{s^{\basind{m}}}{u}
\pdiff{s^{\basind{n}}}{v}\,\D u\,\D v\,
\end{equation*} 
obeys the right-hand rule with respect to the circulation resulting from $s^{\basind{i}}(\phi)$.

In general, we need two observers in order to measure the Sagnac time delay in an invariant way. Whereas the first one is responsible for the construction of proper reference frame coordinates, the second one measures the Sagnac time delay $\Delta\tau_{Sq}$. As indicated in the right picture of fig.~\ref{FigSagnacProper}, we denote the world line of the first observer by ${p^{\basind{\sigma}}(x^{\basind{0}})=(x^{\basind{0}},0,0,0)}$. The coordinate time $x^{\basind{0}}$ is related to his proper time $\tau_p$ by $x^{\basind{0}}=c\,\tau_p$. 
The second observer is at rest at the fixed spatial position $q^{\basind{r}}$ and moves 
along the world line ${q^{\basind{\sigma}}(x^{\basind{0}})=(x^{\basind{0}},q^{\basind{1}},q^{\basind{2}},q^{\basind{3}})}$. 
He finds that his proper time $\tau_q$ is related to the coordinate time $x^{\basind{0}}$ by
\begin{equation}
\tau_q=\frac{1}{c}\int \limits_0^{x^{\basind{0}}} 
\sqrt{g_{\basind{0 0}}(q^{\basind{\sigma}})}\,\D q^{\basind{0}}\,   
\label{eq:tau_q}\,,
\end{equation}
in contrast to the first observer who directly defines the global coordinate time $x^{\basind{0}}$ by his proper time $\tau_p$.

The derivation of the Sagnac time delay presented in \ref{SecDerivationTimeDelay} assumes that the stationary metric does not depend on the time coordinate in the underlying reference frame. For this reason we have to make an important assumption concerning the first observer: his acceleration $a^{\basind{i}}(\tau_p)$ and the rotation vector $\omega^{\basind{i}}(\tau_p)$ of his spatial tetrad {\it should not change considerably} during the time it takes to perform the Sagnac time delay measurement. In this case, eq.~(\ref{eq:tau_q}) reduces to
\begin{equation}
 \tau_q=\frac{1}{c}\,\sqrt{g_{\basind{0 0}}(q^{\basind{r}})}\;x^{\basind{0}}=
\sqrt{g_{\basind{0 0}}(q^{\basind{r}})}\;\tau_p\,,
\label{CompPropertimes}
\end{equation}
which implies that the proper times $\tau_p$ and $\tau_q$ of both observers differ from each other just by the redshift factor 
$\sqrt{g_{\basind{0 0}}(q^{\basind{r}})}$.

\subsection{Leading-order contributions of the Sagnac time delay}

So far, we have illustrated the measurement scheme for the Sagnac time delay. We now continue with a discussion of the first two leading-order contributions of the Sagnac time delay in proper reference frame coordinates. Since we want to focus on the essential results, we have moved the detailed calculations to appendix~\ref{CalcInPropRef}. In this appendix we derive a formally exact expression for $\Delta\tau_{Sq}$ as a series expansion in moments of ``unit fluxes'', and partial derivatives of the metric coefficients evaluated along the world line ${p^{\basind{\sigma}}(x^{\basind{0}})}$.

According to the original formula, eq.~(\ref{famsagnac}), the Sagnac time delay
crucially depends on the scalar product $\mathbf{A}\cdot\boldsymbol{\Omega}$ between the area vector and the angular velocity. In order to establish an analogous expression within the framework of general relativity, we introduce the zeroth and the first moments of the unit fluxes
\begin{equation*}
A_{\basind{a}}= \iint\limits_{\mathcal{A}}  \D \sigma_{\basind{a}}
\quad\text{and}\quad
A_{\basind{a}}^{\;\;\basind{i_1}}= \iint\limits_{\mathcal{A}}\, s^{\basind{i_{1}}}  \D \sigma_{\basind{a}} \,.
\end{equation*}
The definition of the corresponding higher-order moments $A_{\basind{a}}^{\;\;\basind{i_1}\ldots\basind{i_n}}$ is straight forward.
In the general relativistic analogue for the Sagnac time delay the contravariant components $A^{\basind{a}}$ will replace the area vector $\mathbf{A}$ of the original eq.~(\ref{famsagnac}). 

The derivation of the first two leading-order contributions basically relies on (i) the substitution of the metric expansion, eq.~(\ref{ProperMetricExpansion}), into the Sagnac time delay, eq.~(\ref{GenSagnacTimeDelay}), and (ii) on the subsequent application of Stokes' theorem.
As shown in appendix~\ref{CalcInPropRef}, we obtain the invariant characterization of the 
Sagnac time delay
\begin{equation}
\begin{split}
 \Delta\tau_{Sq}=\frac{4}{c^2}\,\sqrt{g_{\basind{00}}(q^{\basind{r}})}\,
&\left[-\omega^{\basind{a}} A_{\basind{a}}+
\frac{2 c}{3}\,\varepsilon^{\basind{0 a j k}}\, 
R_{\basind{0}\{\basind{i_1}\basind{j}\}\basind{k}}(p^{\basind{r}})\,A_{\basind{a}}^{\;\;\basind{i_1}}
\right.\\
&+\left.
\frac{1}{c^2}\left(\omega^{\basind{l}} a_{\basind{l}}\,\delta^{\basind{a}}_{\basind{i_1}}-
3\,\omega^{\basind{a}} a_{\basind{i_1}}\right) A_{\basind{a}}^{\;\;\basind{i_1}}\right]
+\mathcal{O}\Big(A_{\basind{a}}^{\;\;\basind{i_1}\basind{i_2}}\Big)
\label{LeadingOrdersSagnacTime}
\end{split}
\end{equation}
which provides an adequate generalization of the original expression, eq.~(\ref{famsagnac}), within the framework of general relativity. Here we have added the additional subscript $q$ to the Sagnac time delay $\Delta \tau_S$ given by eq.~(\ref{GenSagnacTimeDelay}) to express the fact, that the measurement is performed by the observer with world line $q^{\basind{\sigma}}(x)$.

Clearly, the main contribution arises from the first term in the brackets. We call this term the zeroth-order contribution due to its dependence on the zeroth moment of the unit fluxes $A_{\basind{a}}$. The additional factor $\sqrt{g_{\basind{00}}(q^{\basind{r}})}$ which is not present in the original formula stems from the different proper times, eq.~(\ref{CompPropertimes}), measured by the two observers at different positions.
Moreover, we encounter several components of the curvature tensor, as well as the traceless matrix 
${\omega^{\basind{l}} a_{\basind{l}}\,\delta^{\basind{a}}_{\basind{i_1}}-
3\,\omega^{\basind{a}} a_{\basind{i_1}}}$ containing the acceleration $a^{\basind{i}}$ and the tetrad rotation vector $\omega^{\basind{i}}$. Both of these terms appear within the first-order contribution corresponding to the first moments $A_{\basind{a}}^{\;\;\basind{i_1}}$.

\subsection{Measurement strategies}

Despite of the mathematical machinery built up in the preceding sections, we have not yet given an operational definition of rotation within general relativity. In this subsection we show that the Sagnac time delay, eq.~(\ref{LeadingOrdersSagnacTime}), enables us to achieve this goal. 

For this purpose let us first suppose that we have the ability to Fermi-Walker transport the spatial tetrad vectors $e^\mu_{\,\basind{i}}(\tau_p)$ along the world line of the first observer. In this case the tetrad rotation vector $\omega^{\basind{i}}$ vanishes. Nevertheless, eq.~(\ref{LeadingOrdersSagnacTime}) still predicts a non-vanishing time delay $\Delta\tau_{Sq}$ between the arrival times of the counter-propagating light rays. This delay originates from the curvature of the spacetime itself, as well as from higher-order corrections. But how can we then decide {\it experimentally}, whether the tetrads are Fermi-Walker transported or not?

A first possibility is the method of the ``bouncing photon''~\cite{Synge} introduced by John L.~Synge and reformulated by Felix A.~E.~Pirani~\cite{Pirani65,Laemmerzahl01}, which uses light emitted by an observer and reflected back from a mirror in the immediate neighborhood of the observer. When the direction of the outgoing and incoming light ray coincide, the observer is not rotating.

The measurement of the Sagnac time delay, eq.~(\ref{LeadingOrdersSagnacTime}), using special surface configurations of the Sagnac interferometer, constitutes another approach. In the present section we pursue this idea. Moreover, we show how to change the experimental setup in order to measure several components of the curvature tensor.

\subsubsection{Rotation sensor\label{SecRotSensor}}

\begin{figure}
\centering
\includegraphics[width=0.7\textwidth]{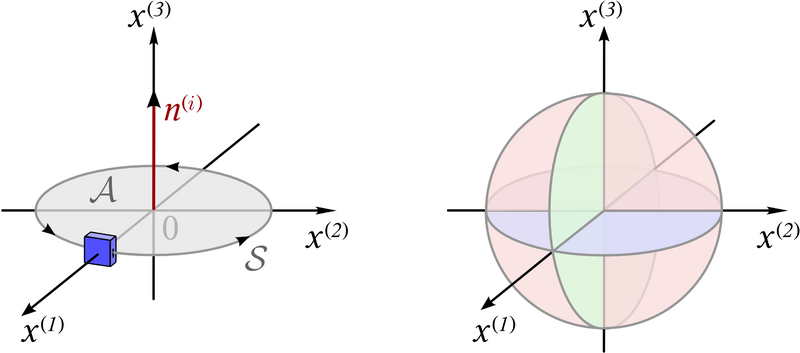}
\caption{Rotation sensor: measurement of the components $\omega^{\basind{i}}$ of the tetrad rotation vector using Sagnac interferometers. On the left we show a single circular interferometer in the $x^{\basind{1}}$-$x^{\basind{2}}$-plane, which allows us to measure the component $\omega^{\basind{3}}$. On the right we depict three orthogonal, circular interferometers. Each one provides information about one component of $\omega^{\basind{i}}$.}
\label{FigSagnacLeadingZerothOrder}
\end{figure}

We now apply eq.~(\ref{LeadingOrdersSagnacTime}) to a specific experimental setup. We assume that the spatial curve $\mathcal{S}$, which encloses a planar surface with area $\mathcal{A}$ in the \mbox{$x^{\basind{1}}$-$x^{\basind{2}}$}-plane, is symmetric under reflection with respect to the \mbox{$x^{\basind{1}}$-$x^{\basind{3}}$-} and \mbox{$x^{\basind{2}}$-$x^{\basind{3}}$}-plane. In the left picture of fig.~\ref{FigSagnacLeadingZerothOrder} this situation is exemplified with a circular path $\mathcal{S}$. 
In this case the contravariant components of the surface normal $\mathbf{n}$ read ${n^{\basind{a}}=(0,0,1)}$, and we obtain for the zeroth and the first moments of the unit fluxes
\begin{equation*}
A_{\basind{a}}=\mathcal{A}\,n_{\basind{a}}
\quad\text{and}\quad
A_{\basind{a}}^{\;\;\basind{i_1}}=n_{\basind{a}}\,\iint\limits_{\mathcal{A}} s^{\basind{i_1}}\; \D s^{\basind{1}} \D s^{\basind{2}}=0\,.
\end{equation*}
We note that, according to the second identity, the first moments of the unit fluxes $A_{\basind{a}}^{\;\;\basind{i_1}}$ allow for a simple interpretation when $\mathcal{A}$ is planar. For if we suppose, that the area $\mathcal{A}$ is filled with a homogeneous mass distribution, $A_{\basind{a}}^{\;\;\basind{i_1}}$ would just correspond to the product of the covariant components $n_{\basind{a}}$ of the surface normal and the contravariant components of the ``center of mass'' of $\mathcal{A}$.

When we now insert the previous expressions into eq.~(\ref{LeadingOrdersSagnacTime}), the Sagnac time delay reduces to
\begin{equation}
 \Delta\tau_{Sq}=-\frac{4}{c^2}\,\sqrt{g_{\basind{00}}(q^{\basind{r}})}\,
\omega^{\basind{a}} n_{\basind{a}}\,\mathcal{A}+
\mathcal{O}\Big(A_{\basind{a}}^{\;\;\basind{i_1}\basind{i_2}}\Big)
\label{SagnacRotSensor}
\,.
\end{equation}
With the help of the identity $-\omega^{\basind{a}} n_{\basind{a}}=\omega^{\basind{3}}$, we are thus able to determine the third component of the tetrad rotation vector $\omega^{\basind{i}}$ from the Sagnac time delay 
$\Delta\tau_{Sq}$, as far as the enclosed area $\mathcal{A}$ is known and the higher-order contributions are negligible. Similarly, by aligning the normal axis of the Sagnac interferometer along the $x^{\basind{1}}$ and $x^{\basind{2}}$-axes, we are able to find the remaining components $\omega^{\basind{1}}$ and $\omega^{\basind{2}}$. In summary, we can determine the tetrad rotation vector by considering three Sagnac interferometers with normal vectors aligned along the three coordinate axes, as depicted in the right part of fig.~\ref{FigSagnacLeadingZerothOrder}. In what follows, we call such a collection of Sagnac interferometers briefly a rotation sensor.

With this rotation sensor we are now in the position to provide an operational characterization of the Fermi-Walker transport of an observer along his world line: the tetrad of the observer undergoes Fermi-Walker transport when the Sagnac time delay $\Delta\tau_{Sq}$ vanishes for all three orthogonal orientations of the interferometer. In other words, we call the observer non-rotating only if all components of the tetrad rotation vector $\omega^{\basind{i}}$ vanish for this measurement procedure. Needless to say, this statement is only correct when we can confine ourselves to the first two leading orders of $\Delta\tau_{Sq}$ in eq.~(\ref{LeadingOrdersSagnacTime}).

\subsubsection{Curvature sensor}

We now turn to the question, how to determine individual components of the curvature tensor using Sagnac interferometry. For this purpose, it is necessary to ensure that the tetrad is Fermi-Walker transported in order to eliminate all contributions which arise from the tetrad rotation. According to the discussion of the previous subsection, we can achieve this goal by measuring the Sagnac time delay induced in the rotation sensor shown in fig.~\ref{FigSagnacLeadingZerothOrder}. Using dynamical feedback, we then appropriately realign the tetrad vectors of the observer in order to maintain $\omega^{\basind{i}}=0$ along the world line of the satellite. 

\begin{figure}[h]
\centering
\includegraphics[width=0.75\textwidth]{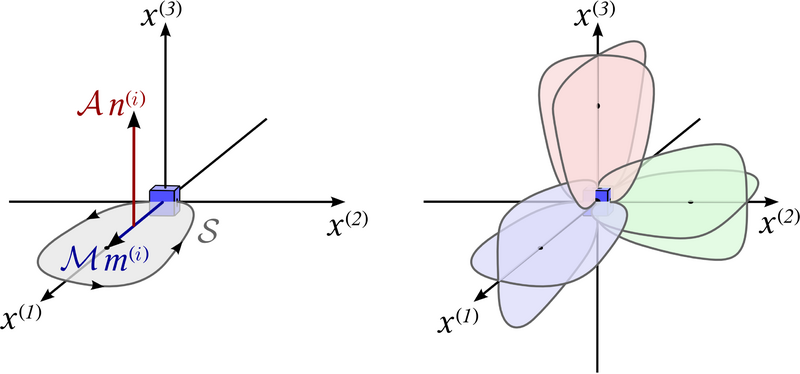}
\caption{Curvature sensor: measurement of a single (left) and six independent components (right) of the curvature tensor with the help of Sagnac interferometry. On the left we show an interferometer with a single loop allowing us to obtain a single component of the curvature tensor. This device requires only a single observer located at the origin. He defines the coordinates and measures the Sagnac time delay. The corresponding tetrad basis is Fermi-Walker transported with the help of the rotation sensor. The spatial curve $\mathcal{S}$ encloses a planar surface in the $x^{\basind{1}}$-$x^{\basind{2}}$-plane with normal vector $n^{\basind{a}}=(0,0,1)$. The curve is reflection symmetric with respect to the $x^{\basind{1}}$-$x^{\basind{3}}$-plane. Moreover, the ``center of mass'' of the area $\mathcal{A}$ is located at $\mathcal{M} m^{\basind{a}}$, with $m^{\basind{a}}$ being the unit vector pointing towards the direction of the position of the ``center of mass''. With this setup we are able to determine the component $R_{\basind{0}\basind{1}\basind{1}\basind{2}}(p^{\basind{r}})$ of the curvature tensor.
On the right we depict six equivalent interferometers aligned along mutually orthogonal directions. They allow us to determine six independent components of the curvature tensor, which are related to the vectors $\mathbf{n}$ and $\mathbf{m}$ in table \ref{SagnacCurvatureComp}.}
\label{FigSagnacLeadingFirstOrder}
\end{figure}

Next, we use an additional interferometer with a closed curve $\mathcal{S}$ which runs through the origin of the proper reference frame, as sketched in the left part of fig.~\ref{FigSagnacLeadingFirstOrder}. 
Moreover, we restrict ourselves to a single observer located at the origin, who defines the coordinates and measures the Sagnac time delay. In this case, we can identify the world line $q^{\basind{\sigma}}(x^{\basind{0}})$ with $p^{\basind{\sigma}}(x^{\basind{0}})$ giving rise to the redshift factor $g_{\basind{00}}(q^{\basind{r}})=1$.
When we take the Fermi-Walker transport of the tetrad into account, the Sagnac time delay~(\ref{LeadingOrdersSagnacTime}) reduces to 
\begin{equation}
 \Delta\tau_{Sp}=\frac{8}{3 c}\,\,
\,\varepsilon^{\basind{0 a j k}}\, 
R_{\basind{0}\{\basind{i_1}\basind{j}\}\basind{k}}(p^{\basind{r}})\,A_{\basind{a}}^{\;\;\basind{i_1}}
+\mathcal{O}\Big(A_{\basind{a}}^{\;\;\basind{i_1}\basind{i_2}}\Big)\,.
\label{SagnacCurvSensor}
\end{equation}
Here we have added the subscript $p$ to the Sagnac time delay $\Delta \tau_S$ since in this case the measurement is performed by the observer along the world line $p^{\basind{\sigma}}(x^{\basind{0}})$. 

Expression~(\ref{SagnacCurvSensor}) allows us to establish a direct connection between the Sagnac time delay and several components of the curvature tensor.
As in the previous subsection, we first want to exemplify the idea by considering a spatial curve $\mathcal{S}$ which encloses a planar area $\mathcal{A}$ in the $x^{\basind{1}}$-$x^{\basind{2}}$-plane. However, in the present case, we only require that $\mathcal{S}$ is symmetric under reflection with respect to the $x^{\basind{1}}$-$x^{\basind{3}}$-plane as sketched on the left of fig.~\ref{FigSagnacLeadingFirstOrder}. Due to this weakened symmetry condition, the first moments of the unit fluxes $A_{\basind{a}}^{\;\;\basind{i_1}}$  will no longer vanish. This feature is in contrast to the previously discussed rotation sensor. As mentioned before, it is possible to relate the first moments of the unit fluxes $A_{\basind{a}}^{\;\;\basind{i_1}}$ to the ``center of mass'' of the planar area $\mathcal{A}$, located at $\mathcal{M}\,m^{\basind{i}}$. Here we have introduced the unit vector $\mathbf{m}$ with components $m^{\basind{i}}=(1,0,0)$, as well as the separation $\mathcal{M}$ of the ``center of mass'' to the origin. Moreover, we again denote the contravariant components of the unit surface normal $\mathbf{n}$ by $n^{\basind{a}}=(0,0,1)$. With these definitions we then obtain for the zeroth and the first moments of the unit fluxes
\begin{equation}
A_{\basind{a}}=\mathcal{A}\,n_{\basind{a}}
\quad\text{and}\quad
A_{\basind{a}}^{\;\;\basind{i_1}}=n_{\basind{a}}\,\iint\limits_{\mathcal{A}} s^{\basind{i_1}}\; \D s^{\basind{1}} \D s^{\basind{2}}=\mathcal{M}\,n_{\basind{a}}\,m^{\basind{i_1}}\,.
\label{ZerothFirstMomentSpecial}
\end{equation}
When we insert these expressions into the Sagnac time delay~(\ref{SagnacCurvSensor}), we obtain
\begin{equation}
 \Delta\tau_{Sp}=\frac{8\mathcal{M}}{3 c}\,n_{\basind{a}}
\,\varepsilon^{\basind{0 a j k}}\, 
R_{\basind{0}\{\basind{i_1}\basind{j}\}\basind{k}}(p^{\basind{r}})\,m^{\basind{i_1}}
+\mathcal{O}\Big(A_{\basind{a}}^{\;\;\basind{i_1}\basind{i_2}}\Big)\,,
\label{Curvformula1}
\end{equation} 
which yields with the current values of $\mathbf{n}$ and $\mathbf{m}$ and the first Bianchi identity 
\begin{equation}
\Delta\tau_{Sp}=\frac{4\mathcal{M}}{c}\,R_{\basind{0}\basind{1}\basind{1}\basind{2}}(p^{\basind{r}})+
\mathcal{O}\Big(A_{\basind{a}}^{\;\;\basind{i_1}\basind{i_2}}\Big)\,.
\label{Curvformula2}
\end{equation} 
Hence, Sagnac interferometry allows us to determine some of the components of the curvature tensor. We emphasize, that this result is only valid provided second and higher order contributions can be neglected. 

We conclude by briefly outlining the scheme how to measure six off all 20 independent components of the curvature tensor.
For this purpose, we change the orientation of the planar Sagnac interferometer without affecting its shape. In particular, we align the unit surface normal $\mathbf{n}$ along one of the coordinate axes, and place the position $\mathcal{M}\,\mathbf{m}$ of the ``center of mass'' onto another coordinate axes orthogonal to $\mathbf{n}$. As shown on the right of fig.~\ref{FigSagnacLeadingFirstOrder}, we are left with six different orientations of our Sagnac interferometer which allow for the determination of six independent components of the curvature tensor. 

In table \ref{SagnacCurvatureComp} we present the connection between these components and the Sagnac time delay for the different orientations of the loops described by the vectors $\mathbf{n}$ and $\mathbf{m}$. Here we have made use of eq.~(\ref{Curvformula1}) and have evaluated the resulting expressions in complete analogy to the example leading to eq.~(\ref{Curvformula2}).
\begin{table}
\caption{Connection between the orientation of the Sagnac interferometer and the component of the curvature tensor  measured by this device. The orientation of the planar interferometer is encoded in the surface normal $n^{\basind{a}}$ and in the position $\mathcal{M}\,m^{\basind{i}}$ of the ``center of mass'' of $\mathcal{A}$. The Sagnac time delay $\Delta\tau_{Sp}$ follows from the components of the curvature tensor by eq.~(\ref{Curvformula1}).}
\label{SagnacCurvatureComp}
\hspace{0.15\textwidth}
\begin{minipage}[t]{0.7\textwidth}
  \begin{narrowtabular}{1cm}{ccc}
\hline
$n^{\basind{a}}$ & $m^{\basind{i}}$ & $\frac{c}{4\mathcal{M}}\cdot\Delta\tau_{Sp}$    \\
\hline
(1,0,0)&(0,1,0)&$-R_{\basind{0}\basind{2}\basind{2}\basind{3}}(p^{\basind{r}})$  \\
(1,0,0)&(0,0,1)&$+R_{\basind{0}\basind{3}\basind{3}\basind{2}}(p^{\basind{r}})$  \\
(0,1,0)&(1,0,0)&$+R_{\basind{0}\basind{1}\basind{1}\basind{3}}(p^{\basind{r}})$  \\
(0,1,0)&(0,0,1)&$-R_{\basind{0}\basind{3}\basind{3}\basind{1}}(p^{\basind{r}})$  \\
(0,0,1)&(1,0,0)&$-R_{\basind{0}\basind{1}\basind{1}\basind{2}}(p^{\basind{r}})$  \\
(0,0,1)&(0,1,0)&$+R_{\basind{0}\basind{2}\basind{2}\basind{1}}(p^{\basind{r}})$\\
\hline
  \end{narrowtabular}
\end{minipage}
\end{table}

We emphasize, that Sagnac interferometry is not capable of reproducing all components of the curvature tensor. However, there exist other methods based e.~g. on geodesic deviation or parallel transport along closed loops, which are capable of providing the remaining components of the curvature tensor and which allow for a deeper understanding of the curvature of spacetime \cite{Ehlers73,Audretsch83}.

\subsubsection{Double eight-Loop interferometer (DELI)}

One might wonder, whether it is really necessary to install a rotation as well as a curvature sensor in order to obtain information about the tetrad rotation and the curvature of spacetime. We now show that indeed a single device suffices. For this purpose we combine the main ideas of the two preceding subsections to construct a single Sagnac interferometer in which we can easily switch between rotation and curvature measurements.

\begin{figure}[h]
\centering
\includegraphics[width=0.35\textwidth]{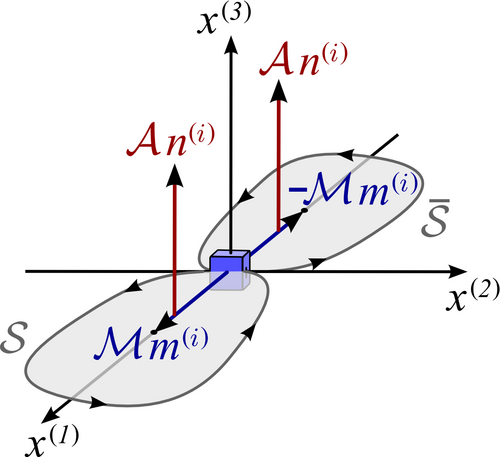}
\caption{Central idea of a device combining tetrad rotation and curvature measurements. Addition or subtraction of the individual Sagnac time delay measurements obtained in two mirror-inverted curvature sensors yields either the tetrad rotation or a specific component of the curvature tensor, respectively.}
\label{FigSagnacCombined}
\end{figure}

\paragraph{General idea} The appropriate combination of the symmetry aspects of the light paths $\mathcal{S}$ used in the rotation and curvature sensor is the key point of our approach. We upgrade the curvature sensor displayed on the left of fig.~\ref{FigSagnacLeadingFirstOrder} by including a mirror-inverted interferometer with ``center of mass'' position $-\mathcal{M}\,m^{\basind{i}}$. We denote the oppositely located loops of both curvature sensors by $\mathcal{S}$ and $\bar{\mathcal{S}}$. The parameterizations of $\mathcal{S}$ and $\bar{\mathcal{S}}$ are both positively oriented as illustrated in fig.~\ref{FigSagnacCombined}.

When we recall the zeroth and first moments of the unit fluxes, eq.~(\ref{ZerothFirstMomentSpecial}), we obtain from the expansion, eq.~(\ref{LeadingOrdersSagnacTime}), the Sagnac time delays
\begin{equation}
 \Delta\tau_{Sp}(\mathcal{S})=-\frac{4}{c^2}\,\omega^{\basind{a}} n_{\basind{a}}\,\mathcal{A}+
 \Phi\,\mathcal{M}+
\mathcal{O}\Big(A_{\basind{a}}^{\;\;\basind{i_1}\basind{i_2}}\Big)
\label{OppositeSagnac1}
\end{equation}
and
\begin{equation}
 \Delta\tau_{Sp}(\bar{\mathcal{S}})=-\frac{4}{c^2}\,\omega^{\basind{a}} n_{\basind{a}}\,\mathcal{A}-
 \Phi\,\mathcal{M}+
\mathcal{O}\Big(A_{\basind{a}}^{\;\;\basind{i_1}\basind{i_2}}\Big)
\label{OppositeSagnac2}
\end{equation}
corresponding to the two loops $\mathcal{S}$ and $\bar{\mathcal{S}}$.
Here, we have introduced the short hand notation 
\begin{equation*}
\Phi=n_{\basind{a}}\left[
\frac{8}{3c}\,\varepsilon^{\basind{0 a j k}}
R_{\basind{0}\{\basind{i_1}\basind{j}\}\basind{k}}(p^{\basind{r}})
+\frac{4}{c^4}\left(\omega^{\basind{l}} a_{\basind{l}}\,\delta^{\basind{a}}_{\basind{i_1}}-
3\,\omega^{\basind{a}} a_{\basind{i_1}}\right)\right] m^{\basind{i_1}}\,.
\end{equation*} 
In contrast to eq.~(\ref{Curvformula1}), where the tetrad attached to the observer was Fermi-Walker transported along the world-line $p^{\basind{\sigma}}(\tau)$, equations (\ref{OppositeSagnac1}) and (\ref{OppositeSagnac2}) include both the tetrad rotation and the curvature of spacetime.

Next, we take the sum  
\begin{equation}
 \Delta\tau_{Sp}(\mathcal{S})+ \Delta\tau_{Sp}(\bar{\mathcal{S}})=-\frac{4}{c^2}\,\omega^{\basind{a}}
 n_{\basind{a}}\cdot 2\mathcal{A}+
\mathcal{O}\Big(A_{\basind{a}}^{\;\;\basind{i_1}\basind{i_2}}\Big)
\label{SumProperTimeDelays}
\end{equation}
of the individual time delays $\Delta\tau_{Sp}(\mathcal{S})$ and $\Delta\tau_{Sp}(\bar{\mathcal{S}})$ and their difference
\begin{equation}
\Delta\tau_{Sp}(\mathcal{S})- \Delta\tau_{Sp}(\bar{\mathcal{S}})=\Phi\cdot 2\mathcal{M}
+\mathcal{O}\Big(A_{\basind{a}}^{\;\;\basind{i_1}\basind{i_2}}\Big)\,.
\label{DiffProperTimeDelays}
\end{equation}
In this way, we have separated the zeroth from the first order contribution of the Sagnac time delay.

We conclude by noting that the standard Sagnac interferometer experiments do not directly measure the proper time difference between two counter-propagating light rays, but rather use its manifestation in phase or frequency differences. For this reason, the proposed method of adding and subtracting the individual Sagnac time delays $\Delta\tau_{Sp}(\mathcal{S})$ and $\Delta\tau_{Sp}(\bar{\mathcal{S}})$ might not be the most convenient experimental approach towards the measurement of the tetrad rotation and the curvature of spacetime with the same device.

\paragraph{Road to DELI}
For this reason, we now pursue a slightly different approach and choose appropriate combinations of the paths $\mathcal{S}$ and $\bar{\mathcal{S}}$. This new measurement scheme can be easily motivated by a reinterpretation of $\Delta\tau_{Sp}(\mathcal{S})\pm\Delta\tau_{Sp}(\bar{\mathcal{S}})$ in terms of the proper arrival times $\tau_\pm(\mathcal{S})$ and $\tau_\pm(\bar{\mathcal{S}})$ of the clockwise~($-$) and counterclockwise~($+$) propagating light rays after one circulation around $\mathcal{S}$ and $\bar{\mathcal{S}}$, respectively. 
The quantities $\tau_\pm(\mathcal{S})$ and $\tau_\pm(\bar{\mathcal{S}})$ follow from eq.~(\ref{IntLightSolutions}) with ${\tau_\pm=\sqrt{g_{\basind{00}}(p^{\basind{r}})}\;t_\pm/c}$.

Since we have required that the metric expressed in proper reference frame coordinates is time independent, the values of the proper times $\tau_\pm(\mathcal{S})$ and $\tau_\pm(\bar{\mathcal{S}})$ do not depend on the moment of the measurement. Thus, we can cast the sum of the proper time delays into the form
\begin{equation}
 \Delta\tau_{Sp}(\mathcal{S})+ \Delta\tau_{Sp}(\bar{\mathcal{S}})=
[\tau_+(\mathcal{S})+\tau_+(\bar{\mathcal{S}})]-[\tau_-(\mathcal{S})+\tau_-(\bar{\mathcal{S}})]
=\Delta\tau_{Sp}(\mathcal{S}_0)\,,
\label{ProperTimeDelaysS0}
\end{equation} 
where we have introduced the total time delay
\begin{equation*}
\Delta\tau_{Sp}(\mathcal{S}_0)=\tau_+(\mathcal{S}_0)-\tau_-(\mathcal{S}_0)
\quad\text{with}\quad
{\tau_\pm(\mathcal{S}_0)=\tau_\pm(\mathcal{S})+\tau_\pm(\bar{\mathcal{S}})}\,.
\end{equation*}

\begin{figure}
\centering
\includegraphics[width=0.7\textwidth]{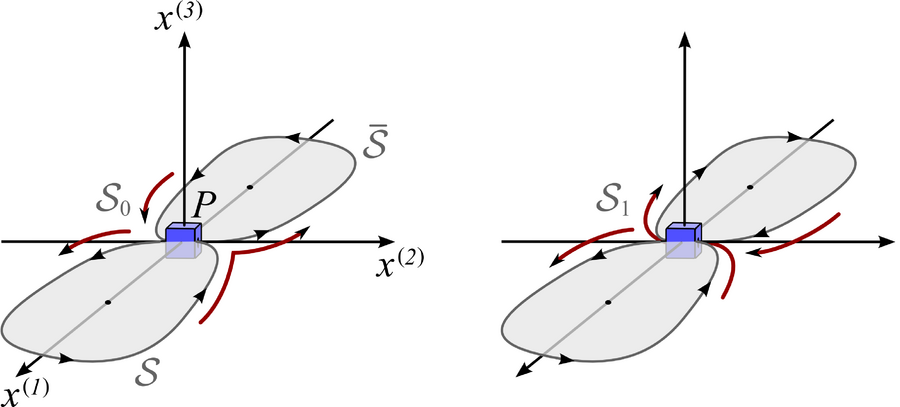}
\caption{Double eight-loop interferometer (DELI):
Measurement of the tetrad rotation (left) and of the curvature of spacetime (right) using two identical, mirror-reflected curvature sensors with loops $\mathcal{S}$ and $\bar{\mathcal{S}}$. In the left picture, the
counterclockwise-propagating light ray emitted at $P$ first travels around the loop $\mathcal{S}$. After its first return at $P$, it gets reflected and continues to propagate in counter-clockwise direction along the closed curve $\bar{\mathcal{S}}$. When the light ray arrives at $P$ for the second time, we take the total time $\tau_+(\mathcal{S}_0)$ elapsed during the circulation of the light ray along $\mathcal{S}_0$. Here $\mathcal{S}_0$ denotes the combination of both loops $\mathcal{S}$ and $\bar{\mathcal{S}}$. In contrast, the right picture summarizes a situation in which the initially counterclockwise-propagating light ray is transmitted when first arriving at $P$. Therefore, it circulates around the second loop $\bar{\mathcal{S}}$ in clockwise direction, giving rise to the total proper time $\tau_+(\mathcal{S}_1)$ at its second arrival at $P$. In order to bring out most clearly the similarities and differences between the two pictures, we have marked in the right picture only those quantities which are different from the ones in the left picture.}
\label{FigSagnacCombinedSetup}
\end{figure}

We illustrate the connection between the total time delay $\Delta\tau_{Sp}(\mathcal{S}_0)$ and the proper times $\tau_+(\mathcal{S}_0)$ and $\tau_-(\mathcal{S}_0)$ in the left picture of fig.~\ref{FigSagnacCombinedSetup}. When we consider the counter\-clock\-wise-propagating light ray which first circulates around $\mathcal{S}$ and, after {\it reflection} at $P$, continues to travel around $\bar{\mathcal{S}}$, we obtain the total proper time $\tau_+(\mathcal{S}_0)$ for the propagation along the positively oriented eight-loop $\mathcal{S}_0$. In accordance, we denote the proper time which results from the circulation of the clockwise-propagating beam along the eight-loop curve $\mathcal{S}_0$ by $\tau_-(\mathcal{S}_0)$ .

In the same spirit we find for the differences of the proper time delays the relation
\begin{equation}
 \Delta\tau_{Sp}(\mathcal{S})- \Delta\tau_{Sp}(\bar{\mathcal{S}})=
[\tau_+(\mathcal{S})+\tau_-(\bar{\mathcal{S}})]-[\tau_-(\mathcal{S})+\tau_+(\bar{\mathcal{S}})]
=\Delta\tau_{Sp}(\mathcal{S}_1)\,,
\label{ProperTimeDelaysS1}
\end{equation} 
where we have defined
\begin{equation*}
\Delta\tau_{Sp}(\mathcal{S}_1)=\tau_+(\mathcal{S}_1)-\tau_-(\mathcal{S}_1)
\quad\text{with}\quad
{\tau_\pm(\mathcal{S}_1)=\tau_\pm(\mathcal{S})+\tau_\mp(\bar{\mathcal{S}})}\,.
\end{equation*} 
The corresponding situation is illustrated on the right of fig.~\ref{FigSagnacCombinedSetup}. Here, the initially counterclockwise-propagating light ray is transmitted at $P$ after its first circulation around the loop $\mathcal{S}$. Thus, it will travel along the second loop $\bar{\mathcal{S}}$ in clockwise direction, giving rise to the definition of a different eight-loop path $\mathcal{S}_1$. Hence, the total time $\tau_+(\mathcal{S}_1)$ follows from the sum of the individual proper times $\tau_+(\mathcal{S})$ and $\tau_-(\bar{\mathcal{S}})$. The proper time $\tau_-(\mathcal{S}_1)$ results from the initially clockwise-propagating light ray which then circulates around $\bar{\mathcal{S}}$ in counter-clockwise direction, respectively.

The crucial difference between the two eight-loops $\mathcal{S}_0$ and $\mathcal{S}_1$ stems from the reversed circulation of the corresponding light rays along $\bar{\mathcal{S}}$. In fact, this difference provides the bedrock for the determination of the zeroth and first order contributions of the Sagnac time delay, eq.~(\ref{LeadingOrdersSagnacTime}), with this method. Indeed, when we compare eq.~(\ref{SumProperTimeDelays}) to eq.~(\ref{ProperTimeDelaysS0}), we obtain the zeroth order contribution 
\begin{equation}
\Delta\tau_{Sp}(\mathcal{S}_0)=-\frac{4}{c^2}\,\omega^{\basind{a}}
 n_{\basind{a}}\cdot 2\mathcal{A}+
\mathcal{O}\Big(A_{\basind{a}}^{\;\;\basind{i_1}\basind{i_2}}\Big)
\label{DELI1}
\end{equation} 
from a Sagnac time delay measurement with the eight-loop $\mathcal{S}_0$, whereas the first order contribution
\begin{equation}
\Delta\tau_{Sp}(\mathcal{S}_1)= \Phi\cdot 2\mathcal{M}
+\mathcal{O}\Big(A_{\basind{a}}^{\;\;\basind{i_1}\basind{i_2}}\Big)  
\label{DELI2}
\end{equation} 
follows from a measurement with $\mathcal{S}_1$ as indicated by a comparison of eq.~(\ref{DiffProperTimeDelays}) with~(\ref{ProperTimeDelaysS1}).

In this way, we have constructed an intuitive and operational method to gain insight into the zeroth and first order contributions of the Sagnac time delay. We can distinguish between purely gravitational and inertial effects in $\Delta\tau_{Sp}(\mathcal{S}_1)$, when we adjust the orientation of our satellite such that $\Delta\tau_{Sp}(\mathcal{S}_0)=0$ is always satisfied, thus giving rise to $\omega^{\basind{i}}=0$. However, we have to be sure that the second and higher order contributions in eq.~(\ref{LeadingOrdersSagnacTime}) are still negligibly small.

Due to the two measurement modes of the eight-loop interferometer, we call the device depicted in fig.~\ref{FigSagnacCombinedSetup} a {\it double eight-loop interferometer} (DELI).

\paragraph{Possible experimental realization}

We conclude this subsection by briefly presenting some ideas for the experimental implementation of the DELI. However, these ideas are preliminary and call for further investigations.

As a first thought, we are tempted to take advantage of the horizontal and vertical polarization of light and use a polarization beam splitter at $P$, which allows for a reflection of the horizontally and a transmission of the vertically polarized light ray after its first circulation around $\mathcal{S}$, see fig.~\ref{FigSagnacCombinedSetup}. In this case, the horizontally polarized component would propagate along the loop $\mathcal{S}_0$, whereas the vertically polarized component would follow the loop $\mathcal{S}_1$. After the second arrival of the polarized beams at $P$, one would have to separate them, e.~g. with a birefringent medium, in order to obtain two separate interference patterns. 

This implementation of a DELI bears an additional complication: spacetime influences the polarization of light. In fact, a rotation of the polarization vector relative to the observer's proper reference frame stems from inertial effects such as the tetrad rotation on the one hand and from the curvature of spacetime in the vicinity of the observer's world line on the other. As a consequence, the initially horizontal and vertical components of the light rays will intermingle after the circulation around $\mathcal{S}_0$ and $\mathcal{S}_1$, resulting in more complex interference patterns at the detector. 

A solution to the problem of polarization mixing must crucially depend on the particular realization of the guiding mechanism for the light rays. For simplicity, let us suppose, that the counter-propagating light rays stay on target with the help of a large number of mirrors. In this case, the light rays freely propagate along null geodesics between two successive mirrors and the polarization vector undergoes parallel transport \cite{MTW}. Taking also into account the change of the polarization vector induced by the guiding mirrors along the eight-loop, we could predict the total change of the polarization vector and try to countervail, were it not for our ignorance of the local metric in the neighborhood of the observer's world line. 

We conclude by emphasizing, that this brief discussion makes a strong case for a thorough analysis of polarization changes due to parallel transport and mirror reflections in order to obtain valuable limits on the influence of the local, unknown metric for a particular realization of the eight-loop. Only then we are able to decide, whether to favor or to reject this polarization-based implementation of a DELI.

\subsection{Rotation in general relativity\label{SectRotInRel}}

We now briefly compare and contrast two concepts of rotation in general relativity. The first one is based on the local definition of rotation using e.~g. a Sagnac interferometer. The second one is connected to the more traditional point of view, which unconsciously relates rotation to the circular motion of the stars in the sky. We close this section with some comments on Mach's principle.

\paragraph{Inertial compass}

In subsection~\ref{SecRotSensor} we have outlined an operational method to determine the inertial effect of tetrad rotation using the rotation sensor. In other words, we have given an absolute meaning to the ``rotation of the observer's coordinate axes relative to a Fermi-Walker transported tetrad''. 

One sometimes refers to a Fermi-Walker transported tetrad as {\it inertial compass}. The use of the word ``compass'' might be slightly misleading in this context, since the inertial compass does not characterize a particular spatial direction, but rather describes a certain state of motion of the observer's coordinate axes. 
The adjective ``inertial'' originates from the analysis of timelike geodesics in the local neighborhood of the observer. Suppose our observer disperses a cloud of freely falling test particles simultaneously in all spatial directions. Using a local approximation of the geodesic equation in proper reference frame coordinates, the congruence of all particle world lines would only be irrotational in the local neighborhood of the observer's world line as long as his coordinate axes were Fermi-Walker transported. Indeed, a non-vanishing tetrad rotation vector $\omega^{\basind{i}}\neq 0$ would lead to Coriolis and centrifugal like contributions as first order corrections to the geodesic equation \cite{Ehlers73,MTW}.

\paragraph{Stellar compass}

Astrometry rests upon the observation of celestial light sources and suggests the use of spatial reference systems such as catalog stars or Very Long Baseline Interferometry (VLBI) \cite{Soffel89}. With these marvelous techniques in mind, one may conclude that rotation could also be understood as a relative concept which characterizes the revolution of the observer's reference frame relative to the distant ``celestial bodies''. Indeed, this point of view is frequently put forward in the context of Mach's criticism concerning Newton's rotating bucket \cite{Mach1893}:
\begin{quotation}
{\it ``Newton's experiment with the rotating vessel simply informs us that the relative motion of the water with respect to the sides of the vessel produces no noticeable centrifugal forces, but that such forces are produced by its relative rotation with respect to the mass of the Earth and the other celestial bodies.''}
\end{quotation}
However, when we analyze such celestial reference systems within the framework of general relativity aiming for a relative meaning of rotation, we should be aware of two intricacies. 

The first one has to do with the concept of motion in general relativity. Suppose that a cloud of test particles travels along a congruence of timelike world lines in a spacetime with fixed metric. In this case, it is always possible to introduce adapted coordinates such that all particles are spatially at rest and all the spatial components of their four-velocities vanish. This simple example shows clearly, that we cannot give a rigorous meaning to the notion of motion, without accepting the metric as a crucial ingredient.

However, the metric is accompanied by the next intricacy, namely the global aspects of curved spacetime. When we examine the position of a star in the sky, we perceive the tangent vector of the incident null geodesic which connects the star with our telescope. It is clear, that the direction of the incident tangent crucially depends on the underlying metric of spacetime, and not only on the initial position and direction of the null geodesic emanating from the star. 

The preferred spatial directions obtained by several such catalog stars define the so-called {\it stellar compass} \cite{Weyl24}. Despite the fact that the stellar compass\footnote{In some literature the stellar compass is called {\it light compass}.} crucially depends on the global aspects of the spacetime metric, it allows for an independent definition of a spatial reference system. Instead of using a Sagnac interferometer or another gyroscope to locally define a Fermi-Walker-transported tetrad, which gives rise to the inertial compass, our observer could likewise establish the orientation of his spatial tetrad axes e.~g. with the help of these catalog stars. Roughly speaking, the Gravity Probe B experiment was designed to compare the time evolutions of the inertial and the stellar compass along the world line of the satellite.

In this spirit, one should {\it not} state that the water in Newton's bucket is rotating relative to the distant ``celestial bodies'', but relative to the inertial compass. However, it is an experimentally well established fact that the inertial and stellar compass do not rotate relative to each other in our present universe. For this reason one might equally well assert that the water in the bucket rotates relative the to the stellar  compass.

\paragraph{Mach's principle}

We conclude this discussion of the concept of rotation with some brief comments on the validity of Mach's principle in the theory of relativity. As already adumbrated by Isenberg's and Wheeler's quote in the introduction, sect.~\ref{SecIntroduction}, of these lectures, it is not an easy matter to decide whether Mach's principle is satisfied in general relativity or not. This difficulty arises mainly due to the vague formulation of Mach's principle which allows for many different interpretations \cite{Pfister95,Ciufolini95}. Here, we will only mention three distinct versions. For additional formulations of Mach's principle we refer to~\cite{Bondi97}.

The first version states that {\it ``the universe, as represented by the average motion of distant galaxies does not appear to rotate relative to local inertial frames''} \cite{Bondi97}. In other words, within our current universe the inertial compass is not rotating relative to the stellar compass. This claim has been tested experimentally to a high accuracy. However, in connection with general relativity, it rules out certain cosmological solutions of Einstein's field equations in favor of other ones. In fact, it can be shown that for a static spacetime which is endowed with a timelike, hypersurface orthogonal Killing vector field, the inertial compass does not rotate relative to the stellar compass provided both, the observer as well as the celestial bodies follow the integral curves of the Killing field \cite{Straumann04}. For stationary spacetimes whose timelike Killing vector field is not hypersurface orthonormal, the inertial compass will rotate with respect to the stellar one. Since static spacetimes constitute a very special class of solutions of Einstein's field equations, the coincidence of inertial and stellar compass is a rather exceptional case in general relativity. Therefore, it is remarkable that this version of Mach's principle fits perfectly with our observations.

A second version, that we would like to mention here, reads: {\it ``local inertial frames are affected by the cosmic motion and distribution of matter''} \cite{Bondi97}. In fact, this formulation of Mach's principle also holds true in general relativity since the cosmic energy and momentum distribution influences the whole spacetime metric directly via Einstein's field equations. The Lense-Thirring effect may serve as a prominent example: in case the Earth would not rotate with respect to the asymptotically flat metric at spatial infinity, the inertial and the stellar compass would coincide. But since the Earth rotates it gives rise to a rotation of the inertial compass relative to the stellar compass, as mentioned above in the context of the Gravity Probe B experiment.

The third version of Mach's principle brings forward a keen and suggestive idea: {\it ``inertial mass is affected by the global distribution of matter''} \cite{Bondi97}. This formulation does not apply to general relativity, since the inertial mass takes the role of an independent quantity in this theory. By no means does general relativity allow for an substitution of (inertial) mass in terms of any other fundamental quantities. 

This brief discussion illustrates the murkiness surrounding the interpretations of Mach's principle, as alluded by the quote in the Introduction. It also demonstrates in a striking way, that some of Mach's ideas did find their way into general relativity, but some others did not.


\section{Rotating frame of reference in flat spacetime\label{SecRotatingFrame}}

After this brief excursion into the concept of rotation, we proceed in the next two sections with the application of the Sagnac time delays for our DELI, eq.~(\ref{DELI1}) and~(\ref{DELI2}), to two very different physical situations: (i) an observer located in a rotating reference frame in Minkowski spacetime, and (ii) an observer at rest with respect to the ideal fluid in G\"odel's Universe.

In the present section we concentrate on the first case. We start by briefly recalling the metric coefficients and some properties of the rotating reference frame in flat spacetime, thereby making use of the corresponding light cone diagram. We then assign a proper reference frame to an observer at rest, analyze his acceleration and the corresponding tetrad rotation and deduce the general Sagnac time delays for our DELI, eq.~(\ref{DELI1}) and~(\ref{DELI2}).

\subsection{Metric}

When expressed in cylindrical coordinates $x^\mu=(t,r,\phi,z)$, the line element of Minkowski spacetime reads 
\begin{equation}
\D s^2=c^2\,\D t^2-\D r^{\,2}-r^{\,2}\,\D \phi^2-\D z^2\,.
\label{eq:dsFlat}
\end{equation}
The coordinates $x'^\mu=(t',r',\phi',z')$ of the rotating reference frame are then established by the coordinate transformation
\begin{equation}
t\equiv t'\,, \quad
r\equiv r'\,, \quad
\phi\equiv \phi' +\Omega_{\mathrm R} t'\,,\quad
z\equiv z'\,,
\label{TrafoRot}
\end{equation}
which corresponds to a rotation around the $z$-axis with rotation rate 
${\Omega_{\mathrm R} >0}$. As a consequence, we obtain the transformed line element
\begin{equation}
\D s^2=(c^2-r'^{\,2}\Omega_{\mathrm R}^2)\,\D t'^2-\D r'^{\,2}-r'^{\,2}\,\D \phi'^2-\D z'^2-2 r'^{\,2}\Omega_{\mathrm R}\,\D t'\,\D \phi' 
\label{eq:dsRot}  \,.
\end{equation}
We emphasize, that the metric coefficients in the rotating reference frame do not depend on time, and therefore satisfy the assumptions used in the derivation of the Sagnac time delay~(\ref{GenSagnacTimeDelay}).

\subsection{Light cone diagram for a rotating reference frame}

In order to give a first impression of the propagation of light within a rotating reference frame, we present a light cone diagram which is the collection of the ``infinitesimal'' light cones attached to every point $P$ in spacetime. The light cones indicate all directions in which a flash of light can propagate when emitted from $P$. The construction of such a diagram is briefly discussed in appendix~\ref{MinkowskiForm}. In short, the ``infinitesimal" light cones are established by the set of all tangents to the null geodesics through $P$, since these null geodesics represent the actual trajectories of all freely propagating light rays through $P$. In general, the ``infinitesimal'' light cones tilt and change their apex angle from point to point due to the curvature of spacetime or simply as a result of the chosen reference frame. 

\begin{figure}
\centering
\includegraphics[width=0.8\textwidth]{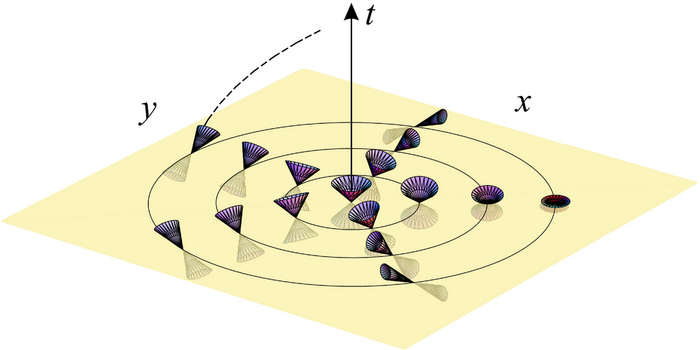}
\caption{Light cone diagram for a reference frame rotating with rate $\Omega_R$ in flat spacetime. We depict two spatial coordinates $x=r'\cos\phi'$ and $y=r'\sin\phi'$, and one time coordinate $t$. The $z$ coordinate has been suppressed. As we increase the separation $r'$ of the light cones from the origin, the cones tilt and the apex angles decrease. For ${c}/{\Omega_{\mathrm R}}<r'$ the tilting is so large that the curve ${x^\mu (\lambda)=(\lambda,r'_0,\phi'_0,z'_0)}$ becomes spacelike. For this reason, we exclude this region in our Sagnac interferometer experiments. The dashed line depicts a particular null geodesic which brings out most clearly the fact that the infinitesimal light cones represent the tangents to the real light paths.}
\label{lightconesrot}
\end{figure}
Figure~\ref{lightconesrot} depicts the light cone diagram for the rotating reference frame in Minkowski spacetime. Here the $z$-axis has been suppressed.
From the line element, eq.~(\ref{eq:dsRot}), we note that the light cone located at the center of the coordinate system coincides with the corresponding light cone in the non-rotating inertial frame in flat spacetime. However, when we increase the radial position of the light cones, they start to tilt and narrow due to the rotation of our chosen reference frame. Formally, this tilting results from the off-diagonal element of the metric which couples the time and the angular coordinate. 

For radii $r'>c/\Omega_{\mathrm R}$, neither a massive particle nor light is able to stay at rest at a fixed position $(r'_0,\phi'_0,z'_0)$ in this rotating reference frame since the spacetime curve ${x^\mu(\lambda)=(\lambda,r'_0,\phi'_0,z'_0)}$ then becomes spacelike. Hence, a rotating reference frame cannot be established globally using massive particles or light rays. It should rather be considered as a purely mathematical construction, which simply labels the spacetime points in accordance to the coordinate transformation~\eqref{TrafoRot}.

\subsection{Sagnac time delay\label{SagRotatingReferenceFrame}}

Next we turn to the discussion of the Sagnac time delays, eq.~(\ref{DELI1}) and~(\ref{DELI2}), for an observer at rest in the rotating reference frame with radial position $0<r'_0<c/\Omega_{\mathrm R}$. 

\subsubsection{World line, four-velocity and acceleration}

When we denote the spatial position of the observer by $(r'_0,\phi'_0,z'_0)$ and parameterize his world line in terms of his measured proper time, we obtain the expression 
\begin{equation*}
p'^{\mu}(\tau)=(\mathcal{N}\tau,r'_0,\phi'_0,z'_0)
\end{equation*}
for his world line and
\begin{equation*}
u'^{\mu}(\tau)=(\mathcal{N},0,0,0)=\mathcal{N}\,\delta^\mu_0
\end{equation*}
for his four-velocity. Here we have introduced $\mathcal{N}=\left(1-\left(r'_0\Omega_R/c\right)^2\right)^{-1/2}$
in order to satisfy the condition $u^\mu(\tau)\, u_\mu(\tau)=c^2$.

Recalling the non-vanishing components
\begin{equation*}
 \Gamma'^1_{\;\;0 0}=-r'\Omega_R^2,\;\;
\Gamma'^1_{\;\;0 2}=-r'\Omega_R,\;\;
\Gamma'^1_{\;\;2 2}=-r',\;\;
\Gamma'^2_{\;\;0 1}=\frac{\Omega_R}{r'},\;\;
\Gamma'^2_{\;\;1 2}=\frac{1}{r'},\;\;
\end{equation*}
of the Christoffel symbols, 
we find the four-acceleration
\begin{equation*}
a'^\mu(\tau)=\frac{d^2 p'^\mu}{d\tau^2}+\Gamma'^\mu_{\;\;\alpha\beta}\,u'^\alpha\, u'^\beta=-\mathcal{N}^2\,r'_0\Omega_R^2\,\delta^\mu_1
\end{equation*}
of the observer which has a non-vanishing component in radial direction.

\subsubsection{Tetrad basis and transport matrix}

We now have to specify the coordinate axes of the observer's proper reference frame in terms of a suitable orthonormal tetrad. For simplicity, we assume that the spacelike tetrad vectors point in the same spatial directions as the spatial coordinate axes of the rotating reference frame. According to condition~(\ref{e0}) in the definition of a proper reference frame, our timelike basis vector is given by
\begin{equation*}
 e'^\mu_{\;\,\basind{0}}(\tau)=\frac{1}{c}\,u'^\mu(\tau)=\frac{\mathcal{N}}{c}\,\delta^\mu_0\,.
\end{equation*}
Two of the corresponding spacelike tetrad vectors can be chosen according to
\begin{equation*}
e'^\mu_{\;\,\basind{1}}(\tau)=\delta^\mu_1\quad\text{and}\quad
e'^\mu_{\;\,\basind{3}}(\tau)=\delta^\mu_3\,.
\end{equation*}
However, the orthogonality condition~(\ref{RelOrthogonality}) imposes a non-vanishing time component on the remaining spacelike tetrad vector, such that 
\begin{equation*}
e'^\mu_{\;\,\basind{2}}(\tau)=\left(\frac{r'_0\Omega_R}{c}\right)\frac{\mathcal{N}}{c}\,\delta^\mu_0+
\frac{1}{r'_0\,\mathcal{N}}\,\delta_2^\mu\,.
\end{equation*}
This vector completes the tetrad basis of our observer along his world line. In this tetrad basis the components of the four-acceleration read
\begin{equation}
a'_{\basind{\alpha}}=e'^\mu_{\;\,\basind{\alpha}}\,a'_\mu=-\mathcal{N}^2 r'_0\Omega_R^2\,\eta_{\basind{1\alpha}}\quad
\text{and}\quad
a'^{\basind{\alpha}}=-\mathcal{N}^2 r'_0\Omega_R^2\,\delta^{\basind{\alpha}}_{\basind{1}}\,.
\label{RotAcc}
\end{equation}

Before we proceed with the determination of the tetrad rotation vector, it is reasonable to examine the transport matrix $\Omega'^{\mu \nu}$, eq.~(\ref{TransportMatrix}). For our given family of tetrads along the world line of the observer, the transport matrix follows directly from the proper transport law, eq.~(\ref{ProperTransport}), by using the orthogonality relation~(\ref{OrthoNice2}). After some minor algebra, we arrive at
\begin{equation*}
\Omega'^{\mu \nu}=-\left(e'^\mu_{\;\;\basind{\alpha};\rho}\, u'^\rho\right)\eta^{\basind{\alpha\beta}}\,e'^\nu_{\;\,\basind{\beta}}=\mathcal{N}\;\frac{\Omega_R}{r'_0}
\left(\delta^\mu_2\delta^\nu_1-\delta^\mu_1\delta^\nu_2\right)\,.
\end{equation*}
When we now express the transport matrix in terms of the corresponding tetrad coefficients, we obtain the slightly more extended expression
\begin{equation*}
\Omega'_{\basind{\alpha\beta}}=\frac{\mathcal{N}^2}{c} r'_0\Omega_R^2\left(\delta^{\basind{0}}_{\basind{\alpha}}
\delta^{\basind{1}}_{\basind{\beta}}-\delta^{\basind{1}}_{\basind{\alpha}}
\delta^{\basind{0}}_{\basind{\beta}}\right)
+\mathcal{N}^2\Omega_R\left(\delta^{\basind{2}}_{\basind{\alpha}}
\delta^{\basind{1}}_{\basind{\beta}}-\delta^{\basind{1}}_{\basind{\alpha}}
\delta^{\basind{2}}_{\basind{\beta}}\right)\,.
\end{equation*} 
We then solve the transport matrix, eq.~\eqref{TransportMatrix}, for the rotation vector and finally obtain
\begin{equation}
\omega'^{\basind{\mu}}=-\frac{1}{2}\varepsilon^{\basind{0\mu\alpha\beta}}\Omega'_{\basind{\alpha\beta}}=\mathcal{N}^2\Omega_R\,
\delta^{\basind{\mu}}_{\basind{3}}\,.
\label{RotVec}
\end{equation}
Hence, we conclude that the third component of the tetrad rotation vector $\omega'^{\basind{\mu}}$ coincides with the rotation rate $\Omega_R$ when the observer is located at $r'_0=0$. However, for increasing radii $0<r'_0<c/\Omega_{\mathrm R}$ we encounter a growth of this third component.

\subsubsection{Sagnac time delays for the DELI}

The tetrad components of the four-ac\-cel\-er\-a\-tion and of the tetrad rotation vector, eqs.~(\ref{RotAcc}) and~(\ref{RotVec}), enable us to finally establish the the Sagnac time delays, eq.~(\ref{DELI1}) and~(\ref{DELI2}), for an observer at rest in the rotating reference frame. For the first measurement mode of the DELI, eq.~(\ref{DELI1}) predicts
\begin{equation}
\Delta\tau_{Sp}(\mathcal{S}_0)=\frac{4}{c^2}\,\mathcal{N}^2\,\Omega_R\,
 n^{\basind{3}} \cdot 2\mathcal{A}+
\mathcal{O}\Big(A_{\basind{a}}^{\;\;\basind{i_1}\basind{i_2}}\Big)\,,
\label{DELI1Rot}
\end{equation} 
which is the relativistic analogue of the classical expression~(\ref{famsagnac}). 
As expected, the Sagnac time delay is non-vanishing as long as the unit normal $\mathbf{n}$ to the planar area $\mathcal{A}$ possesses a non-vanishing component in $z$-direction. 

Moreover, for the second measurement mode of the DELI, eq.~(\ref{DELI2}) reads
\begin{equation}
\Delta\tau_{Sp}(\mathcal{S}_1)=
\frac{12}{c^4}\,\mathcal{N}^4\,r'_0\,\Omega_R^3\,n^{\basind{3}}\,m^{\basind{1}}
\cdot 2\mathcal{M}+
\mathcal{O}\Big(A_{\basind{a}}^{\;\;\basind{i_1}\basind{i_2}}\Big)\,.
\label{DELI2Rot}
\end{equation} 
Here, we have taken advantage of the fact that all components of the curvature tensor vanish in flat spacetime. In this second measurement mode, we obtain a non-vanishing Sagnac time delay, provided the DELI is oriented in such a way, that the unit normal $\mathbf{n}$ again possesses a non-vanishing $z$-component, while at the same time the unit vector $\mathbf{m}$ pointing towards the ``center of mass'' of $\mathcal{A}$ possesses a non-vanishing radial component. 

We conclude by considering the ratio
\begin{equation*}
\frac{\Delta\tau_{Sp}(\mathcal{S}_1)}{\Delta\tau_{Sp}(\mathcal{S}_0)}\approx
3\,\mathcal{N}^2\,\left(\frac{r'_0\,\Omega_R}{c}\right)^2 \;m^{\basind{1}}\, \frac{\mathcal{M}}{r'_0\mathcal{A}}
\end{equation*} 
between the two time delays. Since in a typical experimental situation ${\mathcal{M}/(r'_0\mathcal{A})\approx 1}$ and ${r'_0\,\Omega_R/c\ll 1}$ which leads to $\mathcal{N}\approx 1$, we find 
$\Delta\tau_{Sp}(\mathcal{S}_1) \ll \Delta\tau_{Sp}(\mathcal{S}_0)$.
As a consequence, the time delay obtained in the second measurement mode is in general much smaller than the one of the first measurement mode.


\section{G\"odel's Universe\label{PropertiesGoedel}}

\begin{quotation}
{\it ``Now in his universe one can travel into the past.''}\footnote{Entry in the diary of Oscar Morgenstern on May 12, 1949.}
\end{quotation}

In this section, we give a brief introduction into G\"odel's Universe, its spacetime structure and some of its peculiar features. We then study the resulting Sagnac time delay and compare it to the corresponding expression found for the rotating reference frame.

\subsection{Why a rotating universe?}

How did it happen, that the ingenious mathematician and logician Kurt G\"odel busied himself studying cosmology and rotating universes? It is difficult to answer this question. Nevertheless, we offer some reasonable arguments why G\"odel might have become interested in this topic \cite{Schuecking03,Jung06}.

After G\"odel had left Austria in January 1940, he moved to Princeton and became an ordinary member of the ``Institute for Advanced Studies'', which he had already visited in 1933. In Princeton he became a close friend of Albert Einstein, with whom he frequently shared the walks home from the Institute. 

In September 1946 Einstein received a letter from George Gamov who argued that the galaxies seem to have more angular momentum than predicted. The underlying model is based on the assumption that galaxies were formed by fluctuations of an initial uniform matter distribution. Gamov suggested as a possible explanation a rotation of the universe \cite{Gamov46}. Moreover, he wondered whether anyone had considered anisotropic solutions of Einstein's field equations which would describe a rotating and expanding universe. Einstein approached this idea with skepticism and in his answer to Gamov he pondered: ``What does it mean, that the Universe as a whole possesses an angular momentum?''\footnote{Einstein's letter was written in German. The original phrase reads: ``Was soll es bedeuten, da{\ss} das Universum {\it als Ganzes} ein angular momentum hat?'' \cite{Jung06}}.
It is very likely, that Einstein discussed the ideas of Gamov with G\"odel on one of their walks. 

Nearly three years later, on May 7th, 1949 G\"odel reported in his ``Lecture on rotating universes'' \cite{LectureGoedel} at the Institute for Advanced Studies about his results concerning a new type of cosmological solution of Einstein's field equations. He summarized his results in a seminal paper published in ``Review of Modern Physics'' \cite{Goedel49} on the occasion of Albert Einstein's 70th birthday, as well as in the ``Library of Living Philosophers'' \cite{GoedelPhil}. Moreover, G\"odel improved his original model by including an expansion of the universe while preserving intrinsic rotation \cite{Goedel50}.

\subsection{Metric}

We start our discussion of G\"odel's Universe by introducing its metric. In contrast to the line element presented by G\"odel in his original paper \cite{Goedel49}, we take advantage of the slightly modified expression \cite{Kajari04}
\begin{equation}
\D s^2=c^2\D t^2-\frac{\D r^2}{1+\left(\frac{r}{2a}\right)^2}
-r^2\left(1-\left(\frac{r}{2a}\right)^2\right)\D \phi^2-\D z^2+2r^2\frac{c}{\sqrt{2}a}\D t\,\D \phi\,,
\label{metric}
\end{equation}
where the parameter $a>0$ has the dimension of a length. This form brings out most clearly the interesting property that G\"odel's spacetime approaches the line element of Minkowski spacetime \eqref{eq:dsFlat} in the limit of small radii ${r/\sqrt{a}\rightarrow 0}$. 

A rather lengthy calculation shows that the metric coefficients arising from the line element, eq.~(\ref{metric}), indeed solve Einstein's field equations 
\begin{equation*}
R_{\mu\nu}-\frac{1}{2}g_{\mu\nu}\,R=\kappa\, T_{\mu\nu}+\Lambda\,g_{\mu\nu}
\end{equation*} 
with the energy-momentum tensor
\begin{equation*}
   T_{\mu\nu}\equiv\left(\rho+\frac{p}{c^2}\right) u_\mu u_\nu-p \, g_{\mu\nu}
\end{equation*} 
of an ideal fluid.
As a result, the mass density $\rho$, the pressure $p$ and the cosmological constant $\Lambda$ are directly coupled to the parameter $a$ via the two constraint equations 
\begin{equation*}
\kappa\left(\rho+\frac{p}{c^2}\right)=\frac{1}{a^2c^2}\quad \text{and}\quad
\kappa p=\Lambda+\frac{1}{2a^2}\,.
\end{equation*} 

The particles of the ideal fluid travel along the world lines $x^\mu(\tau)=(\tau,r_0,\phi_0,z_0)$, with $\tau$ being their proper time and $(r_0,\phi_0,z_0)$ their fixed spatial positions. Therefore, the coordinates used in the line element (\ref{metric}) are adapted to the reference frame in which the particles of the ideal fluid are at rest. As a result, the four-velocity of the fluid particles which appears in the energy-momentum tensor $T_{\mu\nu}$ simply reads $u^\mu(\tau)=(1,0,0,0)$. An important characterization of timelike vector fields such as $u^\mu$ can be obtained by the scalar volume expansion, the shear tensor and the rotation tensor \cite{Ehlers61,Hawking06}. Although we will make no further use of these important quantities within our lecture notes, we want to briefly mention, that the volume expansion as well as the shear tensor vanish for the velocity field of the ideal fluid in G\"odel's Universe. However, the rotation tensor is non-zero and it displays a constant rotation of the ideal fluid around the $z$-axis. The magnitude of this rotation is most conveniently expressed in terms of the so-called rotation scalar, which reads in G\"odel's Universe
\begin{equation}
\Omega_{\mathrm G}=\frac{c}{\sqrt{2}a}>0 \,.
\label{rotskalar}
\end{equation}

Moreover, it can be shown that G\"odel's Universe admits five independent Killing vector fields, which we specify in appendix~\ref{AppKillings}. In particular, the existence of a timelike Killing vector field implies the stationarity of G\"odel's spacetime. A more detailed analysis of the Killing vectors shows that G\"odel's metric is spacetime homogeneous and rotationally symmetric, but neither static nor isotropic. The latter feature becomes obvious when we recall the rotation of the ideal fluid about the $z$-axis.

\subsection{Light cone diagram}

\begin{figure}[h]
\centering
\includegraphics[width=0.95\textwidth]{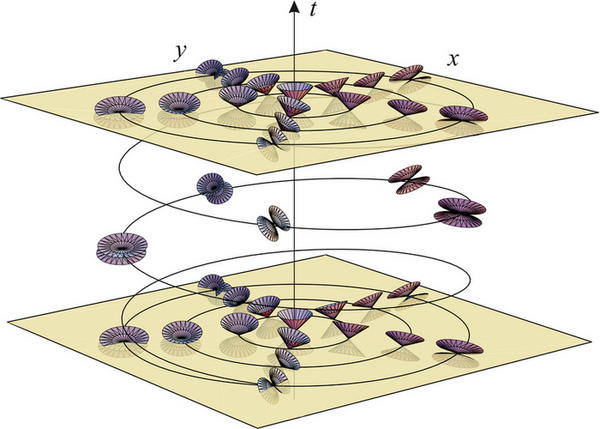}
\caption{Light cone diagram of G\"odel's Universe. We depict two spatial coordinates ${x=r\cos\phi}$ and ${y=r\sin\phi}$, and one time coordinate $t$. The $z$-coordinate has been suppressed. On two planes of constant coordinate time we show the light cones for three typical separations $r$ from the origin. Within the inner circle the light cones behave approximately like in a rotating reference frame in flat spacetime. The next circle corresponds to the G\"odel radius $r=2a$, which divides the plane into two different domains. Here the light cones touch the plane of constant coordinate time. On the outer circle the light cones fall through this plane and therefore allow for closed timelike curves. Time travel is possible as indicated by the spiral connecting the two planes.}
\label{LightconeDiagGoedel}
\end{figure}

As in the case of the rotating reference frame, we take advantage of a light cone diagram to gain insight into the causal structure of G\"odel's Universe. Figure~\ref{LightconeDiagGoedel} displays infinitesimal light cones attached to several spacetime points. Again, we have suppressed the $z$-axis in the figure.

At the origin $r=0$ the light cone coincides with the one in flat spacetime. This feature is in agreement with the limit $r/\sqrt{a}\rightarrow 0$ in eq.~\eqref{metric}, as mentioned in the previous subsection. Moreover, the G\"odel Universe possesses a non-diagonal metric element which couples the angular and the time coordinate, thus leading to a tilting of the light cones for increasing radius. We have noticed this behavior already for the light cones in the rotating reference frame. However, now the light cones do not narrow with increasing radius, but broaden. Indeed, this behavior is of crucial importance for the world lines the fluid particles to be timelike \cite{Schuecking03}.

This combination of broadening and tilting has a dramatic consequence illustrated by the outer circle in the fig.~\ref{LightconeDiagGoedel}. It enables the light cones to intersect the plane of constant coordinate time $t$ along a closed circle in the $x$-$y$-plane, and thus leads to the formation of closed timelike world lines
\begin{equation*}
x^\mu(\tau)=(t_0,r_0,\alpha\,\tau,z_0)\,\quad\text{with}\quad \alpha=
\frac{c}{r_0\sqrt{\left(\frac{r_0}{2a}\right)^2-1}}
\end{equation*} 
for $r_0>2a$. 

The transition between the inner region, where the light cones do not cut through the plane of constant coordinate time $t$ and the outer region where they do, happens at the so-called critical G\"odel radius $r=2a$ which is indicated by the middle circle. 

Even more mind-boggling is the possibility to travel along a timelike world line from any spacetime point to any other one. For example, if we start our journey at the origin $r=0$ of the upper plane of constant coordinate time shown in fig.~\ref{LightconeDiagGoedel} and seek to travel back to the origin $r=0$ of the lower plane, we first follow a timelike world line leading from the origin to a point beyond the critical G\"odel radius. The coordinate time along this world line will of course increase during the first part of our journey. After crossing the critical G\"odel radius, we take a sharp turn to the ``left'' in the spacetime diagram, fig~\ref{LightconeDiagGoedel}, and follow the spiral curve down to earlier coordinate times. Having completed our journey into the ``past of coordinate time", we take another ``left'' turn which brings us back to the origin $r=0$, but now in the lower plane of constant coordinate time. 

Although G\"odel was the first to point out the existence of closed timelike world lines within the framework of general relativity, his universe is not the only model which allows for such peculiar features. The rotating Kerr black hole~\cite{Kerr63,Carter68} and the van Stockum dust cylinder~\cite{Stockum37} are two out of many other prominent examples which admit closed timelike world lines. 

The existence of closed timelike world lines implies the breakdown of our familiar notions of time, chronology and causality. But in most cases it is possible to bring up reasonable arguments which may appease our concerns. They might even lead us to abandon these peculiar solutions of Einstein's field equations and mark them as toy models. However, a general theorem which rules out the possibility of creating a time machine is still missing, although much effort has gone into achieving this goal from first principles. In this context a time machine has to work within a local region of spacetime starting from a well-behaved metric such as the Robertson-Walker metric. Moreover, it has to result from a physically reasonable energy-momentum tensor. Stephen~W.~Hawking established the so-called chronology protection conjecture \cite{Hawking92} which states, that ``the laws of physics do not allow the appearance of closed timelike curves''. Hawking closes his seminal paper \cite{Hawking92} with the penetrative observation:
\begin{quotation}
{\it``There is also strong experimental evidence in favor of the conjecture from the fact that we have not been invaded by hordes of tourists from the future.'' }
\end{quotation}
For a brief and instructive survey on chronology protection we refer to \cite{Visser02}.

\subsection{Sagnac effect}

After this short introduction into G\"odel's spacetime, we proceed with the derivation of the corresponding Sagnac time delays of the DELI, eqs.~(\ref{DELI1}) and~(\ref{DELI2}). We suppose that the observer is at rest with respect to the ideal fluid of G\"odel's Universe, and we denote his spatial position by $(r_0,\phi_0,z_0)$. 

\subsubsection{World line, four-velocity and acceleration}

When we parameterize the world line of the observer in terms of his proper time, we obtain
\begin{equation*}
p^{\mu}(\tau)=(\tau,r_0,\phi_0,z_0)
\end{equation*}
which yields 
\begin{equation*}
u^{\mu}(\tau)=(1,0,0,0)=\delta^\mu_0
\end{equation*}
as corresponding four-velocity in accordance with $u^\mu(\tau)\,u_\mu(\tau)=c^2$.

Recalling the line element~(\ref{metric}) as well as the rotation scalar~(\ref{rotskalar}), the non-vanishing components of the Christoffel symbols read
\begin{equation*}
\begin{split}
\Gamma^0_{\;0 1}&=\frac{\frac{1}{a}\left(\frac{r}{2a}\right)}{1+\left(\frac{r}{2a}\right)^2},\quad
\Gamma^0_{\;1 2}=\frac{\frac{1}{a\Omega_G}\left(\frac{r}{2a}\right)^3}{1+\left(\frac{r}{2a}\right)^2},\quad
\Gamma^1_{\;0 2}=r\,\Omega_G\,\left(1+\left(\frac{r}{2a}\right)^2\right),\\
\Gamma^1_{\;1 1}&=-\frac{\frac{1}{2a}\left(\frac{r}{2a}\right)}{1+\left(\frac{r}{2a}\right)^2},\quad
\Gamma^1_{\;2 2}=-r\left(1+\left(\frac{r}{2a}\right)^2\right)\left(1-2\left(\frac{r}{2a}\right)^2\right),\;\;\\
\Gamma^2_{\;0 1}&=-\frac{\frac{\Omega_G}{r}}{1+\left(\frac{r}{2a}\right)^2},\quad
\Gamma^2_{\;1 2}=\frac{\frac{1}{r}}{1+\left(\frac{r}{2a}\right)^2} \,.
\end{split}
\end{equation*}
They lead to a vanishing four-acceleration of the observer
\begin{equation*}
a^\mu(\tau)=\frac{d^2 p^\mu}{d\tau^2}+\Gamma^\mu_{\;\;\alpha\beta}\,u^\alpha\, u^\beta=0\,.
\end{equation*}

\subsubsection{Tetrad basis and transport matrix}

Next, we introduce the tetrad attached to the observer. As in the case of the rotating reference frame, we assume that the spacelike tetrad vectors point in the same spatial directions as the spatial coordinate axes. Thus, the spatial axes of our observer in G\"odel's Universe will not rotate relative to the stellar compass defined by the null geodesics which connect the luminous particles of the ideal fluid with the observer. Note, that we will analyze the null geodesics in G\"odel's Universe in subsect.~\ref{SecViewinner}.

Once more, we make use of condition~(\ref{e0}) to define the timelike basis vector
\begin{equation*}
 e^\mu_{\;\basind{0}}(\tau)=\frac{1}{c}\,u^\mu(\tau)=\frac{1}{c}\,\delta^\mu_0\,.
\end{equation*}
The first and third spacelike tetrad vectors are now chosen according to
\begin{equation*}
e^\mu_{\;\basind{1}}(\tau)=\sqrt{1+\left(\frac{r_0}{2a}\right)^2}\;\delta^\mu_1\quad\text{and}\quad
e^\mu_{\;\basind{3}}(\tau)=\delta^\mu_3\,.
\end{equation*}
In analogy to the rotating reference frame, we have to include a non-vanishing time component in the second spacelike tetrad vector in order to satisfy the orthogonality condition~(\ref{RelOrthogonality}). The resulting tetrad vector reads
\begin{equation*}
e^\mu_{\;\basind{2}}(\tau)=\frac{1}{\sqrt{1+\left(\frac{r_0}{2a}\right)^2}}\left(-\frac{r_0\Omega_G}{c^2}\,\delta^\mu_0+\frac{1}{r_0}\,\delta_2^\mu\right)\,.
\end{equation*}
Having defined our tetrad basis, we mention, that the tetrad components of the four-acceleration also vanish,
\begin{equation}
a^{\basind{\alpha}}=0\,.
\label{RotAccGeodel}
\end{equation}

With the complete tetrad basis at hand, we are now in the position to consider the transport matrix $\Omega^{\mu \nu}$, eq.~(\ref{TransportMatrix}), in the case of G\"odel's Universe. Using the same formula for the transport matrix as in the rotating reference frame, we end up with
\begin{equation*}
\Omega^{\mu \nu}=-\left(e^\mu_{\;\basind{\alpha};\rho}\, u^\rho\right)\eta^{\basind{\alpha\beta}}\,e^\nu_{\;\basind{\beta}}=
\frac{r_0}{2 a^2}\left(\delta^\mu_0\delta^\nu_1-\delta^\mu_1\delta^\nu_0\right)
-\frac{\Omega_G}{r_0}\left(\delta^\mu_2\delta^\nu_1-\delta^\mu_1\delta^\nu_2\right)\,.
\end{equation*}
In terms of the covariant components this result yields
\begin{equation*}
\Omega_{\mu \nu}=-r_0\,\Omega_G\left(\delta_\mu^2\delta_\nu^1-\delta_\mu^1\delta_\nu^2\right)\,,
\end{equation*} 
and the corresponding tetrad coefficients reduce to
\begin{equation*}
\Omega_{\basind{\alpha\beta}}=\
-\Omega_G\left(\delta^{\basind{2}}_{\basind{\alpha}}\delta^{\basind{1}}_{\basind{\beta}}
-\delta^{\basind{1}}_{\basind{\alpha}}\delta^{\basind{2}}_{\basind{\beta}}\right)\,.
\end{equation*} 
The tetrad rotation vector follows from the transport matrix according to
\begin{equation}
\omega^{\basind{\mu}}=-\frac{1}{2}\varepsilon^{\basind{0\mu\alpha\beta}}\Omega_{\basind{\alpha\beta}}=-\Omega_G\,
\delta^{\basind{\mu}}_{\basind{3}}\,.
\label{RotVecGoedel}
\end{equation}
There are two points, which are of interest concerning the tetrad rotation vector~(\ref{RotVecGoedel}). The first one is the difference in sign when comparing the Sagnac time delay~(\ref{RotVecGoedel}) to the time delay~(\ref{RotVec}) for the observer in the rotating reference frame. This sign difference manifests itself most prominently in the opposite tilting directions of the infinitesimal light cones in the rotating reference frame, fig.~\ref{lightconesrot}, and in G\"odel's spacetime, fig.~\ref{LightconeDiagGoedel}. 

The second point of interest is the relation of inertial and stellar compass for the observer at rest with respect to the ideal fluid in G\"odel's Universe. As mentioned earlier, the stellar compass coincides with the axes of the observer's proper reference frame. However, since the tetrad rotation vector~(\ref{RotVecGoedel}) is non-zero, the tetrad basis rotates relative to the inertial compass. With the discussion of subsect.~\ref{SectRotInRel} in mind, we conclude that G\"odel's spacetime represents a simple example of a stationary, but non-static spacetime in which the inertial compass rotates with respect to the stellar compass.

\subsubsection{Non-vanishing components of the curvature tensor}

Before we provide the Sagnac time delays, eq.~(\ref{DELI1}) and~(\ref{DELI2}), for an observer at rest in G\"odel's Universe, we have to determine the covariant components of the curvature tensor. Apart from the components which follow from the symmetries of the curvature tensor, all non-zero components are given by
\begin{equation*}
\begin{split}
R_{0101}&=-\frac{\Omega_G^2}{1+\left(\frac{r}{2a}\right)^2}\,,\quad
R_{0112}=\frac{2\,\Omega_G\,\left(\frac{r}{2a}\right)^2}{1+\left(\frac{r}{2a}\right)^2}\\
R_{0202}&=-r^2\,\Omega_G^2\left(1+\left(\frac{r}{2a}\right)^2\right)\,,\quad
R_{1212}=-2\left(\frac{r}{2a}\right)^{\hspace*{-0.5ex}2}\,\frac{1+3\left(\frac{r}{2a}\right)^2}{1+\left(\frac{r}{2a}\right)^2}\,.
\end{split}
\end{equation*}
When we now investigate the corresponding tetrad components 
\begin{equation*}
R_{\basind{\alpha}\basind{\beta}\basind{\gamma}\basind{\delta}}=
e^\mu_{\;\basind{\alpha}}\,e^\nu_{\;\basind{\beta}}\,e^\lambda_{\;\basind{\gamma}}\,e^\rho_{\;\basind{\delta}}\,
R_{\mu \nu \lambda \rho}
\end{equation*}
of the covariant curvature tensor, we find a surprisingly simple result, namely
\begin{equation}
R_{\basind{0}\basind{1}\basind{0}\basind{1}}=R_{\basind{0}\basind{2}\basind{0}\basind{2}}
=R_{\basind{1}\basind{2}\basind{1}\basind{2}}=-\frac{1}{2a^2}\,.
\label{TetradRiemann}
\end{equation}
All other tetrad components which do not follow from the symmetry properties of the curvature tensor vanish.

\subsubsection{Sagnac time delays with the DELI}

We now provide the Sagnac time delays, eqs.~(\ref{DELI1}) and~(\ref{DELI2}), for an observer which is at rest with respect to the ideal fluid in G\"odel's Universe. According to the expression for the tetrad rotation vector, eq.~(\ref{RotVecGoedel}), the first measurement mode of the DELI, eq.~(\ref{DELI1}), takes the form
\begin{equation}
\Delta\tau_{Sp}(\mathcal{S}_0)=-\frac{4}{c^2}\,\Omega_G\;
 n^{\basind{3}} \cdot 2\mathcal{A}+
\mathcal{O}\Big(A_{\basind{a}}^{\;\;\basind{i_1}\basind{i_2}}\Big)\,.
\label{DELI1Goedel}
\end{equation} 
Thus, the Sagnac time delay~(\ref{DELI1Goedel}) in G\"odel's Universe differs from the time delay~(\ref{DELI1Rot}) in the rotating reference frame by just the additional factor $\mathcal{N}^2$ and a minus sign. And again, the time delay~(\ref{DELI1Goedel}) does not vanish as long as the unit normal $\mathbf{n}$ to the planar area $\mathcal{A}$ possesses a non-vanishing component in $z$-direction.

In order to establish the Sagnac time delay, eq.~(\ref{DELI2}), we take advantage of the vanishing four-acceleration, eq.~(\ref{RotAccGeodel}), which eliminates all inertial terms within the first-order contribution~(\ref{DELI2}).
Moreover, when we compare the non-vanishing tetrad components of the curvature tensor, eq.~(\ref{TetradRiemann}), with the components listed in table~\ref{SagnacCurvatureComp}, we immediately realize that the first-order contribution, eq.~(\ref{DELI2}), completely vanishes. Whatever orientation we choose for the surface normal $\mathbf{n}$ and for the vector $\mathbf{m}$, we always measure a vanishing tetrad component of the curvature tensor in G\"odel's Universe. For this reason, we simply obtain 
\begin{equation*}
\Delta\tau_{Sp}(\mathcal{S}_1)=
\mathcal{O}\Big(A_{\basind{a}}^{\;\;\basind{i_1}\basind{i_2}}\Big)\,.
\end{equation*} 
This is a quite interesting result since it tells us that we cannot detect any effect of spacetime curvature with a Sagnac interferometer at rest in G\"odel's Universe. Needless to say, that this statement holds true only as long as we restrict ourselves to the first order contribution of the Sagnac time delay.


\section{How things look like in G\"odel's Universe\label{SecHowThings}}

During the last decades many techniques to create realistic pictures with the aid of computers have been developed. This enormous increase in sophisticated methods in computer graphics
has been made possible by a steadily growing computer power. In this section we give a short introduction into ray tracing, a simple but very fundamental technique in computer aided visualization. Standard ray tracing is done in flat spacetime and can be performed with ease. Since in general relativity spacetime is curved
ray tracing of such phenomena is more complicated. For example, Ertl \etal \cite{Ertl89} demonstrated how to use ray tracing to visualize aspects in astrophysics, whereas Weiskopf \etal \cite{Weiskopf00} have developed a ray tracing method to visualize caustic surfaces generated by a gravitational lens. In this section we visualize several scenarios of the G\"odel Universe \cite{Buser08}. In order to gain better insight into the origin of the physical phenomena brought to light in these visualizations, we also discuss some important aspects of the null geodesics in the G\"odel's Universe.

\subsection{Fundamentals of ray tracing\label{SecFundRaytracing}}

\enlargethispage*{5ex}
The way how we perceive our environment visually is determined by the way how light propagates and interacts with the surrounding objects. A complete simulation of such a problem would take a great effort even on modern computers and thus cannot be carried out. 
However, we can obtain acceptable results if we perform various approximations in the light propagation and the interaction of the light with the scattering objects. For example, it may suffice to describe the propagation of light within ray optics. From this point of view, light can be regarded as a bundle of rays, emitted from a light source and bouncing around between the objects of the scenery. Unfortunately, very few of these light rays eventually hit the eye of an observer or a camera.

For the simulation it would be more efficient to only take that light into account which finally passes through the aperture of the camera. Therefore, it is advantageous to consider the light rays as originating from the camera and then meandering through the scenery. This approach is in accordance with Fermat's principle. The path of the light between two points $A$ and $B$ is given by the shortest distance between $A$ and $B$ measured in wavelength. Since the distance between $A$ and $B$ is equal to the distance between $B$ and $A$ the paths of propagation are identical. For this reason we can trace the light back from the camera via the scattering objects to the light source.

\begin{figure}
  \centering
  \begin{minipage}[b]{0.46\textwidth}
    \includegraphics[width=0.9\textwidth]{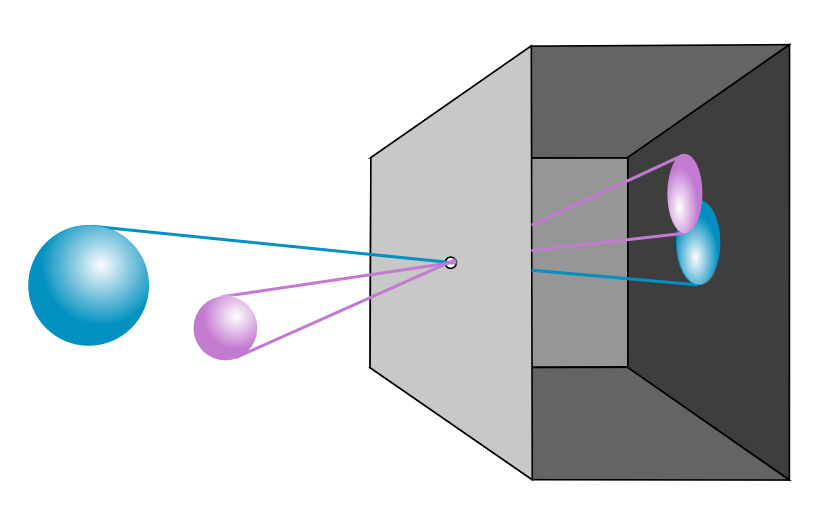} 
  \end{minipage}
\hfill
  \begin{minipage}[b]{0.46\textwidth}
    \includegraphics[width=0.9\textwidth]{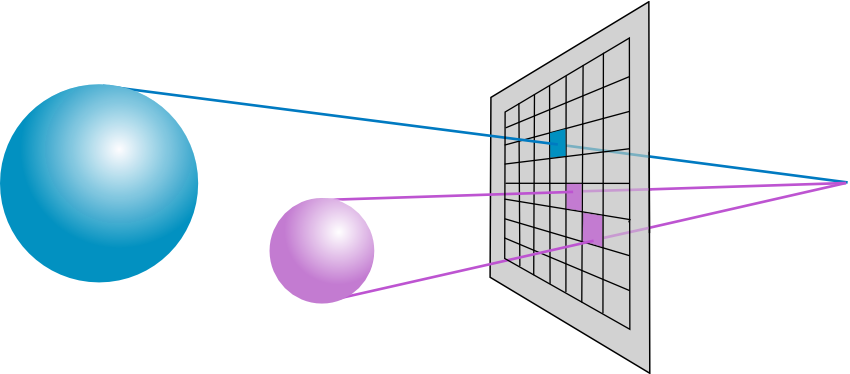}
  \end{minipage}
\caption{A simple model of a camera for visualization. Two spheres are mapped onto the image plane in the back of a pinhole camera (left). We can also use a virtual plane located anywhere between the objects and the camera (right) as an image plane. All light rays which are important for rendering the picture converge in the pinhole. Since the camera is redundant in this situation it is not depicted anymore.}
\label{pinhole_camera}
\end{figure}

In order to bring out the essential ingredients of this idea we consider a pinhole camera. This device is a dark box with a little hole in the front as depicted on the left in fig.~\ref{pinhole_camera}. The light emitted or scattered by the objects propagates through the pinhole and illuminates the film in the back of the box at the corresponding positions. A few light rays on the left of fig.~\ref{pinhole_camera} illustrate the functioning of the pinhole camera.  

In ray tracing we follow the light from the camera to the object rather than from the object to the camera. We construct rays from inside the box through the pinhole and extend them to the scenery. The intersections of such rays with an object define the surface patch which emits the light rays emerging through the pinhole.
An appropriate but finite amount of rays is necessary to obtain an image. However, in order to improve the quality of the image we can take into account the surface properties of the objects such as the color. 

An even simpler version of this method emerges by considering the image plane to be located between the objects and the camera as shown on the right of fig.~\ref{pinhole_camera}. With this virtual plane an image of the scenery is rendered by tracing back a bundle of rays originating from a given point. This source of rays is reminiscent of the hole of the camera. Since the camera itself is dispensable, it is omitted from the picture. 

Since a computer can only trace a limited number of rays, the image plane is divided into small rectangular domains which can be regarded as the pixels of a computer screen. During the rendering process an individual ray is constructed and traced back through each of these tiles. Moreover, we determine how this particular tile needs to be colored. 

This method of composing an approximate image of the scenery is the most elementary version of ray tracing. For this reason it is not surprising that it contains several deficiencies. Shadows as well as diffuse light are missing. Moreover, the transmission through and the reflection from objects need to be included in order to yield more realistic pictures.

Already in 1980 Whitted \cite{Whitted80} presented computer generated pictures which included such effects. He took these phenomena into account by applying a proper illumination model and tracing back additional rays emerging from the intersection points. The illumination model simulates the interaction of the light with the surfaces. However, this approach cannot be applied in a straightforward way to visualize relativistic models. Therefore, for our visualization of the G\"odel Universe we settle for images without those effects.

\begin{figure}[t]
\begin{center}
\includegraphics[width=0.5\textwidth]{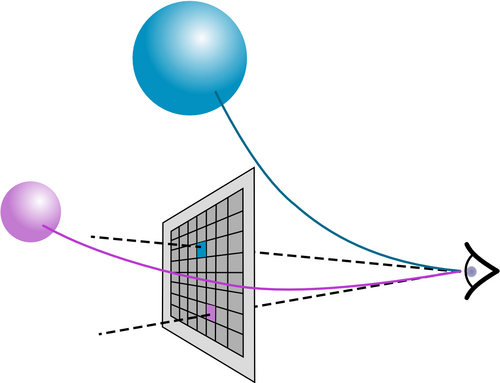}
\end{center}
\caption{Ray tracing in the presence of a gravitational field. Since the light rays are bent an observer at the position of the former pinhole views the two spheres to be straight ahead. However, their real positions differ considerably from his perception, since geometrical optics in curved spacetime is far more complicated than in flat spacetime.}
\label{raytracing_goedel}
\end{figure}

\subsection{Two examples of visualizations of the G\"odel Universe\label{LightPropagationGoe}}

Even the most elementary version of ray tracing has to be adapted for rendering relativistic situations such as the G\"odel Universe. Due to the curvature of spacetime light is propagating differently compared to flat space. In terms of geometrical optics the light rays are bent and coincide with so-called null geodesics discussed in appendix~\ref{Intnullgeo}. We have to obtain them before we can render a picture by ray tracing. The intersection of the light rays with an object can then be found by inching forward along the null geodesics. Unfortunately this procedure is rather time consuming. 

Figure \ref{raytracing_goedel} depicts ray tracing in the presence of gravity with bent light rays originating at the observer's eye. The two objects exemplified here by spheres are above the observer. However, he perceives them to be located straight in front of him.

This example demonstrates the impact of the curvature of spacetime on the visualization. We now choose two scenarios for our visualization and discuss the related aspects of light rays in curved spacetime in these two situations. In order to focus on the visualization we have moved all the mathematics into the appendix~\ref{Intnullgeo}.

\subsubsection{View of the inner surface of a sphere\label{SecViewinner}}

We start our visualization of the G\"odel Universe with a scenario in which an observer looks at the inner surface of a sphere and in which the light rays propagate under the influence of the G\"odel metric. As a consequence the design of the surface gets distorted. The degree of distortion is determined by the null geodesics of the G\"odel metric which we discuss in the next subsection.

\paragraph{Propagation of light in G\"odel's Universe}

According to the general relativistic version of geometrical optics, light rays travel along null geodesics, which correspond to straight lines in the special case of flat spacetime.

However, in curved spacetime they are not straight anymore but get bent by the curvature of spacetime. In mathematical terms the geodesics follow from the differential equation 
\begin{equation}
u^{\mu}_{\;\;;\nu}\, u^{\nu}=0   \label{eq:geodesiceq}
\end{equation}
where 
\begin{equation*}
u^{\mu}(\lambda) \equiv \frac{\D x^{\mu}(\lambda)}{\D \lambda}
\end{equation*}
denotes the tangent vector along the geodesic parameterized by $\lambda$. Since we are dealing with light, $u^{\mu}$ has to fullfill the requirement for being lightlike as expressed by the condition
\begin{equation}
u^{\mu}u_{\mu}=0. 
\label{eq:condition_fourvelocity}
\end{equation}

We now apply eq. (\ref{eq:geodesiceq}) to the G\"odel metric (\ref{metric}). Moreover, we consider an observer located at the origin. As a consequence we are interested in the null geodesics which start from there. Since in this case we have the initial conditions $r(0)=0$, $z(0)=0$ and $t(0)=0$, eq. (\ref{eq:condition_fourvelocity}) reduces to
\begin{equation}
c^2 (u^0(0))^2=(u^1(0))^2+(u^3(0))^2\,. \label{eq:condition_light}
\end{equation}
This expression shows that only the components $0 \le u^1(0)$ and $u^3(0)$ for the initial vector $u^{\mu}(0)$ can be freely chosen. The component $u^0(0)$ is then automatically determined by eq. (\ref{eq:condition_light}). The fact that the initial angular velocity denoted by the component $u^2(0)$ is not well defined at the origin does not influence the calculations.

For an arbitrary spacetime defined by a given metric the differential equation for the null geodesic cannot be solved analytically. In such cases numerical algorithms need to be employed in order to obtain the light rays. Fortunately, in the case of G\"odel's spacetime analytical solutions for the null geodesics starting in the origin exist. They are derived in appendix~\ref{Intnullgeo}.

\begin{figure}
  \centering
  \begin{minipage}[b]{0.54\textwidth}
    \includegraphics[width=0.98\textwidth]{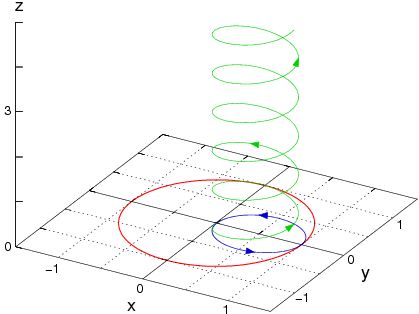} 
  \end{minipage}
\hfill
  \begin{minipage}[b]{0.40\textwidth}
    \includegraphics[width=0.98\textwidth]{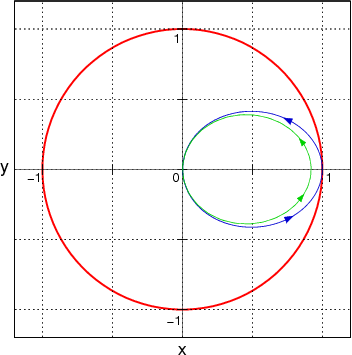}
  \end{minipage}
\caption{Two different null geodesics of the G\"odel metric originating from the center of the coordinate system and represented in the $x$-$y$-$z$-subspace (left) and in the $x$-$y$-subspace (right). The red circle constitutes the G\"odel radius of value 1.0. It represents the maximal radial distance a geodesic, which starts at the origin, can reach. The blue geodesic is defined by the initial velocities $u^1(0)=1.0$ and $u^3(0)=0$. Therefore, it is bound to the $z=0$ plane. In contrast, the green geodesic has a non-zero initial velocity in the $z$-direction. It corresponds to the initial conditions $u^1(0)=1.0$ and $u^3(0)=0.3$ and is of the shape of a spiral aligned along the $z$-axis. In the $x$-$y$-subspace both geodesics are ``ellipses''. However, due to the different values in $u^3(0)$ the major axis of the ellipsoidal green geodesic is smaller than that of the blue one.}
  \label{plot_geodaeten}
\end{figure}

Figure~\ref{plot_geodaeten} shows two null geodesics of the G\"odel Universe corresponding to two different initial velocity vectors $u^{\mu}(0)$. We gain deeper insight into the nature of these geodesics by recalling that the G\"odel spacetime can be decomposed into two independent subspaces. The first one is the $z$-space where the G\"odel Universe is like flat space. As a result, light propagates along the $z$-axis with constant velocity. The second one is formed by the three Cartesian coordinates $x,\,y$ and $t$, or in polar coordinates $r,\,\phi$ and $t$. For the sake of simplicity we omit the time coordinate $t$ in fig.~\ref{plot_geodaeten}.

On the left of fig.~\ref{plot_geodaeten} we display the two null geodesics in three-dimensional space. The first one lies in the $z=0$ plane and is of elliptical shape. The second one conspicuously forms a spiral aligned along the $z$-axis. On the right of fig.~\ref{plot_geodaeten} we show these null geodesics in the $x$-$y$-subspace which brings out most clearly how the rotation contained in the G\"odel spacetime affects the appearance of the geodesics. All geodesics form ``ellipses'' whose areas are determined by the initial velocity component $u^3(0)$ in the $z$-direction. 

Moreover, they are encompassed by a cylinder in the three-dimensional space or a circle in the $x$-$y$-plane with G\"odel radius $2a$. Only geodesics with a vanishing velocity $u^3(0)$ in the $z$-direction can reach the G\"odel radius. With increasing $u^3(0)$ the ``ellipses'' get smaller as discussed in appendix~\ref{Intnullgeo}.

Since the G\"odel radius is the outermost radial point a geodesic can reach, we can also associate with it an optical horizon. Light emitted from the origin cannot cross this border by free propagation. With the same argument we can state that light beyond the G\"odel horizon never reaches the origin. Therefore, the view of an observer in the G\"odel Universe is limited to the area encompassed by a cylinder with the G\"odel radius. Examples for this intriguing effect are provided in the following subsections.

\paragraph{Visualization}

The first scenario gives an impression, how the curvature of spacetime in G\"odel's Universe affects the visual appearance of objects for an observer. We demonstrate this influence of gravity on light by considering a hollow sphere which is viewed by an observer located at its center. We use a three-dimensional sphere with its center located at the origin of the coordinate system. In order to bring out most vividly the distortion of the inner surface of the sphere as it is seen by the observer, we put on it tiles in the form of a checkerboard. 
\begin{figure}[p]
\centering
\begin{minipage}[c]{0.8\textwidth}
 \begin{minipage}[c]{0.45\textwidth}
 \includegraphics[width=\textwidth]{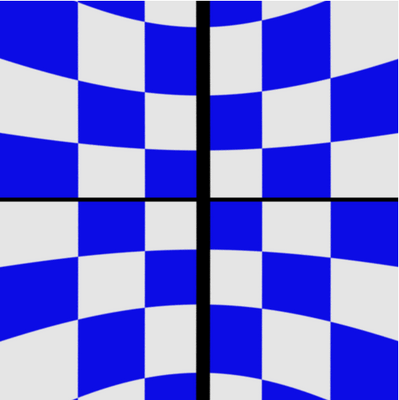}
 \end{minipage}
  \hfill
  \begin{minipage}[c]{0.45\textwidth}
  \includegraphics[width=\textwidth]{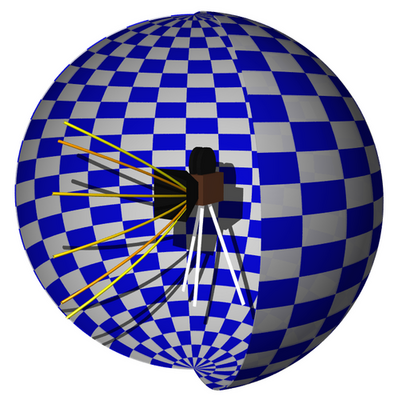}
  \end{minipage}
\end{minipage}
\\[0.2cm]
\begin{minipage}[c]{0.8\textwidth}
 \begin{minipage}[c]{0.45\textwidth}
 \includegraphics[width=\textwidth]{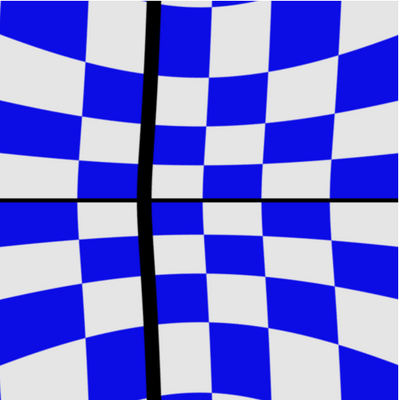}
 \end{minipage}
  \hfill
  \begin{minipage}[c]{0.45\textwidth}
  \includegraphics[width=\textwidth]{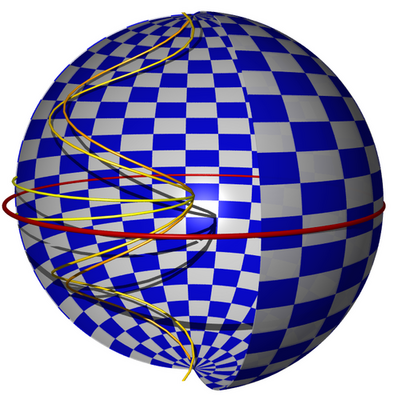}
  \end{minipage}
\end{minipage}
\\[0.2cm]
\begin{minipage}[c]{0.8\textwidth}
 \begin{minipage}[c]{0.45\textwidth}
 \includegraphics[width=\textwidth]{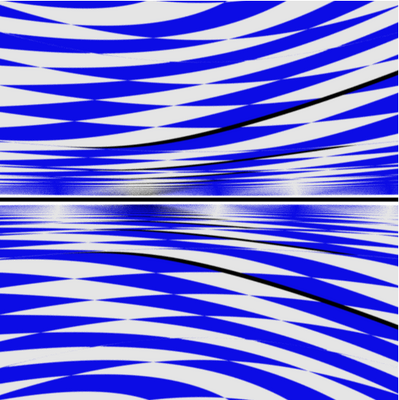}
 \end{minipage}
 \hfill
 \begin{minipage}[c]{0.45\textwidth}
  \includegraphics[width=\textwidth]{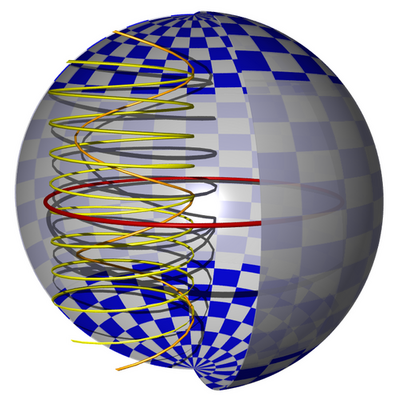}
  \end{minipage}
\end{minipage}
  \caption{
View of the inner surface of a hollow sphere from the center of the G\"odel Universe for three different G\"odel radii. Due to the gravitational field induced by the G\"odel metric the light rays are deflected as shown by the three examples in the right column. As a consequence the inner surface of the sphere tiled with a checkerboard appears distorted as indicated by the left column. Both, the horizontal and the vertical aperture take the value of $70^\circ$. The G\"odel radius is in units of the radius of the sphere. When the G\"odel radius is 5.0 (top) the light rays propagate almost on straight lines. In the case of the G\"odel radius represented by a red circle to be identical to the radius of the sphere (middle) the light rays follow spirals and the checkerboard is substantially distorted. When the G\"odel radius again represented by a red circle is smaller than the radius of the sphere (bottom) the light rays cannot even reach the surface in the neighborhood of the equator as depicted by the gray belt. This part of the surface is inaccessible to the observer. However, the light rays can reach the northrn and the southern hemisphere. On their way they undergo many revolutions on the spiral. As a result the final point on the surface depends sensitively on the exiting angle of the ray at the camera leading to a almost self-similar picture of the checkerboard.}
  \label{hohlkugel}
\end{figure}

Figure \ref{hohlkugel} depicts the scenario for three different G\"odel radii. In the right column of this figure we show typical light rays illustrating paths on which light from the surface of the sphere reaches the observer. The pictures in the left column give an indication, how the observer would perceive the sphere in the particular situation. In all situations we have used a horizontal and vertical aperture angle of the camera of $70^\circ$. The equator is at the center of the aperture and indicated in the left row of fig.~\ref{hohlkugel} by a black horizontal line.

In the top scenario the G\"odel radius is five times larger than the radius of the sphere. Hence, the curvature of spacetime is almost negligible within the sphere and the light rays are bent only slightly. This feature leaves the view of the observer almost unchanged compared to the one in flat space. Longitudes on the sphere remain vertical as exemplified by the black bar shown in the top left figure.

In the next scenario the G\"odel radius is equal to that of the sphere. In this case the light rays are bent considerably. For example, light from the poles propagates along a spiral until it reaches the origin. In contrast light emitted in the neighborhood of the equator is bent due to the small velocity in $z$-direction. The black vertical bar in the middle left figure representing a longitude confirms this statement and now displays a bump at the equator. 

The scenario on the bottom corresponds to a situation where the G\"odel radius is smaller than the one of the sphere. As a result the equatorial area depicted in the bottom right figure as a gray belt around the sphere is located beyond the G\"odel radius. Since the G\"odel radius acts as an optical horizon the belt is out of sight for the observer. Instead he sees several copies of the remaining visible part of the sphere. Indeed, rays close to the equatorial plane need to rotate up or down on a spiral several times before they intersect the sphere. As a consequence the observer has several copies of the visible parts of the sphere in his field of view and the checkerboard is warped with increasing frequency around the equator. The vertical bar is broken up numerously close to the equator. It shows an almost chaotic behavior which can be explained by the enourmous sensitivity of the light rays to small changes of the initial velocity component in the $z$-direction.

We conclude our discussion of this scenario by noting that in a small vicinity of the equator a white stripe appears. The stripe is the result of numerical limitations because after inching along the light ray for some time this procedure has to be abandoned.

\enlargethispage*{5ex}

\subsubsection{View on a small object in the G\"odel Universe\label{SecViewsmallObj}}

\begin{figure}[t]
\begin{center}
\begin{minipage}{0.8\textwidth}
\includegraphics[width=0.4\textwidth]{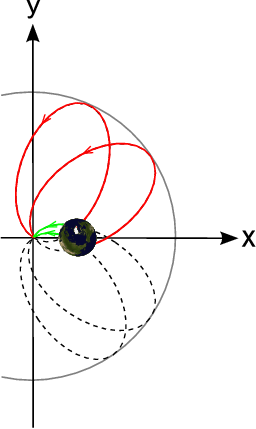}
\hfill
\includegraphics[width=0.4\textwidth]{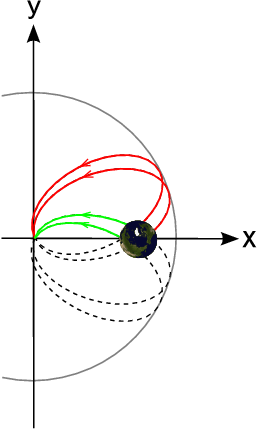}
\end{minipage}
\end{center}
\caption{Physical origin of the two images of a single object such as a terrestrial globe at rest in the G\"odel Universe. We depict the paths of the light rays in the $x$-$y$-subspace. The first image originates from light rays emitted from the front surface of the object and depicted by green solid lines. The null geodesics of those light rays are continued as dotted curves and are of typical elliptical shape as discussed in fig.~\ref{plot_geodaeten}. The second image is due to light rays emitted from the back of the globe and reflected from the G\"odel radius depicted by the gray circle. These light rays indicated by the red lines enable us to observe the back of the globe. However, the reflection from the G\"odel horizon creates a mirror image of the original surface.
}
\label{goedel1}
\end{figure}

\begin{figure}[p]
\centering

 \begin{minipage}[c]{\textwidth}
 \includegraphics[width=\textwidth]{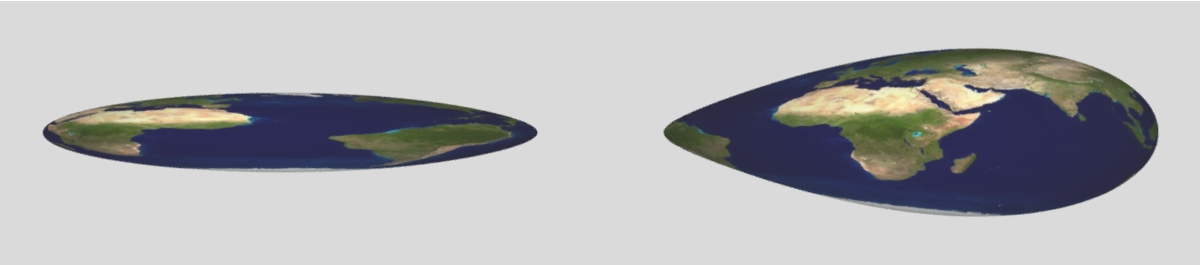}
 \label{r08}
 \end{minipage}

  \begin{minipage}[c]{\textwidth}
  \includegraphics[width=\textwidth]{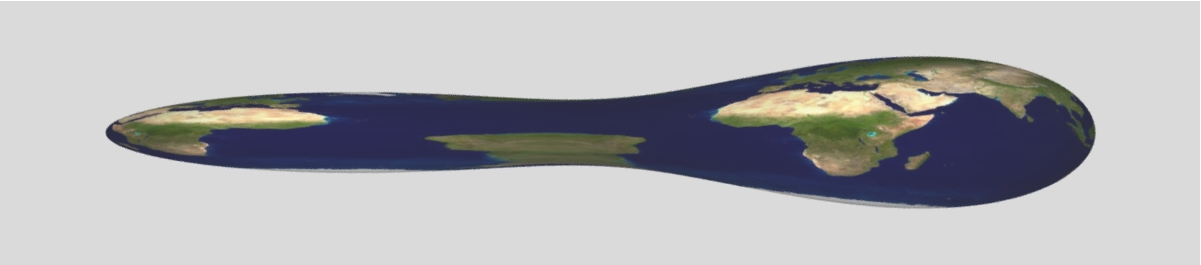}
  \label{r09}
  \end{minipage}

  \begin{minipage}[c]{\textwidth}
  \includegraphics[width=\textwidth]{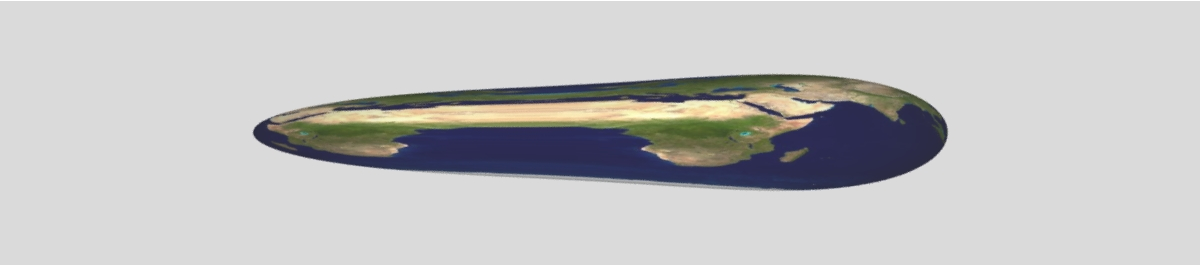}
  \label{r10}
  \end{minipage}

  \begin{minipage}[c]{\textwidth}
  \includegraphics[width=\textwidth]{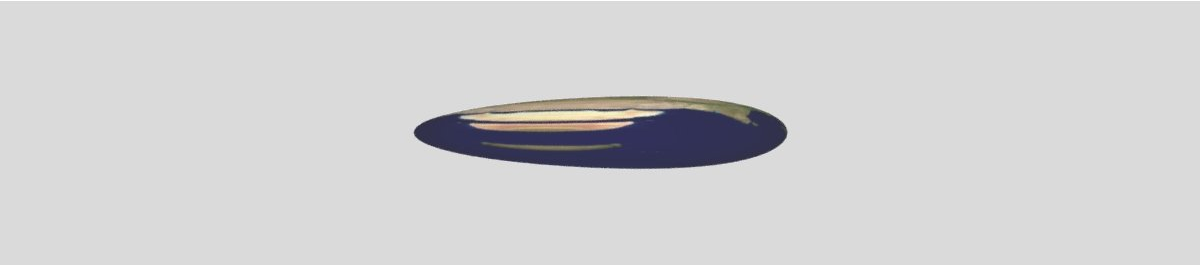}
  \label{r11}
  \end{minipage}
  \caption{
The view of an observer at the origin of the G\"odel Universe on a terrestrial globe located at different positions (top to bottom) on the $x$-axis. In all four situations the sphere has a radius of 0.15 expressed in units of the G\"odel radius. The observer is looking in the $y$-direction with a vertical aperture of $20^\circ$ and a horizontal aperture of $90^\circ$. The top picture shows the sphere at $x=0.8$. Since in this case it is completely within the G\"odel radius, two separate images of the sphere are visible. The right one originates from light rays emitted at the front of the sphere. The left one shows part of its back and relies on rays reflected from the G\"odel horizon. As a result this image is the mirror image of the backside. The next three pictures show the sphere located at $x=0.9$, $x=1.0$ and $x=1.1$, respectively. At these positions the angle between the direct and the reflected light rays decreases and the two images move closer together and merge. Moreover, the sphere reaches beyond the optical horizon and the parts cut off by the G\"odel radius are out of sight.}
  \label{erde_goedel}
\end{figure}

The existence of an optical horizon in G\"odel's Universe manifests itself in another compelling optical effect. A single object at rest has two images. Our second scenario discussed in the present section visualizes this effect.

\paragraph{The G\"odel horizon as a mirror}

Figure \ref{goedel1} illustrates the physical origin of the two images and their dependence on the location of the object relative to the horizon. For this purpose we consider the emission of light from a terrestrial globe. We recognize two fundamentally different type of light rays propagating from the globe towards the origin. On the green paths the light comes in a direct way from the part of the globe facing towards the observer. In contrast the red paths make a detour and go first to the G\"odel horizon before they arrive at the origin. In this way we can observe the back side of the globe. However, due to the reflection of the light rays at the G\"odel radius, the second image of the globe is mirror-inverted. Furthermore, the different lengths of the paths make both images appear in different sizes.

\paragraph{Visualization}

In fig.~\ref{erde_goedel} we now visualize the phenomenon of the two images. The globe is positioned on the $x$-axis and has a radius of 0.15 expressed in units of the G\"odel radius. As suggested by fig.~\ref{goedel1} the rays emitted by the globe arrive at the origin propagating in the neighborhood of the $y$-axis. For this reason the observer is looking in the $y$-direction. His vertical aperture is $20^\circ$ and his horizontal aperture is $90^\circ$.

Figure \ref{erde_goedel} shows the view of the observer when the sphere is located at different positions. In the top scenario the separation of the center of the sphere to the origin is 0.87. Consistent with fig.~\ref{goedel1}, the globe has two images. The right image arise from the short paths and shows the front part of the sphere. In contrast, the left image is due to the long paths and is mirror inverted. 

With increasing separation of the globe from the origin the angle between the light rays corresponding to the direct and the reflected paths decreases as shown in fig.~\ref{goedel1}. This feature manifests itself in the subsequent pictures of fig.~\ref{erde_goedel} where the globe has a separation of 0.9, 1.0 and 1.1, respectively. The two images move closer together and even start to merge. In these scenarios parts of the globe are invisible since they are located beyond the G\"odel radius.


\acknowledgments
We would like to thank \textsc{J.~Bolte}, \textsc{I.~Ciufolini}, \textsc{J.~Ehlers}, \textsc{J.~Frauendiener}, \textsc{D.~Giulini}, \textsc{R.~F.~O'Connell} and \textsc{H.~Pfister} for many helpful discussions and useful hints on several references. Special thanks go to \textsc{C.~L\"ammerzahl} for critical reading of the manuscript and numerous fruitful suggestions. We have enjoyed the wonderful atmosphere at the ``Enrico Fermi'' summer school in Varenna and appreciated the financial support from the ``Wilhelm und Else Heraeus-Stiftung''. Moreover, we want to thank \textsc{F.~Grave}, \textsc{T.~M\"uller}, \textsc{H.~Ruder}, \textsc{G.~Wunner} for the fruitful collaboration and the DFG (German Research Foundation) for the financial support of the project ``Visualisierung geschlossener zeitartiger Kurven in der Allgemeinen Relativit\"atstheorie''. 
As part of the QUANTUS project, we appreciate the support of the German Space Agency DLR with funds provided by the Federal Ministry of Economics and Technology (BMWi) under grant number DLR 50 WM 0837. Finally, we thank the European Space Agency (ESA) for the coordination of the Space Atom Interferometer (SAI) project under grant number AO-2004-64/082. One of the authors, E.~Kajari, would also like to thank \textsc{M.~Eckart} and \textsc{S.~Kajari-Schr\"oder} for their assistance in the preparation of this work.



\appendix

\section{Basic concepts of general relativity: A brief review\label{AppendixBasicConcept}}

This appendix is addressed towards readers with a background mainly in quantum optics. However, we assume that they have already attended some introductory lectures in tensor calculus \cite{Lovelock} and general relativity. The purpose of this appendix is (i) to provide the definitions and properties of several tensorial quantities used throughout this article, and (ii) to offer a rudimentary review of some basic concepts in general relativity. In~\ref{AppConvention} we provide the definitions of some important tensorial quantities. We dedicate appendix~\ref{MinkowskiForm} to transform the metric at a point of spacetime to its Minkowski form. A useful application of this method is the construction of light cone diagrams. We then briefly mention the connection of symmetries and Killing vectors in \ref{AppSymmetries} and proceed in~\ref{WorldLineGeodesic} with the introduction of geodesics. In this context we expose the relation between the constants of motion of the geodesics and continuous symmetries of a given spacetime. In~\ref{AppParalleltransport} we consider parallel and Fermi-Walker transport.

\subsection{Tensors and all that jazz \label{AppConvention}}

A quantity $V^{\nu_1 \ldots \nu_n}_{\;\;\;\;\;\;\;\;\;\;\;\;\,\mu_1\ldots\mu_m}$ constitutes the components of a tensor of type $(n,m)$ if it transforms under a coordinate transformation $x^\mu =x^\mu(x'^\nu)$ according to
\begin{equation*}
V'^{\,\alpha_1 \ldots \alpha_n}_{\;\;\;\;\;\;\;\;\;\;\;\;\,\beta_1\ldots\beta_m}= \pdiff{x'^{\alpha_1}}{x^{\nu_1}}\cdot\ldots\cdot\pdiff{x'^{\alpha_n}}{x^{\nu_n}}\, \pdiff{x^{\mu_1}}{x'^{\beta_1}}\cdot\ldots\cdot\pdiff{x^{\mu_m}}{x'^{\beta_m}}
\, V^{\nu_1 \ldots \nu_n}_{\;\;\;\;\;\;\;\;\;\;\;\;\,\mu_1\ldots\mu_m}\,.
\end{equation*}
Here $n$ and $m$ denote the number of contravariant and covariant indices, respectively.
In particular, we find the transformations
\begin{align*}
V'^\mu=\pdiff{x'^{\mu}}{x^{\nu}}\,V^\nu  \quad \text{and}\quad
V'_\mu=\pdiff{x^{\nu}}{x'^{\mu}}\,V_\nu   
\end{align*}
for a contravariant vector $V^\mu$ and a covariant vector $V_\mu$. \par
The Christoffel symbols
\begin{equation*}
\Gamma^\mu_{\;\alpha\beta}\equiv
\frac{1}{2}\,g^{\mu\nu}\left(g_{\nu\alpha,\beta}+g_{\nu\beta,\alpha}-g_{\alpha\beta,\nu}\right)
\end{equation*} 
given in terms of the metric coefficients $g_{\mu\nu}$ are symmetric in the lower indices, that is  $\Gamma^\mu_{\;\alpha\beta}=\Gamma^\mu_{\;\beta\alpha}$. However, they do not constitute the components of a tensorial quantity.

In this article we denote the ordinary partial derivative by a comma. Since partial derivatives of tensors are not  tensors anymore, it is convenient to define a generalized derivative which preserves the tensorial character. We indicate the so-called covariant derivative
\begin{align*}
V^{\nu_1 \ldots \nu_n}_{\;\;\;\;\;\;\;\;\;\;\;\;\,\mu_1\ldots\mu_m\,;\alpha}=V^{\nu_1 \ldots \nu_n}_{\;\;\;\;\;\;\;\;\;\;\;\;\,\mu_1\ldots\mu_m\,,\alpha} &+\sum_{j=1}^n \Gamma^{\nu_j}_{\;\rho\alpha}\,
V^{\nu_1\ldots\nu_{j-1}\,\rho\,\nu_{j+1} \ldots \nu_n}_{\;\;\;\;\;\;\;\;\;\;\;\;\;\;\;\;\;\;\;\;\;\;\;\;\;\;\;\;\;\;\;\;\,\mu_1\ldots\mu_m} \\
&-\sum_{j=1}^m \Gamma^{\rho}_{\;\mu_j\alpha}\,
V^{\nu_1 \ldots \nu_n}_{\;\;\;\;\;\;\;\;\;\;\;\;\,\mu_1\ldots\mu_{j-1}\,\rho\,\mu_{j+1}\ldots\mu_m} \nonumber
\end{align*} 
of a tensor $V^{\nu_1 \ldots \nu_n}_{\;\;\;\;\;\;\;\;\;\;\;\;\,\mu_1\ldots\mu_m}$ by a semicolon.
In particular, we find
\begin{align*}
V^\alpha_{\,\;;\beta} \equiv V^\alpha_{\;\,,\beta}+\Gamma^\alpha_{\;\mu\beta}\,V^\mu \quad\text{and}\quad
V_{\alpha;\beta} \equiv V_{\alpha,\beta}-\Gamma^\mu_{\;\alpha\beta}\,V_\mu
\end{align*}
for a contravariant vector $V^\alpha$ and a covariant vector $V_\alpha$. 

In general, the covariant derivative of a tensor is more complicated than the partial derivative.  However, in the case of the metric $g_{\mu\nu}$ the covariant derivative vanishes, that is ${g_{\mu\nu;\alpha}=0}$. Moreover, the covariant derivative of a product of two tensor fields satisfies a product rule familiar from partial differentiation.

The curvature tensor  
\begin{equation}
R^\mu_{\;\,\alpha\beta\gamma}\equiv
\Gamma^\mu_{\;\alpha\gamma,\beta}-\Gamma^\mu_{\;\alpha\beta,\gamma}
+\Gamma^\mu_{\;\rho\beta}\Gamma^\rho_{\;\alpha\gamma}-
\Gamma^\mu_{\;\rho\gamma}\Gamma^\rho_{\;\alpha\beta}
\label{DefCurvatureTensor}
\end{equation}
possesses the covariant components ${R_{\mu\alpha\beta\gamma}\equiv g_{\mu\nu}\, R^\nu_{\;\,\alpha\beta\gamma}}$ given by 
\begin{equation*}
 R_{\mu\alpha\beta\gamma}
=\frac{1}{2}\left(g_{\mu\gamma,\alpha,\beta}+g_{\alpha\beta,\mu,\gamma}-
g_{\mu\beta,\alpha,\gamma}-g_{\alpha\gamma,\mu,\beta} \right)
+g_{\nu\rho}\left( \Gamma^\nu_{\;\mu\gamma} \Gamma^\rho_{\;\alpha\beta}-
\Gamma^\nu_{\;\mu\beta}\Gamma^\rho_{\;\alpha\gamma}\right)\,.
\end{equation*}
This expression reveals the symmetry relations
\begin{equation}
R_{\mu\alpha\beta\gamma}=-R_{\alpha\mu\beta\gamma}=-R_{\mu\alpha\gamma\beta}\quad\text{and}\quad
R_{\mu\alpha\beta\gamma}=R_{\beta\gamma\mu\alpha}
\label{RiemannTensorCovSymm}  
\end{equation}
as well as the Bianchi identities
\begin{align*}
R_{\mu\alpha\beta\gamma} +R_{\mu\beta\gamma\alpha}+R_{\mu\gamma\alpha\beta} &=0 
\end{align*}
and 
\begin{align*}
R_{\mu\alpha\beta\gamma;\nu}+R_{\mu\alpha\nu\beta;\gamma}+R_{\mu\alpha\gamma\nu;\beta}&=0 \,.
\end{align*}
We obtain the Ricci tensor 
\begin{equation*}
R_{\alpha\beta}\equiv g^{\mu\nu}\, R_{\mu\alpha\nu\beta}= R^\mu_{\;\,\alpha\mu\beta}
\end{equation*}
and the Ricci scalar  
\begin{equation*}
R\equiv R^\mu_{\;\,\mu}= g^{\mu\nu}\, R_{\mu\nu}
\end{equation*}
from the curvature tensor by contraction.

The four-dimensional Levi-Civita-Symbol 
\begin{equation}
\Delta^{\alpha\beta\gamma\delta}\equiv\left\{\begin{array}{l} \phantom{-}1\quad \mbox{for an even permutation}\\
-1\quad \mbox{for an odd permutation}\\
\phantom{-}0\quad\mbox{otherwise}\end{array}\right.
\label{LeviCevita}
\end{equation}
is used in the definition
\begin{equation}
\varepsilon^{\alpha\beta\gamma\delta}\equiv\frac{1}{\sqrt{-g}}\Delta^{\alpha\beta\gamma\delta}
\label{EpsilonTensor}
\end{equation}
of the four-dimensional antisymmetric tensor $\varepsilon^{\alpha\beta\gamma\delta}$.

It is convenient to introduce the shorthand notation 
\begin{equation}
V_{\{\alpha_1\ldots\alpha_n\}}\equiv\frac{1}{n!}\sum_{\pi \in \mathcal{S}_n} V_{\alpha_{\pi(1)}\ldots\alpha_{\pi(n)}}
\label{SymmBrackets}
\end{equation} 
and 
\begin{equation}
V_{[\alpha_1\ldots\alpha_n]}\equiv\frac{1}{n!}\sum_{\pi \in \mathcal{S}_n} \op{sign}(\pi) \, V_{\alpha_{\pi(1)}\ldots\alpha_{\pi(n)}}
\label{AntiBrackets}
\end{equation} 
for the totally symmetric and the totally antisymmetric part of a tensor, respectively.
Here the sums are taken over all permutations $\mathcal{S}_n$ of $(1,\ldots,n)$ and $\op{sign}(\pi)$ corresponds to the value  $+1$ for even and $-1$ for odd permutations. 

It will also be necessary to consider the totally symmetric part of a tensor $V_{\alpha_1\ldots\alpha_n}$ with respect to a certain subset $\alpha_1\ldots\alpha_k$ of indices, where $k\leq n$. The notation
\begin{equation}
\underset{(\alpha_1\ldots\alpha_k)}{\op{Symm}}\left[V_{\alpha_1\ldots\alpha_n}\right]\equiv
\frac{1}{k!}\sum_{\pi \in \mathcal{S}_k} V_{\alpha_{\pi(1)}\ldots\alpha_{\pi(k)}\alpha_{k+1}\ldots\alpha_n}=
V_{\{\alpha_1\ldots\alpha_k\}\alpha_{k+1\ldots}\alpha_n}
\label{SubsetSymmetrization}
\end{equation}
indicates, that the symmetrization is performed only with respect to the indices which are listed in parentheses below the symmetrization symbol ``$\op{Symm}$''.

The tangent 
\begin{equation*}
u^\mu(\lambda)\equiv\frac{d x^\mu}{d\lambda}
\end{equation*} 
of a spacetime curve $x^\mu(\lambda)$ with the evolution parameter $\lambda$ can be classified by the three categories
\begin{align}
 g_{\mu\nu}\,u^\mu \,u^\nu \left\{\begin{array}{l} <0 \quad \mbox{spacelike}\\ =0 \quad \mbox{lightlike}\\
 > 0 \quad \mbox{timelike.}\\
\end{array} \right.
\end{align}
We call a spacetime curve $x^\mu(\lambda)$ {\it timelike} if its tangent is timelike for all parameters $\lambda$. Null curves or spacelike curves are defined analogously.

\subsection{The Minkowski form of a metric at a fixed spacetime point\label{MinkowskiForm}}

The metric $g_{\mu\nu}$ of a given spacetime has the signature $-2$. As a consequence the number of positive eigenvalues minus the number of negative eigenvalues of the matrix $g_{\mu\nu}(p^\sigma)$ at any fixed point $P$ in spacetime with coordinates $p^\sigma$ is equal to $-2$. This property allows us to transform the metric $g_{\mu\nu}(p^\sigma)$ at any fixed event $P$ to the Minkowski metric 
\begin{equation*}
(\eta_{\mu\nu})=\op{diag}(1,-1,-1,-1) \,,
\end{equation*}
as discussed in what follows.

\subsubsection{Transformation to Minkowski metric}

When we perform a change of coordinates $x^\alpha=x^\alpha(\bar x^\beta)$, we arrive at the transformed metric 
\begin{equation}
\bar g_{\mu\nu}(\bar p^\sigma)=\pdiff{x^\alpha}{\bar x^\mu}\bigg|_P\pdiff{x^\beta}{\bar x^\nu}\bigg|_P g_{\alpha\beta}(p^\sigma(\bar p^\varrho))=T^\alpha_{\;\;\mu}(p^\sigma)\, T^\beta_{\;\;\nu}(p^\sigma)\, g_{\alpha\beta}(p^\sigma)
\label{metrictransform}
\end{equation}
at $P$. Here we have introduced the transformation matrix
\begin{equation}
T^\alpha_{\;\;\mu}(p^\sigma)\equiv\pdiff{x^\alpha}{\bar x^\mu}\bigg|_P\,,
\end{equation}
which is not a tensor even though it is written in a tensor like manner. 

In order to avoid a confusing index notation we now turn to a matrix representation and denote the metric components by the matrices ${g\equiv (\,g_{\mu\nu}(p^\sigma)\,)}$ and ${\bar g\equiv (\,\bar g_{\mu\nu}(\bar p^\sigma)\,)}$, and the transformation matrix by ${T\equiv(T^\alpha_{\;\;\mu}(p^\sigma))}$. In this representation eq.~(\ref{metrictransform}) reads 
\begin{equation*}
\bar g=T^\tp g\,T\,.
\end{equation*}
Since $g$ is a symmetric matrix, it can be diagonalized by an orthogonal matrix ${O=(O^\alpha_{\;\;\mu})}$, resulting in the diagonal matrix $D\equiv\op{diag}(\lambda_0,-\lambda_1,-\lambda_2,-\lambda_3)=O^\tp g\, O$. Here we have assumed $\lambda_i>0$ according to the signature $-2$. Moreover, we have arranged the orthonormal eigenvectors in $O$ such that the orientation in the spatial subspace remains right-handed. 

Furthermore, we define the diagonal matrix  $$C\equiv\op{diag}((\lambda_0)^{-\frac{1}{2}},(\lambda_1)^{-\frac{1}{2}},(\lambda_2)^{-\frac{1}{2}},(\lambda_3)^{-\frac{1}{2}})  \\=(C^\alpha_{\;\;\mu})$$
 with the help of the eigenvalues of $g$ and introduce the Lorentz transformation matrix $\Lambda=(\Lambda^\mu_{\;\;\alpha})$ which satisfies the condition $\Lambda^\tp \eta\, \Lambda=\eta$ with the Minkowski metric $\eta=(\eta_{\mu\nu})$. The transformation matrix $T=O\,C\,\Lambda$ enables us to reduce the transformed metric $\bar g$ to the Minkowski spacetime $\eta$ which yields
\begin{equation*}
\bar g=(O\,C\,\Lambda)^\tp g\,(O\,C\,\Lambda)=\Lambda^\tp \eta\,\Lambda=\eta\,.
\end{equation*}
We emphasize that the matrices $O$ and $C$ are determined by the values of the metric components $g_{\mu\nu}(p^\sigma)$ at the fixed point $P$, whereas the matrix $\Lambda$, which corresponds to homogeneous Lorentz transformations can be chosen arbitrarily within the limits of its defining equation $\Lambda^\tp \eta\, \Lambda=\eta$. Throughout this article we restrict ourselves to proper Lorentz transformations \cite{Weinberg}, whose coefficients obey $\Lambda^0_{\;\;0}\geq 1$ and $\op{det}\,\Lambda=1$.

Thus, we have shown that a coordinate transformation $x^\alpha=x^\alpha(\bar x^\beta)$ with transformation matrix
\begin{equation}
T^\alpha_{\;\;\beta}(p^\sigma)=\pdiff{x^\alpha}{\bar x^\beta}\bigg|_P=O^\alpha_{\;\;\mu}\,C^\mu_{\;\;\nu}\,\Lambda^\nu_{\;\;\beta}\,,
\label{TransformationMatrix}
\end{equation}
at $P$ transforms $g_{\mu\nu}(p^\sigma)$ to the Minkowski metric $\bar g_{\mu\nu}(\bar p^\sigma)=\eta_{\mu\nu}$ at this point. 

The inverse Matrix $S=T^{-1}$ is also of importance, since it appears in the transformation law of a contravariant vector $V^\mu$. According to the decomposition of $T$, eq.~(\ref{TransformationMatrix}), we find
\begin{equation}
S^\alpha_{\;\;\beta}(p^\sigma)=\pdiff{\bar x^\alpha}{x^\beta}\bigg|_P\,\quad\text{with}\quad S=T^{-1}=\Lambda^{-1}\,C^{-1}\,O^{-1} \,.
\label{InverseTransformationMatrix}
\end{equation}
Hence, the vectors in the tangent space at $P$ undergo a four-dimensional rotation $O^{-1}$, a rescaling $C^{-1}$ and a Lorentz transformation $\Lambda^{-1}$ by the corresponding coordinate transformation. \\

\subsubsection{Construction of light cone diagrams}

We now show that this transformation method provides a valuable tool for gaining insight into the causal structure of a given spacetime. In general, the question whether two points $P_1$ and $P_2$ in spacetime can be connected by a timelike or null curve is not a trivial question. However, light cone diagrams, such as figs.~\ref{lightconesrot} and~\ref{LightconeDiagGoedel}, may be helpful in finding an answer to this question. For that reason we now explain their construction in more detail.

All null curves $x^\mu(\lambda)$ which pass through the point $P$ satisfy the relation
\begin{equation}
 g_{\mu\nu}(p^\sigma)\,u^\mu\,u^\nu=0\,.
\label{DefNullvector}
\end{equation}
When we perform a coordinate transformation with transformation matrix $T$, eq.~(\ref{TransformationMatrix}), and inverse $S$, eq.~(\ref{InverseTransformationMatrix}), at the point $P$, we find 
\begin{equation}
\bar u^\alpha=S^\alpha_{\;\;\mu}(p^\sigma)\,u^\mu\quad\text{and}\quad
u^\mu=T^\mu_{\;\;\beta}(p^\sigma)\,\bar u^\beta
\label{utransform}
\end{equation}
for the contravariant components of the tangent vector. Hence, eq.~(\ref{DefNullvector}) turns into
\begin{equation*}
 \bar g_{\mu\nu}(\bar p^\sigma)\,\bar u^\mu\,\bar u^\nu= \eta_{\mu\nu}\,\bar u^\mu\,\bar u^\nu=0\,.
\end{equation*}
As in special relativity, the last equation is fulfilled by all vectors which are element of the light cone in Minkowski spacetime. Thus, we take this set of vectors $\bar u^\mu$ and transform them back to the components $u^\mu=T^\mu_{\;\;\nu}(p^\sigma)\,\bar u^\nu$ of the tangent vector expressed in the original coordinates $x^\mu$. This transformation leads effectively to a rescaling of the apex angle and a tilting of the light cone depending on the metric coefficients at the point $P$. We then attach this transformed light cone to the point $P$, always keeping in mind that these light cones are objects which live in the tangent space and only illustrate the tangents of the null curves passing through $P$. By repeating this procedure for many spacetime points, we finally arrive at a light cone diagram.

We conclude by briefly pointing out that the freedom in the choice of the Lorentz transformation matrix $\Lambda$ within the transformation matrix $T$, eq.~(\ref{TransformationMatrix}), does not have any effect on the construction of the light cone diagram. Indeed, the set of all vectors constituting the light cone is mapped onto itself under a homogeneous Lorentz transformation.

\subsection{Symmetries and Killing vectors \label{AppSymmetries}}

A spacetime manifold endowed with a metric $g_{\mu\nu}$ possesses a symmetry, if there exists a special class of coordinate transformations $x'^\alpha=x'^\alpha(x^\beta)$ which does not change the functional dependence of the metric on the coordinates, that is
\begin{equation}
g'_{\mu\nu}(x'^\sigma)=\pdiff{x^\alpha}{x'^\mu}\pdiff{x^\beta}{x'^\nu}\, g_{\alpha\beta}(x^\sigma)\overset{!}{=}g_{\mu\nu}(x'^\sigma) \,.
\label{iso}
\end{equation}
Such a coordinate transformation is called an {\it isometry}.

\subsubsection{Killing equations}

Of special interest are infinitesimal isometries
\begin{equation}
x'^\alpha=x^\alpha+\varepsilon\xi^\alpha(x^\beta)
\label{infiso}
\end{equation}
with $|\varepsilon|\ll1$, since every continuous symmetry transformation can be reassembled successively by them. 

In order to make use of the infinitesi{\-}mal isometry~(\ref{infiso}) we bring the invariance condition~(\ref{iso}) into the form
\begin{equation*}
g_{\alpha\beta}(x^\eta)=\pdiff{x'^\mu}{x^\alpha}\pdiff{x'^\nu}{x^\beta}g_{\mu\nu}(x^\eta+\varepsilon\xi^\eta(x^\sigma))\,.
\end{equation*}
From eq.~(\ref{infiso}) we find 
\begin{equation*}
 \pdiff{x'^\mu}{x^\alpha}=\delta^{\mu}_{\;\,\alpha}+\varepsilon\xi^\mu_{\;,\alpha}(x^\sigma)
\end{equation*}
and with the Taylor expansion of $g_{\mu\nu}(x^\eta+\varepsilon\xi^\eta(x^\sigma))$ around $x^\eta$ up to first order in $\varepsilon$, we obtain
\begin{equation}
g_{\alpha\beta,\nu}(x^\eta)\, \xi^\nu(x^\eta)+ g_{\mu\beta}(x^\eta)\,\xi^\mu_{\,,\alpha}(x^\eta) +g_{\mu\alpha}(x^\eta)\,\xi^\mu_{\,,\beta}(x^\eta)=0\,.  \label{Killing0}
\end{equation}
When we make use of the fact that the covariant derivative of the metric $g_{\alpha\beta;\nu}$ vanishes, we finally arrive at the Killing equations
\begin{equation}
\xi_{\alpha;\beta}+\xi_{\beta;\alpha}=0
\label{Killinggl}
\end{equation}
for the Killing vector field $\xi^\alpha(x^\beta)$. The solutions $\xi^\alpha$ of this linear system 
of partial differential equations fully characterize the continuous symmetries of a given metric.

\subsubsection{Symmetries}

Our first important example of a symmetry is the time independence of a metric. If a metric possesses a timelike Killing vector field, we call the metric {\it stationary}, since this particular vector field allows the introduction of special coordinates in which the metric is no longer time dependent.

Moreover, we call a metric {\it homogeneous in space and time} if there exist four independent Killing vector fields as solutions of eq.~\eqref{Killinggl} which are linearly independent in the tangent space of every point of the spacetime manifold. In this case it is always possible to connect two arbitrary events by a continuous isometry such that any spacetime point is equivalent to any other with respect to the metric. This feature provides a natural generalization of the translational invariance property of space and time in flat Minkowski spacetime. The concepts of isotropy and purely spatial homogeneity would need a more sophisticated treatment which cannot be given here. Instead, we would like to refer to \cite{Weinberg,Wald}.

\subsection{World lines and geodesics of test particles and of light\label{WorldLineGeodesic}}

Particles with a non-vanishing mass as well as observers travel on timelike curves through spacetime. Hence, a timelike curve is often referred to as {\it world line}. It is sometimes convenient to use a reference frame that is at rest with respect to an observer moving along a world line. The time measured by such an observer is called {\it proper time} $\tau$ and it is connected to the line element $\D s^2$ according to
\begin{equation}
\D s^2=g_{\mu\nu}\,\D x^\mu\,\D x^\nu\equiv c^2\,\D \tau^2\,. \label{propertime}
\end{equation} 
The proper time is typically used as evolution parameter for a world line $x^\sigma(\tau)$.

When we divide the last equation by $d\tau^2$, we find the fundamental condition
\begin{equation*}
g_{\mu\nu}(x^\sigma(\tau))\,u^\mu(\tau)\,u^\nu(\tau)= c^2
\end{equation*}
for the four-velocity $u^\mu(\tau)$. 

Light, on the other hand, evolves on lightlike curves $x^{\mu}(\lambda)$, also called "null curves", corresponding to the line element $ds^2=0$. Therefore, it is not possible to define a proper time for light in the same way as for massive particles. Instead, we consider the evolution parameter $\lambda$ just as a label for the spacetime point of the light ray with no deeper physical meaning.

Since we would like to have a formulation combining the calculations for light and particles, we introduce the short notation
\begin{equation}
g_{\mu\nu}\,u^\mu \,u^\nu=u_\mu\,u^\mu=\epsilon^2 \,,  \label{ucontraction}
\end{equation} 
with $\epsilon=0$ corresponding to light, and with $\epsilon=c$ and $\lambda=\tau$ for massive particles. \par

Freely falling particles and freely propagating light rays follow a special class of spacetime curves called {\it geodesics}, which obey the geodesic equation
\begin{equation}
 \frac{\D^2x^\mu}{\D \lambda^2}+\Gamma^\mu_{\;\alpha\beta}(x^\sigma)\frac{\D x^\alpha}{\D \lambda}
     \frac{\D x^\beta}{\D \lambda}=u^\mu_{\;\,;\nu}\,u^\nu=0  \,.
\label{GeodesicEq}
\end{equation}
To obtain a particular geodesic one has to solve this equation for the initial values of position $x^\mu(\lambda_0)$ and four-velocity $u^\mu(\lambda_0)$. We emphasize that the initial four-velocity has to satisfy the condition~(\ref{ucontraction}). 

It is important to note that by contracting eq.~(\ref{GeodesicEq}) with $u_\mu$, one can show that the scalar $u_\mu\,u^\mu=\epsilon^2$, eq.~(\ref{ucontraction}), is a constant of motion and does not change its value along the geodesic.

In general it is not easy to find analytical solutions of the geodesic equation.
However, intrinsic symmetries of the metric can ease this task since they correspond to constants of motion. In order to find them we observe, that the contraction of the geodesic eq.~(\ref{GeodesicEq}) by a Killing vector field $\xi_\mu$ on both sides yields
\begin{equation*}
0=\xi_\mu u^\mu_{\;\,;\nu}u^\nu=\left(\xi_\mu u^\mu\right)_{;\nu} u^\nu-\xi_{\mu;\nu}u^\mu u^\nu.
\end{equation*}
By virtue of the Killing equations~(\ref{Killinggl}) the covariant derivative of the Killing vector $\xi_{\mu;\nu}$
is antisymmetric and thus the term $\xi_{\mu;\nu}u^\mu u^\nu$ vanishes. 
Moreover, the expression $\xi_{\mu}u^{\mu}$ is a scalar. Therefore, its covariant derivative can 
be rewritten as a partial derivative, that is 
\begin{equation*}
0=\left(\xi_\mu u^\mu\right)_{;\nu} u^\nu = 
 u^{\nu}\frac{\partial}{\partial x^{\nu}}(\xi_{\mu}u^{\mu})=\diff{}{\lambda}(\xi_{\mu}u^{\mu}) \,.
\end{equation*}
Hence, we can assign to every Killing vector a constant of motion 
\begin{equation}
u_{\mu}(\lambda)\,\xi^{\mu}(x^\sigma(\lambda))=C=u_{\mu}(\lambda_0)\,\xi^{\mu}(x^\sigma(\lambda_0))
\label{ConstMotion}
\end{equation} 
of the geodesic equation. 

Finally, we point out that the geodesic equation~(\ref{GeodesicEq}) is only valid in the presence of gravitational forces. In the case of an additional, non-gravitational four-force $K^\mu(x^\sigma)$ acting on a massive particle, such as the electromagnetic force, we must extend the geodesic equation to the general equation of motion
\begin{equation}
m \,u^\mu_{\;\,;\nu}\,u^\nu=m\,a^\mu=K^\mu  \,.
\label{fourforcemotion}
\end{equation}
Here we have introduced the four-acceleration $a^\mu\equiv u^\mu_{\;\,;\nu}\,u^\nu$ and the rest mass $m$ of the particle. 

We emphasize that condition~(\ref{ucontraction}) still holds true, since we have chosen the proper time $\tau$ as the curve parameter for massive particles. When we take the covariant derivative of eq.~(\ref{ucontraction}) we find
\begin{equation}
u_\mu\,u^\mu_{\,\;;\nu}=0 \label{diffucontraction}\,.
\end{equation} 
Moreover, when we contract the equation of motion~(\ref{fourforcemotion}) with $u_\mu$ we conclude with the help of 
eq.~(\ref{diffucontraction}) that the four-acceleration $a^\mu$ and the four-force $K^\mu$ have to satisfy the additional constraint
\begin{equation}
u_\mu\,a^\mu=0 \quad \text{and} \quad u_\mu\,K^\mu=0\,.
\label{constraintonaccel}
\end{equation}
Hence, the definition of the four-velocity $u^\mu$ in terms of the proper time $\tau$ is sufficient to imprint the constraint eqs.~(\ref{constraintonaccel}) on the four-acceleration and the four-force appearing in the equation of motion (\ref{fourforcemotion}).

\subsection{Parallel transport versus Fermi-Walker transport\label{AppParalleltransport}}

In contrast to Euclidean geometry, there exists no absolute concept of parallelism in curved spacetime. There is only a definition of parallelism of vectors along a curve $\mathcal{C}$. To be specific, let the curve $\mathcal{C}$ be defined parametrically by $x^\mu(\lambda)$ with corresponding tangent vector $u^\mu(\lambda)$. We then call a vector $V^\mu(\lambda_0)$ located at the spacetime point $x^\mu(\lambda_0)$ {\it parallel} to another vector $V^\mu(\lambda_1)$ at $x^\mu(\lambda_1)$ along $\mathcal{C}$, if there exists for all $\lambda_0\leq\lambda\leq\lambda_1$ a solution $V^\mu(\lambda)=V^\mu(x^\sigma(\lambda))$ of the system of linear differential equations
\begin{equation}
V^\mu_{\;\;;\nu}\,u^\nu=\diff{V^\mu(\lambda)}{\lambda}
+\Gamma^\mu_{\;\alpha\beta}(x^\sigma(\lambda))\,u^\alpha(\lambda)\,V^\beta(\lambda)=0\,, 
\label{paralleltransport}
\end{equation} 
which coincides with $V^\mu(\lambda_0)$ at $\lambda_0$ and with $V^\mu(\lambda_1)$ at $\lambda_1$.

The differential equation~(\ref{paralleltransport}) is usually referred to as the {\it equation of parallel transport}, since one can invert the above statement: given an initial vector $V^\mu(\lambda_0)$ and a spacetime curve $\mathcal{C}$, then the unique solution of eq.~(\ref{paralleltransport}) for these initial conditions determines all vectors parallel to $V^\mu(\lambda_0)$ along $\mathcal{C}$. One important consequence of this fact is, that for a given closed curve $\mathcal{C}$ with initial and final point $x^\mu(\lambda_0)=x^\mu(\lambda_1)$, the initial and the parallel transported vector do in general not coincide, that is $V^\mu(\lambda_0)\neq V^\mu(\lambda_1)$. 

The parallel transport possesses two pleasant properties:  it preserves \mbox{(i) the} ``spacetime length'' ${V^\mu(\lambda)\,V_\mu(\lambda)}$ of the initial vector $V^\mu(\lambda_0)$ as can be seen by contraction of eq.~(\ref{paralleltransport}) with $V_\mu$, and (ii) the ``spacetime angle'' $W_\mu(\lambda)\,V^\mu(\lambda)$ when the two vector fields $V^\mu$ and $W^\mu$ are both parallel transported along the same curve $\mathcal{C}$. The latter statement follows directly from a contraction of the parallel transport equation~(\ref{paralleltransport}) with $W_\mu$.

\begin{figure}
\centering
\includegraphics[width=0.8\textwidth]{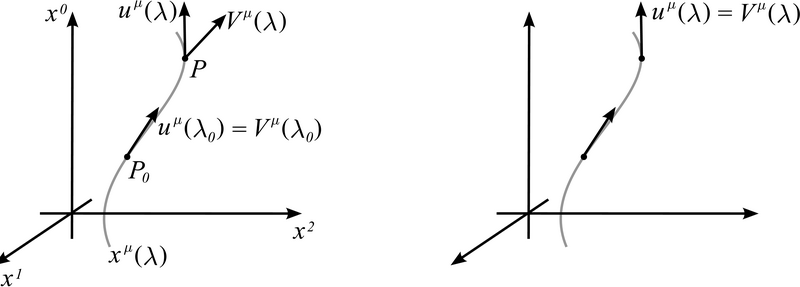}
\caption{Comparison between parallel transport (left) and Fermi-Walker transport (right) of a vector $V^\mu(\lambda)$ along a curve $x^\mu(\lambda)$. In both cases we start at the point $P_0$ with the tangent vector $V^\mu(\lambda_0)=u^\mu(\lambda_0)$ and end up at the point $P$. In general, the parallel transported vector (left) at $P$ differs from the tangent vector $u^\mu(\lambda)$ in $P$. However, for the Fermi-Walker transport (right), the vector $V^\mu(\lambda)$ still coincides with $u^\mu(\lambda)$.
In order to bring out the similarities and differences between the two pictures most clearly, we have marked in the right picture only those quantities which are different from the ones in the left picture.}
\label{picparalleltransport}
\end{figure}
When we choose the tangent vector $u^\mu(\lambda_0)$ as the initial vector $V^\mu(\lambda_0)$ at the point $P_0$, we can observe that in general the parallel transported vector $V^\mu(\lambda)$ does not coincide anymore with the tangent vector $u^\mu(\lambda)$ at the point $P$, as indicated by the left picture of fig.~\ref{picparalleltransport}, unless the spacetime curve $x^\mu(\lambda)$ is a geodesic. In the special case of a geodesic, the tangent vector $V^\mu=u^\mu$ is parallel propagated along the geodesic itself. For this reason geodesics are sometimes called {\it autoparallel curves}. 

However, often one would like to have a transport law for vectors $V^\mu$ along arbitrary curves, which maps the initial tangent vector $V^\mu(\tau_0)=u^\mu(\tau_0)$ located at $P_0$ to the tangent vector $V^\mu(\tau)=u^\mu(\tau)$ at $P$, as depicted in the right drawing of fig.~\ref{picparalleltransport}. The so-called Fermi-Walker transport incorporates this special feature, and it is defined by the generalization
\begin{equation}
V^\mu_{\;\;;\nu}\,u^\nu+\frac{1}{c^2}\left(u^\mu a_\beta-a^\mu u_\beta\right)V^\beta=
\diff{V^\mu}{\tau}+\left[\Gamma^\mu_{\;\alpha\beta}\,u^\alpha+\frac{1}{c^2}
\left(u^\mu a_\beta-a^\mu u_\beta \right)\right]V^\beta=0
\label{FermiWalkertransport}
\end{equation}
of eq.~(\ref{paralleltransport}).
Here we have used the definition $a^\mu=u^\mu_{\;\;;\nu}\,u^\nu$ of the four-acceleration and have parameterized our world line $x^\mu(\tau)$ with the proper time $\tau$ of the observer. Indeed, by inserting the four-velocity $V^\mu=u^\mu$ into eq.~(\ref{FermiWalkertransport}) and making use of eq.~(\ref{constraintonaccel}), which holds true for any world line $x^\mu(\tau)$, we can verify that $u^\mu$ is a solution of the Fermi-Walker transport equation. As for the parallel transport, we can show without much effort, that the Fermi-Walker transport eq.~(\ref{FermiWalkertransport}) preserves the ``spacetime length'' ${V^\mu(\tau)\,V_\mu(\tau)}$ 
and the ``spacetime angle'' ${W_\mu(\tau)\,V^\mu(\tau)}$ for two vectors, which are Fermi-Walker transported along the same world line $\mathcal{C}$.
Finally, we would like to mention that for $\mathcal{C}$ being a geodesic with four-acceleration $a^\mu=0$, the Fermi-Walker transport, eq.~(\ref{FermiWalkertransport}), reduces to the parallel transport law~(\ref{paralleltransport}).

\section{Tetrads and their orthonormal transport\label{TetradsAndTransport}}

In this appendix, we briefly introduce the concept of orthonormal tetrads in general relativity and examine several important properties. Since this topic is often omitted in introductory courses on general relativity, we discuss it extensively following the spirit of~\cite{Synge}. 
Finally, we analyze the general orthonormal transport of tetrads and motivate a natural generalization of the Fermi-Walker transport.

\subsection{Orthonormal tetrads\label{AppTetrad}}

Let $e^\mu_{\,\basind{\alpha}}$ be four linearly independent basis vectors spanning the tangent space at $P$. Here we have placed the index $(\alpha)\in\{0,1,2,3\}$ in parenthesis to indicate that it is not a tensor index, but a label for  the particular basis vector. If these four basis vectors satisfy the relativistic orthonormality condition
\begin{equation}
 e^\mu_{\,\basind{\alpha}}\, e^\nu_{\,\basind{\beta}}\,g_{\mu\nu}=\eta_{\basind{\alpha\beta}}\,,
\label{OrthonormalTetrad}
\end{equation}
we call the four contravariant vectors $e^\mu_{\,\basind{\alpha}}$ an {\it orthonormal tetrad}.

The invariant matrix $\eta_{\basind{\alpha\beta}}=\op{diag}(1,-1,-1,-1)$ is defined in analogy to the flat spacetime metric. However, it is important to keep in mind, that its indices $(\alpha\beta)$ do not denote tensor indices\footnote{Throughout this article indices in parenthesis appear only in connection with tetrad indices. We emphasize that this notation is in contrast to the standard literature. 
In order to avoid ambiguities in the notation, we have denoted the symmetrization brackets, defined by eq.~(\ref{SymmBrackets}), by curly brackets.}.
Therefore, the matrix $\eta_{\basind{\alpha\beta}}$ remains invariant under a coordinate transformation. We can verify this statement by  performing a coordinate change $x'^\mu=x'^\mu(x^\sigma)$. Since the tetrad vectors  $e^\mu_{\,\basind{\alpha}}$ transform like ordinary contravariant vectors, the orthonormality condition reads in the new coordinates
\begin{equation*}
 e'^\mu_{\,\basind{\alpha}}\, e'^\nu_{\,\basind{\beta}}\,g'_{\mu\nu}=\eta_{\basind{\alpha\beta}}\,.
\end{equation*}
Furthermore, we label the vectors of the tetrad in a way such that $e^\mu_{\,\basind{0}}$ is timelike and the remaining spacelike vectors $e^\mu_{\,\basind{i}}$ provide a right-handed basis of the spatial subspace.

\subsubsection{Construction}

The problem of finding an orthonormal tetrad in the tangent space at a certain point $P$ of a given spacetime with metric coefficients $g_{\mu\nu}$ is solved by simply taking advantage of the transformation matrix $T$, eq.~(\ref{TransformationMatrix}), which brings the metric at $P$ into its Minkowski form. Thus, the contravariant components of an orthonormal tetrad at $P$ are simply found by identifying them with the corresponding elements 
\begin{equation}
  e^\mu_{\,\basind{\alpha}}\equiv T^\mu_{\;\,\basind{\alpha}}(p^\sigma)=
 O^\mu_{\;\basind{\nu}}\,C^{\basind{\nu}}_{\;\,\basind{\rho}}\,\Lambda^{\basind{\rho}}_{\;\,\basind{\alpha}}
\label{TetradTransformationMatrix}
\end{equation}
of the transformation matrix. 

An orthonormal tetrad constructed in this way is not uniquely defined, since the transformation matrix $T$ contains an arbitrary Lorentz transformation matrix $\Lambda$. For this reason, two arbitrary orthonormal tetrads $\tilde e^\mu_{\,\basind{\alpha}}$ and $e'^\mu_{\,\basind{\alpha}}$ can always be connected by a Lorentz transformation matrix $\Lambda$. This property holds true, because the orthogonal matrix $O$ and the scaling matrix $C$ in the decomposition, eq.~(\ref{TetradTransformationMatrix}), are determined by the metric coefficients at $P$, whereas the Lorentz matrix $\Lambda$ can be freely chosen. If we decompose the two tetrads according to
\begin{equation*}
 \tilde e^\mu_{\,\basind{\alpha}}\equiv  \tilde T^\mu_{\;\,\basind{\alpha}}(p^\sigma)=
 O^\mu_{\;\basind{\nu}}\,C^{\basind{\nu}}_{\;\,\basind{\rho}}\,\tilde
 \Lambda^{\basind{\rho}}_{\;\,\basind{\alpha}}
\end{equation*}
and
\begin{equation*}
  e'^\mu_{\,\basind{\alpha}}\equiv T'^\mu_{\;\,\,\basind{\alpha}}(p^\sigma)=
  O^\mu_{\;\basind{\nu}}\,C^{\basind{\nu}}_{\;\,\basind{\rho}}\,\Lambda'^{\basind{\rho}}_{\;\;\basind{\alpha}}
\end{equation*}
and use the fact, that the Lorentz transformations form a group and thus can be decomposed according to $\tilde\Lambda^{\basind{\rho}}_{\;\,\basind{\alpha}}=\Lambda'^{\basind{\rho}}_{\;\;\basind{\beta}}\,
\Lambda^{\basind{\beta}}_{\;\,\basind{\alpha}}$, we find
\begin{equation}
 \tilde e^\mu_{\,\basind{\alpha}}
=\Lambda^{\basind{\beta}}_{\;\,\basind{\alpha}}\, e'^\mu_{\,\,\basind{\beta}}\,.
\label{TetradConnection}
\end{equation}
The defining equation of a Lorentz matrix reads in tetrad index notation
\begin{equation}
\Lambda^{\basind{\alpha}}_{\;\;\basind{\mu}}\,\Lambda^{\basind{\beta}}_{\;\;\basind{\nu}}\,\eta_{\basind{\alpha\beta}}=\eta_{\basind{\mu\nu}}\,, 
\label{DefLorentz}
\end{equation} 
and the inverse $(\Lambda^{-1})^{\basind{\beta}}_{\;\,\basind{\alpha}}$ of a Lorentz matrix follows herefrom according to
\begin{equation}
(\Lambda^{-1})^{\basind{\beta}}_{\;\;\basind{\alpha}}=
\eta_{\basind{\alpha\gamma}}\,\Lambda^{\basind{\gamma}}_{\;\;\basind{\rho}}\,\eta^{\basind{\rho\beta}}\,.
\label{InverseLorentz}
\end{equation} 

\begin{figure}
\centering
\includegraphics[width=0.8\textwidth]{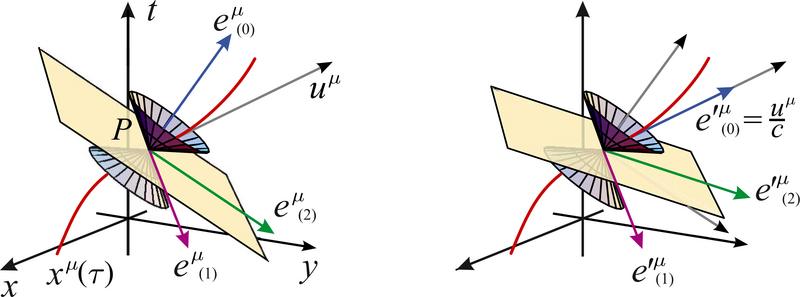}
\caption{Construction of an orthonormal tetrad at a point $P$ in spacetime. The observer passes through $P$ along the world line $x^\mu(\tau)$. When we choose a decomposition (\ref{TetradTransformationMatrix}) of the tetrad $e^\mu_{\,\basind{\alpha}}$ with the special Lorentz matrix $\Lambda^{\basind{\rho}}_{\;\,\basind{\alpha}}=\delta^{\basind{\rho}}_{\basind{\alpha}}$,
the timelike basis vector $e^\mu_{\;\basind{0}}(\tau)$ is not parallel (left) to the four-velocity $u^\mu(\lambda)$ at $P$. However, it is always possible to determine a special Lorentz boost $\Lambda^{\basind{\rho}}_{\;\,\basind{\alpha}}$ in the decomposition (\ref{TetradTransformationMatrix}) which aligns the two vectors (right).
In order to bring out the similarities and differences between the two pictures most clearly, we have marked in the right picture only those quantities which are different from the ones in the left picture.}
\label{LocalLightcone}
\end{figure}

We emphasize that the proper choice of the Lorentz matrix $\Lambda$ in (\ref{TetradTransformationMatrix}) is crucial when we desire a particular orthonormal tetrad. We illustrate this circumstance in fig.~\ref{LocalLightcone}, where we have suppressed the $z$-axis. The two images show two different orthonormal tetrads together with the infinitesimal light cone in the tangent space at the spacetime point $P$. Moreover, a timelike curve $x^\mu(\tau)$ with four-velocity $u^\mu$ at $P$ is included. The left picture displays the special case of an  orthonormal tetrad $e^\mu_{\,\basind{\alpha}}$ with the identity as Lorentz matrix 
${\Lambda^{\basind{\rho}}_{\;\,\basind{\alpha}}=\delta^{\basind{\rho}}_{\basind{\alpha}}}$ in its decomposition~(\ref{TetradTransformationMatrix}). 
However, we can reduce the number of possible tetrads by demanding, that the timelike basis vector 
$e'^\mu_{\;\basind{0}}$ of the tetrad $e'^\mu_{\;\basind{\alpha}}$ should coincide with the normalized four-velocity $u^\mu/c$ in $P$. Indeed, this requirement leads to a particular Lorentz boost. The right picture in fig.~\ref{LocalLightcone} serves as an example of this special case. Here, the additional Lorentz boost maps the basis vectors $e^\mu_{\,\basind{0}}$ and $ e^\mu_{\,\basind{2}}$ onto $e'^\mu_{\,\,\basind{0}}=u^\mu/c$ and $e'^\mu_{\,\,\basind{2}}$ just as in the Minkowski diagrams of special relativity, thereby leaving $e^\mu_{\,\basind{1}}=e'^\mu_{\,\basind{1}}$ invariant. Nevertheless, a rotation in the three-dimensional subspace orthogonal to $e'^\mu_{\,\,\basind{0}}$ can still be subjoined as an additional Lorentz matrix $\Lambda$ in the decomposition~(\ref{TetradTransformationMatrix}). Thus the restriction $e'^\mu_{\,\,\basind{0}}=u^\mu/c$ provides us with enough freedom in the choice of the spacelike basis vectors $e'^\mu_{\,\,\basind{i}}$.

\subsubsection{Decomposition into tetrad components}

Before we turn over to the general, orthonormal transport law of tetrads, we introduce some notations, which considerably shorten the subsequent calculations. 

\paragraph{Co- and Contravariant Components of Tetrads}

We denote the covariant components of the tetrad $e^\mu_{\,\basind{\alpha}}$ by lowering the tensor index as usual according to $e_{\mu\basind{\alpha}}\equiv g_{\mu \nu}\, e^\nu_{\,\basind{\alpha}}$. Thus, the relativistic orthonormality condition~(\ref{OrthonormalTetrad}) reads  
\begin{equation*}
e^\mu_{\,\basind{\alpha}}\, e_{\mu\basind{\beta}}=\eta_{\basind{\alpha\beta}} \,.
\end{equation*}
With the help of the invariant inverse $\eta^{\basind{\alpha\beta}}=\op{diag}(1,-1,-1,-1)$ of $\eta_{\basind{\alpha\beta}}$ and an analogous summation convention, we can adopt the lowering and raising also for the parenthesized indices by defining
\begin{equation}
 e^{\mu\basind{\alpha}}\equiv\eta^{\basind{\alpha\beta}} e^\mu_{\,\basind{\beta}}\quad\text{and}\quad
e_{\mu}^{\;\basind{\alpha}}\equiv\eta^{\basind{\alpha\beta}} e_{\mu\basind{\beta}}\,.
\label{Indexraising}
\end{equation}
When we multiply the last expressions by $\eta_{\basind{\alpha\gamma}}$, we obtain
\begin{equation*}
e^\mu_{\,\basind{\gamma}}=\eta_{\basind{\gamma\alpha}} e^{\mu\basind{\alpha}}
\quad\text{and}\quad
e_{\mu\basind{\gamma}}=\eta_{\basind{\gamma\alpha}} e_{\mu}^{\;\basind{\alpha}} \,.
\end{equation*}
After multiplying eq.~(\ref{OrthonormalTetrad}) by $\eta^{\basind{\beta\gamma}}$ and using eq.~(\ref{Indexraising}), we can reformulate the orthonormality condition in a much more intuitive fashion
\begin{equation}
 e^\mu_{\,\basind{\alpha}}\, e_\mu^{\;\basind{\gamma}}=\delta^{\basind{\gamma}}_{\basind{\alpha}}\,.
\label{OrthoNice1}
\end{equation}
This equation suggests that $e_\mu^{\;\basind{\gamma}}=g_{\mu \nu}\eta^{\basind{\gamma\alpha}}\, e^\nu_{\,\basind{\alpha}}$ is the inverse matrix to our tetrad $ e^\mu_{\,\basind{\alpha}}$. But this interpretation implies furthermore, that the tensorial relation
\begin{equation}
 e_\mu^{\;\basind{\alpha}}\,e^\nu_{\,\basind{\alpha}}=\delta^\nu_\mu
\label{OrthoNice2}
\end{equation}
holds as well.

\paragraph{Tetrad indices}

A useful application of the tetrad formalism is that every vector $V^{\mu}$ and tensor $W^{\mu\nu}$, can be decomposed into tetrad components 
\begin{equation*}
V^{\mu}=V^{\basind{\alpha}}\, e^\mu_{\,\basind{\alpha}}\quad\text{and}\quad
W^{\mu\nu}=W^{\basind{\alpha\beta}}\, e^\mu_{\,\basind{\alpha}}\, e^\nu_{\,\basind{\beta}}
\end{equation*} 
along the tetrad basis $e^\mu_{\,\basind{\alpha}}$.
The inverse relations 
\begin{equation*}
V^{\basind{\alpha}}=V^{\mu}\,e_\mu^{\;\basind{\alpha}}\quad\text{and}\quad
W^{\basind{\alpha\beta}}=W^{\mu\nu}\,e_\mu^{\;\basind{\alpha}}\, e_\nu^{\,\basind{\beta}}
\end{equation*}
are obtained with the help of eq.~(\ref{OrthoNice1}).
We can also decompose covariant vector and tensor components 
\begin{equation*}
V_{\mu}=V_{\basind{\alpha}}\,e_\mu^{\;\basind{\alpha}}\quad\text{and}\quad
W_{\mu\nu}=W_{\basind{\alpha\beta}}\,e_\mu^{\;\basind{\alpha}}\,e_\nu^{\;\basind{\beta}}\,.
\end{equation*} 
By substituting the decompositions of $V^\mu$ and $V_\mu$ in $V_\mu=g_{\mu\nu}V^\nu$ we furthermore conclude that 
\begin{equation*}
V_{\basind{\alpha}}=\eta_{\basind{\alpha\beta}}\,V^{\basind{\beta}} \,.
\end{equation*}
In the same way, we can show that
\begin{equation*}
W_{\basind{\alpha\beta}}=\eta_{\basind{\alpha \mu}}\,\eta_{\basind{\beta\nu}}\,W^{\basind{\mu\nu}}\,.
\end{equation*}

\paragraph{Antisymmetric tensor}

Let us consider as short and useful example the tetrad decomposition of the total antisymmetric tensor $\varepsilon^{\alpha\beta\gamma\delta}$, eq.~(\ref{EpsilonTensor}), with $\varepsilon^{0123}=1/\sqrt{-g}$. Its tetrad components 
\begin{equation*}
\varepsilon^{\basind{\alpha\beta\gamma\delta}}=e_\mu^{\;\basind{\alpha}}\,e_\nu^{\;\basind{\beta}}\,
e_\rho^{\;\basind{\gamma}}\,e_\sigma^{\;\basind{\delta}}\,\varepsilon^{\mu\nu\rho\sigma}
\end{equation*} 
can be rewritten in terms of the non-tensorial Levi-Civita symbol (\ref{LeviCevita})
\begin{equation}
\varepsilon^{\basind{\alpha\beta\gamma\delta}}=
e_\mu^{\;\basind{\alpha}}\,e_\nu^{\;\basind{\beta}}\,
e_\rho^{\;\basind{\gamma}}\,e_\sigma^{\;\basind{\delta}}\frac{\Delta^{\mu\nu\rho\sigma}}{\sqrt{-g}}=
\frac{\op{det}(e_\mu^{\;\basind{\alpha}})}{\sqrt{-g}}\Delta^{\alpha \beta \gamma \delta}\,,
\label{EpsilonTetradTrafo}
\end{equation}
where the right hand side of the last equation follows directly from the definition of the determinant.

From eq.~(\ref{OrthonormalTetrad}) we obtain for the determinant of the tetrad ``matrix'' $(e^\mu_{\;\basind{\alpha}})$ the relation\footnote{Throughout this article, we assume an ordering of the four tetrad vectors $e^\mu_{\;\basind{\alpha}}$ which preserves the orientation of the manifold and thus leads to $\op{det}(e^\mu_{\;\basind{\alpha}})>0$.}
\begin{equation}
\op{det}(e^\mu_{\;\basind{\alpha}})=\frac{1}{\sqrt{-g}}\,.
\label{DetTetradMatrix}
\end{equation}
Making use of the relativistic orthonormality condition (\ref{OrthoNice1}), we thus conclude for the determinant of the inverse tetrad ``matrix'' ${\op{det}(e_\mu^{\;\basind{\alpha}})=(\op{det}(e^\mu_{\;\basind{\alpha}}))^{-1}=\sqrt{-g}}$. But this result implies that the tetrad coefficients of the total antisymmetric tensor just reduce to the Levi-Civita symbol according to
\begin{equation}
\varepsilon^{\basind{\alpha\beta\gamma\delta}}=\Delta^{\alpha \beta \gamma \delta}\,.
\label{EpsilonTetradForm1}
\end{equation}
The correspondence 
\begin{equation}
 \varepsilon_{\basind{\alpha\beta\gamma\delta}}=\eta_{\basind{\alpha\mu}}\,\eta_{\basind{\beta\nu}}\,
\eta_{\basind{\gamma\rho}}\,\eta_{\basind{\delta\sigma}}\,\varepsilon^{\basind{\mu \nu \rho \sigma}}=-\Delta^{\alpha \beta \gamma \delta}
\label{EpsilonTetradForm2}
\end{equation}
between the lower tetrad indices and the Levi-Civita symbol could have also be found by deriving $\varepsilon_{\basind{\alpha\beta\gamma\delta}}$ in analogy to the above argumentation.

We conclude by pointing out the useful identity
\begin{equation}
\varepsilon_{\basind{0 i j k}}\;\varepsilon^{\basind{0 i p q}}= -\delta^{\basind{p}}_{\basind{j}}\delta^{\basind{q}}_{\basind{k}}+\delta^{\basind{p}}_{\basind{k}}\delta^{\basind{q}}_{\basind{j}} \,,
\label{epsilonDelta}
\end{equation} 
which we will frequently use throughout this article.

\paragraph{Transformation properties}

Finally, we would like to recall that the tensor components $V^\alpha$ change under coordinate transformations, but do not depend on the choice of the tetrad. On the other hand, the tetrad components $V^{\basind{\alpha}}$ are invariants under coordinate transformations, but they crucially depend on the choice of the tetrad vectors, as can be easily understood by the following argument. We start from two tetrads $\tilde e^\mu_{\,\basind{\alpha}}$ and $e'^\mu_{\,\basind{\beta}}$ related to each other via eq.~(\ref{TetradConnection}). Then by virtue of the decompositions 
\begin{equation}
\Lambda^{\basind{\beta}}_{\;\;\basind{\alpha}}= \tilde e^\mu_{\,\basind{\alpha}}\, e_\mu'^{\;\basind{\beta}}
\quad\text{and}\quad
(\Lambda^{-1})^{\basind{\beta}}_{\;\;\basind{\alpha}}=  e'^{\mu}_{\;\basind{\alpha}}\, \tilde e_\mu^{\,\basind{\beta}}
\label{LorentzDecomposition}
\end{equation} 
of the Lorentz matrix and its inverse, which follow from~(\ref{TetradConnection}) and~(\ref{OrthoNice1}), we obtain the connection
\begin{equation*}
\tilde V^{\basind{\alpha}}=\tilde e_\mu^{\,\,\basind{\alpha}}\,V^\mu=
\tilde e_\mu^{\,\,\basind{\alpha}}\left(e'^\mu_{\,\;\basind{\beta}}\,V'^{\,\basind{\beta}}\right)
=(\Lambda^{-1})^{\basind{\alpha}}_{\;\;\basind{\beta}}\,V'^{\,\basind{\beta}}
\end{equation*}
between the tetrad components $\tilde V^{\basind{\alpha}}$ and $V'^{\basind{\alpha}}$.

\subsection{General orthonormal transport of tetrads and the proper transport\label{AppTetradTransport}}

We now turn to the formulation of the general orthonormal transport law of tetrads~\cite{Synge,Krause75}. For this purpose we consider a timelike curve $\mathcal{C}$ expressed by $x^\mu(\tau)$ with four-velocity $u^\mu(\tau)$ and acceleration vector ${a^\mu=u^\mu_{\;\,;\nu}\,u^\nu}$. Moreover,  $e^\mu_{\;\basind{\alpha}}(\tau)$ denotes a one-parameter family of tetrads which satisfies the orthonormality conditions~(\ref{OrthoNice1}) and~(\ref{OrthoNice2}) in the tangent space of every point $x^\mu(\tau)\in\mathcal{C}$.

\subsubsection{Transport matrix}

In order to arrive at the general transport law, we make use of the fact that any vector located at $x^\mu(\tau)$ can be decomposed into a given tetrad basis $e^\mu_{\;\basind{\alpha}}(\tau)$. By applying this recipe to the vector $e^\mu_{\;\basind{\alpha};\nu}(\tau) u^\nu(\tau)$, which appears in the parallel transport law, we arrive at the decomposition
\begin{equation}
 e^\mu_{\;\basind{\alpha};\nu}(\tau)\, u^\nu(\tau)
 =-J^{\basind{\beta}}_{\;\;\,\basind{\alpha}}(\tau)\,e^\mu_{\;\basind{\beta}}(\tau)\,,
\label{GeneralTransport}
\end{equation}
where we have added the minus sign for reasons of conventions. 

The coefficients $J^{\basind{\beta}}_{\;\;\,\basind{\alpha}}(\tau)$ have to satisfy certain conditions in order to guarantee that the relativistic orthonormality conditions~(\ref{OrthoNice1}) and (\ref{OrthoNice2}) are preserved by the general transport law~(\ref{GeneralTransport}). These conditions follow by taking the covariant derivative of eq.~(\ref{OrthoNice1}) and contraction with the four-velocity $u^\nu$ which gives
\begin{equation*}
 e^\mu_{\,\basind{\alpha};\nu}\,u^\nu \, e_\mu^{\;\basind{\gamma}}+
e_{\mu\basind{\alpha}}\,\eta^{\basind{\gamma\beta}} e_{\;\basind{\beta};\nu}^{\mu}\,u^\nu
=\diff{}{\tau}\left(\delta^{\basind{\gamma}}_{\basind{\alpha}}\right)=0\,.
\end{equation*}
Substitution of the general transport law~(\ref{GeneralTransport}) into the last expression yields
\begin{equation*}
 -J^{\basind{\gamma}}_{\;\;\,\basind{\alpha}}-
\eta_{\basind{\alpha\sigma}}\,\eta^{\basind{\gamma\beta}}\,J^{\basind{\sigma}}_{\;\;\,\basind{\beta}}=0\,.
\end{equation*}
Further contraction with $\eta_{(\rho\gamma)}$ reveals that the coefficients $J^{\basind{\beta}}_{\;\;\,\basind{\alpha}}$ have to obey the anti-symmetry relation 
\begin{equation}
 J_{\basind{\rho\alpha}}=-J_{\basind{\alpha\rho}}\,,
 \label{TransportMatrixCondition}
\end{equation}
in order to preserve the orthonormality conditions. 

We call every matrix $J^{\basind{\beta}}_{\;\;\,\basind{\alpha}}$ which appears in a transport law of the form~(\ref{GeneralTransport}) and which satisfies the antisymmetry relation~(\ref{TransportMatrixCondition}) a {\it transport matrix}. The parallel transport of a tetrad 
\begin{equation}
 e^\mu_{\;\basind{\alpha};\nu}(\tau)\, u^\nu(\tau) =0
\label{Tetradparalleltrans}
\end{equation} 
is included as the special case for which $J^{\basind{\beta}}_{\;\;\,\basind{\alpha}}=0$. 

As discussed in the last subsection every tensor can be decomposed into tetrad coefficients. Vice versa, it is also possible to assign tensor components to a given set of tetrad coefficients. Hence, the transport law (\ref{GeneralTransport}) can be reformulated in a tensorial form with the so-called transport tensor $J^\mu_{\;\;\nu}$ which is connected to the transport matrix via
\begin{equation}
J^\mu_{\;\;\nu}=e^\mu_{\;\basind{\alpha}}\,e_\nu^{\;\basind{\beta}}
\,J^{\basind{\alpha}}_{\;\;\,\basind{\beta}}\quad\text{and}\quad
J^{\basind{\alpha}}_{\;\;\,\basind{\beta}}=
e_\mu^{\;\basind{\alpha}}\,e^\nu_{\;\basind{\beta}}\,J^\mu_{\;\;\nu}\,.
\label{DefTransportTensor}
\end{equation} 
Inserting eq.~(\ref{DefTransportTensor}) into the transport law (\ref{GeneralTransport}) yields its tensorial form
\begin{equation}
 e^\mu_{\;\basind{\alpha};\nu}(\tau)\, u^\nu(\tau)
 =-J^\mu_{\;\;\nu}(\tau)\,e^\nu_{\;\basind{\alpha}}(\tau)\quad\text{with}\quad
J_{\mu\nu}=-J_{\nu\mu}\,.
\label{TensorTransport}
\end{equation} 
This equation, which holds for the different tetrad vectors, can be easily translated into a tensorial transport law for any contravariant vector $V^\mu(\tau)$. Suppose, the decomposition of the vector along the world line $x^\mu(\tau)$ reads ${V^\mu(\tau)=V^{\basind{\alpha}}\,e^\mu_{\;\basind{\alpha}}(\tau)}$, where the whole change of the vector along the world line is incorporated in the tetrad and the coefficients $V^{\basind{\alpha}}$ do not depend on $\tau$. Then a multiplication of eq.~(\ref{TensorTransport}) by $V^{\basind{\alpha}}$ provides us with the transport law for vectors $V^\mu_{\;\;;\nu}u^\nu=-J^\mu_{\;\;\nu}\,V^\nu$. An analogous argument leads to a transport equation for arbitrary tensors.

\subsubsection{Correspondence between distinct transport laws}

We now try to gain a deeper insight into the general orthogonal transport eq.~(\ref{GeneralTransport}) and the transport matrix $J^{\basind{\alpha}}_{\;\;\,\basind{\beta}}$. For this purpose it is reasonable to consider the relationship between the transport laws of two different families of tetrads $e^\mu_{\;\basind{\alpha}}(\tau)$ and $\tilde e^\mu_{\;\basind{\alpha}}(\tau)$, which both live in the tangent spaces along the world line $x^\mu(\tau)$.
By virtue of eq.~(\ref{TetradConnection}) it is always possible to bring them together with the help of a specific family of Lorentz transformation matrices, say
\begin{equation}
e^\mu_{\;\basind{\alpha}}(\tau)=\Lambda^{\basind{\beta}}_{\;\,\basind{\alpha}}(\tau)\,\tilde e^\mu_{\;\basind{\beta}}(\tau)\,.
\label{TetradConnect2}
\end{equation} 
We assume, that the tetrad $e^\mu_{\;\basind{\alpha}}(\tau)$ satisfies the general transport law (\ref{GeneralTransport}). In order to derive the corresponding transport law for $\tilde e^\mu_{\;\basind{\alpha}}(\tau)$ we simply substitute eq.~(\ref{TetradConnect2}) into eq.~(\ref{GeneralTransport}). Together with the inverse Lorentz matrix~(\ref{InverseLorentz}), we find after some algebra the transport law 
\begin{equation}
\tilde e^\mu_{\;\basind{\alpha};\nu}\, u^\nu
=-\tilde J^{\basind{\beta}}_{\;\;\,\basind{\alpha}}\,\tilde e^\mu_{\;\basind{\beta}}\;\,, \quad
\tilde J^{\basind{\beta}}_{\;\;\,\basind{\alpha}}=
\Lambda^{\basind{\beta}}_{\;\;\basind{\gamma}}\,J^{\basind{\gamma}}_{\;\;\,\basind{\rho}}\,
(\Lambda^{-1})^{\basind{\rho}}_{\;\;\basind{\alpha}}+
\diff{\Lambda^{\basind{\beta}}_{\;\;\basind{\rho}}}{\tau}(\Lambda^{-1})^{\basind{\rho}}_{\;\;\basind{\alpha}}
\label{CorrespondTransport}
\end{equation} 
for the tetrad $\tilde e^\mu_{\;\basind{\alpha}}(\tau)$. These formulas bring out most clearly the role of the Lorentz matrix in the relation between the old transport matrix $J^{\basind{\beta}}_{\;\;\,\basind{\alpha}}$ and the new one $\tilde J^{\basind{\beta}}_{\;\;\,\basind{\alpha}}$. 

In particular, it is interesting to compare the general orthonormal transport to the parallel transport. For this reason, we assume that the tetrad $e^\mu_{\;\basind{\alpha}}(\tau)$ is initially congruent with 
$\tilde e^\mu_{\;\basind{\alpha}}(\tau)$, in other words $e^\mu_{\;\basind{\alpha}}(\tau_0)=\tilde e^\mu_{\;\basind{\alpha}}(\tau_0)$ at the initial point $x^\mu(\tau_0)$. Moreover, we define the one-parameter family $e^\mu_{\;\basind{\alpha}}(\tau)$ by virtue of parallel transport of the initial tetrad $e^\mu_{\;\basind{\alpha}}(\tau_0)$ along $x^\mu(\tau)$. On the other hand $\tilde e^\mu_{\;\basind{\alpha}}(\tau)$
shall be based on the general orthonormal transport~(\ref{CorrespondTransport}) along the same world line $x^\mu(\tau)$. As a consequence, the Lorentz matrix $\Lambda^{\basind{\beta}}_{\;\;\basind{\alpha}}$ which connects both tetrads as in eq.~(\ref{TetradConnect2}), is specified by the linear differential equation 
\begin{equation}
 \diff{\Lambda^{\basind{\beta}}_{\;\;\basind{\rho}}}{\tau}=
\tilde J^{\basind{\beta}}_{\;\;\,\basind{\alpha}}(\tau)
\,\Lambda^{\basind{\alpha}}_{\;\;\basind{\rho}}(\tau)
\label{GeneratorofLorentz}
\end{equation}
following from eq.~(\ref{CorrespondTransport}). Due to ${e^\mu_{\;\basind{\alpha}}(\tau_0)=\tilde e^\mu_{\;\basind{\alpha}}(\tau_0)}$, the corresponding initial condition simply reads
\begin{equation*}
\Lambda^{\basind{\alpha}}_{\;\;\basind{\beta}}(\tau_0)=\delta^{\basind{\alpha}}_{\basind{\beta}}\,.
\end{equation*}
Hence, the coefficients $\tilde J^{\basind{\beta}}_{\;\;\,\basind{\alpha}}(\tau)$ of the transport matrix determine the Lorentz transformation which connects the parallel transported tetrad with the generally transported one. In particular, we observe for small $\varepsilon$ that the infinitesimal Lorentz transformation 
$\Lambda^{\basind{\alpha}}_{\;\;\basind{\beta}}(\tau_0+\varepsilon)=
\delta^{\basind{\alpha}}_{\basind{\beta}}+\varepsilon\,\tilde J^{\basind{\alpha}}_{\;\;\basind{\beta}}(\tau_0)$, 
which directly links both tetrads at $x^\mu(\tau_0+\varepsilon)$, is generated by the elements of the transport matrix. Hence, one usually refers to the elements of the transformation matrix as the generators of infinitesimal Lorentz transformations, see e.\,g. \cite{Sexl01}.

\subsubsection{Proper transport as natural generalization of the Fermi-Walker transport}

We conclude this appendix by discussing another prominent transport equation, the so-called proper transport, which will be of crucial importance in the definition of proper reference frame coordintates analyzed in appendix~\ref{AppProperRef}. They may serve as coordinates well suited for the theoretical description of future satellite experiments testing the local curvature of spacetime by various interferometric devices. In order to appreciate the implications of this proper transport, we provide its motivation in two steps: first we impose certain restrictions on the transformation matrix by demanding, that the timelike tetrad vector $e^\mu_{\;\basind{0}}(\tau)$ should be equal to the scaled four-velocity $u^\mu(\tau)/c$, and second, we derive its most general form by taking advantage of an intuitive illustration.

\paragraph{Fermi-Walker transport}
We start our motivation of the proper transport by recalling fig.~\ref{LocalLightcone}.
As indicated in the right picture it is always possible to find a tetrad in the tangent space of a given point in such a way, that the timelike vector $e^\mu_{\;\basind{0}}$ is parallel to the four-velocity $u^\mu$ at $P$. Stimulated by this fact, we want to examine the transport laws for which the timelike tetrad vector $e^\mu_{\;\basind{0}}(\tau)$ is equal to the scaled four-velocity vector $u^\mu(\tau)/c$ on the whole world line $x^\mu(\tau)$ for all proper times $\tau$. For this reason we suppose, that initially $e^\mu_{\;\basind{0}}(\tau_0)= u^\mu(\tau_0)/c$ and that $e^\mu_{\;\basind{0}}(\tau)=u^\mu(\tau)/c$ represents a solution of the transport equation
\begin{equation}
e^\mu_{\;\basind{\alpha};\nu}(\tau)\, u^\nu(\tau)
 =-\Omega^{\basind{\beta}}_{\;\;\,\basind{\alpha}}(\tau)\,e^\mu_{\;\basind{\beta}}(\tau)\,.
\label{FermiWalkerTetrad}
\end{equation} 
The coefficients $\Omega^{\basind{\beta}}_{\;\;\,\basind{\alpha}}(\tau)$ of the new transport matrix will now be specified in order to fulfill the desired requirement. Inserting $e^\mu_{\;\basind{0}}(\tau_0)=u^\mu(\tau_0)/c$ into the zeroth component of eq.~(\ref{FermiWalkerTetrad}) we find 
\begin{equation*}
e^\mu_{\;\basind{0};\nu}(\tau)\, u^\nu(\tau)=\frac{1}{c}\,a^\mu(\tau)
 =-\Omega^{\basind{\beta}}_{\;\;\,\basind{0}}(\tau)\,e^\mu_{\;\basind{\beta}}(\tau)\,,
\end{equation*} 
where we have made use of the definition $a^\mu=u^\mu_{\;;\nu}\,u^\nu$ of the four-acceleration.
When we now contract the last expression with $e_\mu^{\;\,\basind{\alpha}}$ and introduce the invariant tetrad coefficients $a^{\basind{\alpha}}(\tau)=a^\mu(\tau)\,e_\mu^{\;\,\basind{\alpha}}(\tau)$ of the four-acceleration, we arrive at the condition
\begin{equation}
\Omega_{\basind{\gamma 0}}
=-\frac{1}{c}\,\eta_{\basind{\gamma\alpha}}\,a^{\basind{\alpha}}=-\frac{1}{c}\,a_{\basind{\gamma}}\,.
\label{FermiWalkerCondition}
\end{equation} 
In particular, we obtain $-\Omega_{\basind{00}}=\frac{1}{c}\,a_{\basind{0}}
=\frac{1}{c}a^\mu\,e_{\mu\basind{0}}=\frac{1}{c^2}\,a^\mu\,u_\mu=0$, which follows directly from eq.~(\ref{constraintonaccel}). Indeed, eq.~(\ref{FermiWalkerCondition}) and the antisymmetry relation ${\Omega_{\basind{0 \gamma}}=-\Omega_{\basind{\gamma 0}}}$ are the only conditions for the transport matrix $\Omega^{\basind{\beta}}_{\;\;\,\basind{0}}$ to guarantee 
${e^\mu_{\;\basind{0}}(\tau)=u^\mu(\tau)/c}$ as a natural solution of eq.~(\ref{FermiWalkerTetrad}). If we assume for simplicity, that all the other elements $\Omega_{\basind{i k}}=0$ with $i,k\in\{1,2,3\}$ vanish, then the transport matrix reads 
\begin{equation}
\Omega_{\basind{\gamma\alpha}}=-\frac{1}{c}\left(a_{\basind{\gamma}}\eta_{\basind{\alpha 0}}-a_{\basind{\alpha}}\eta_{\basind{\gamma 0}}\right)\,,
\label{FermiWalkerTetradForm}
\end{equation} 
in accordance with eq.~(\ref{FermiWalkerCondition}) and the antisymmetry condition. 

In this case we arrive at the corresponding transport equation
\begin{equation*}
 e^\mu_{\;\basind{\alpha};\nu}\, u^\nu=\frac{1}{c}\,
 \eta^{\basind{\beta\gamma}}\left(a_{\basind{\gamma}}\eta_{\basind{\alpha 0}}-
 a_{\basind{\alpha}}\eta_{\basind{\gamma 0}}\right)e^\mu_{\;\basind{\beta}}
=\frac{1}{c}\,a^\mu\,\eta_{\basind{\alpha 0}}-\frac{1}{c^2}\,u^\mu a_\nu\, e^\nu_{\;\basind{\alpha}}\,.
\end{equation*}
Its tensor form follows in analogy to eqs.~(\ref{DefTransportTensor}) and (\ref{TensorTransport}) with the help of the transport tensor
\begin{equation*}
\Omega^\mu_{\;\;\nu}=\Omega_{\basind{\gamma\alpha}}\,e^{\mu\basind{\gamma}}\,e_\nu^{\;\,\basind{\alpha}}
=-\frac{1}{c^2}\left(a^\mu u_\nu-u^\mu a_\nu\right)
\end{equation*} 
and thus reads
\begin{equation}
e^\mu_{\;\basind{\alpha};\nu}\, u^\nu=\frac{1}{c^2}\left(a^\mu u_\nu-u^\mu a_\nu\right)\,e^\nu_{\;\basind{\alpha}}\,,
\label{FermiWalkertransport2}
\end{equation} 
which is just the Fermi-Walker transport of our tetrad $e^\mu_{\;\basind{\alpha}}(\tau)$, as can be seen by comparing it with eq.~(\ref{FermiWalkertransport}). 

Before we turn to a natural generalization of eq.~(\ref{FermiWalkertransport2}), we briefly compare the Fermi-Walker transport in its tetrad form, eq.~(\ref{FermiWalkerTetrad}), to the parallel transport, eq.~(\ref{Tetradparalleltrans}). Again, we denote the initial tetrad located at the point $x^\mu(\tau_0)$ by $e^\mu_{\;\basind{\alpha}}(\tau_0)$. We transport the tetrad to the adjacent point $x^\mu(\tau_0+\epsilon)$ in two different ways: (i) by parallel transport, and (ii) by Fermi-Walker transport. The resulting tetrads are connected by an {\it infinitesimal Lorentz boost} $\Lambda^{\basind{\alpha}}_{\;\;\basind{\beta}}(\tau_0+\varepsilon)=
\delta^{\basind{\alpha}}_{\basind{\beta}}+\varepsilon\, \Omega^{\basind{\alpha}}_{\;\;\basind{\beta}}(\tau_0)$ whose generators are the coefficients of the corresponding transport matrix ${\Omega^{\basind{\beta}}_{\;\;\,\basind{\alpha}}
=\eta^{\basind{\beta\gamma}}\Omega_{\basind{\gamma\alpha}}}$, given by eq.~(\ref{FermiWalkerTetradForm}). 

When we compare the two tetrads at a point $x^\mu(\tau)$ which does not lie in the immediate neighborhood of $x^\mu(\tau_0)$, the connection between them will no longer be given by a simple Lorentz boost. This feature reflects the fact that in general the combination of two infinitesimal Lorentz boosts does not lead to another Lorentz boost since spatial rotations have to be taken into account.

\begin{figure}
\centering
\includegraphics[width=0.88\textwidth]{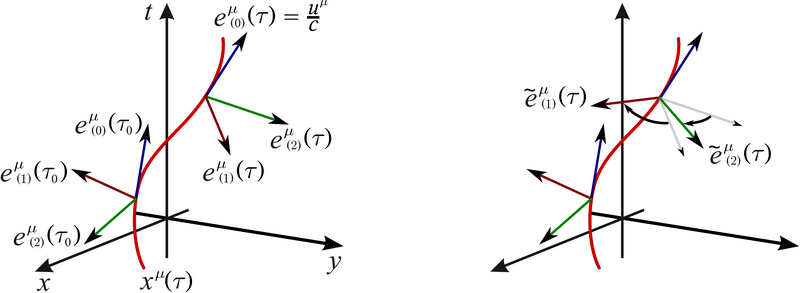}
\caption{Two different orthonormal transport laws applied to the same initial tetrad.
Both possess the property, that the timelike basis vector $e^\mu_{\,\basind{0}}(\tau)=\tilde e^\mu_{\,\basind{0}}(\tau)$ remains tangential to the world line $x^\mu(\tau)$. However, they differ in the orientation of the spacelike basis vectors. A spatial rotation of the tetrad vectors
$e^\mu_{\,\basind{i}}(\tau)$ (left), which were obtained by transporting the initial tetrad $e^\mu_{\;\basind{\alpha}}(\tau_0)$ according to the Fermi-Walker transport~(\ref{FermiWalkerTetrad}), yields the tetrad (right) with the basis vectors $\tilde e^\mu_{\;\basind{i}}(\tau)$.
In order to bring out the similarities and differences between the two pictures most clearly, we have marked in the right picture only those quantities which are different from the ones in the left picture.
}
\label{FigProperTransport}
\end{figure}

\paragraph{Proper transport}
So far, we have shown, that the Fermi-Walker transport can be derived by simply imposing the condition (\ref{FermiWalkerCondition}) on the transport matrix, whereas all other components $\Omega_{\basind{i k}}$ have been set to zero. However, the choice $\Omega_{\basind{i k}}=0$ was haphazard. Condition (\ref{FermiWalkerCondition}) was the only necessary restriction to provide a transport law which keeps the timelike tetrad ${e^\mu_{\;\basind{0}}(\tau)=u^\mu(\tau)/c}$ tangential to the world line $x^\mu(\tau)$ for all $\tau$. As illustrated in fig.~\ref{FigProperTransport}, the presented transport law is not the only one which satisfies this condition. In the left picture we sketched the Fermi-Walker transport of an initial tetrad 
$e^\mu_{\;\basind{\alpha}}(\tau_0)$ at the point $x^\mu(\tau_0)$. However, it is possible to additionally rotate the Fermi-Walker-transported spacelike vectors $\tilde e^\mu_{\;\basind{i}}(\tau)$ in every point $x^\mu(\tau)$, without affecting the property of the timelike vector $e^\mu_{\;\basind{0}}(\tau)$ to remain tangential to the world line $x^\mu(\tau)$. The right picture of fig.~\ref{FigProperTransport} shall serve as an illustration of this statement. The transport equation which includes also this spatial rotation is exactly the proper transport we are seeking.

Hence, in order to find the proper transport equation we make use of eqs.~(\ref{TetradConnect2}) and (\ref{CorrespondTransport}). We start from an initial tetrad $e^\mu_{\;\basind{\alpha}}(\tau_0)=\tilde e^\mu_{\;\basind{\alpha}}(\tau_0)$ which is located at the point $x^\mu(\tau_0)$ of the world line $x^\mu(\tau)$. The first family of tetrads $e^\mu_{\;\basind{\alpha}}(\tau)$ shall arise from the Fermi-Walker transport (\ref{FermiWalkerTetrad}) of the initial tetrad along $x^\mu(\tau)$, whereas the second family of tetrads 
$\tilde e^\mu_{\;\basind{\alpha}}(\tau)$ shall be defined by adding an arbitrary, spatial rotation to the Fermi-Walker transported tetrad. Hence, we choose the ansatz 
\begin{equation*}
e^\mu_{\;\basind{\alpha}}(\tau)=(\Lambda_R)^{\basind{\beta}}_{\;\,\basind{\alpha}}(\tau)\,\tilde e^\mu_{\;\basind{\beta}}(\tau)\,
\end{equation*} 
for the tetrad $\tilde e^\mu_{\;\basind{\alpha}}(\tau)$. Here we have introduced the Lorentz matrix $(\Lambda_R)^{\basind{\beta}}_{\;\,\basind{\alpha}}(\tau)$ which relates the two tetrads by a simple spatial rotation of their basis vectors $e^\mu_{\;\basind{i}}$ and $\tilde e^\mu_{\;\basind{k}}$. Apart from eq.~(\ref{DefLorentz}), it is specified by the relations
\begin{equation}
(\Lambda_R)^{\basind{0}}_{\;\,\basind{\alpha}}=\delta^{\basind{0}}_{\basind{\alpha}}\,,\quad 
(\Lambda_R)^{\basind{\alpha}}_{\;\,\basind{0}}=\delta^{\basind{\alpha}}_{\basind{0}}\,,\quad 
(\Lambda_R)^{\basind{i}}_{\;\,\basind{k}}=R^{\basind{i}}_{\;\,\basind{k}}\,,
\label{LorentzRotation}
\end{equation} 
with the rotation matrix $R^{\basind{i}}_{\;\,\basind{k}}$. 

When we substitute the defining equation of $\tilde e^\mu_{\;\basind{\alpha}}(\tau)$ into the Fermi-Walker transport (\ref{FermiWalkerTetrad}) we arrive in analogy to eq.~(\ref{CorrespondTransport}) at the corresponding transport law for the basis vectors $\tilde e^\mu_{\;\basind{\alpha}}(\tau)$
\begin{equation}
\tilde e^\mu_{\;\basind{\alpha};\nu}\, u^\nu
=-\tilde \Omega^{\basind{\beta}}_{\;\;\,\basind{\alpha}}\,\tilde e^\mu_{\;\basind{\beta}}\,,
\label{ProperTransportTetrad}
\end{equation} 
with the new transport matrix
\begin{equation}
\tilde \Omega^{\basind{\beta}}_{\;\;\,\basind{\alpha}}=
(\Lambda_R)^{\basind{\beta}}_{\;\;\basind{\gamma}}\,\Omega^{\basind{\gamma}}_{\;\;\,\basind{\rho}}\,
(\Lambda^{-1}_R)^{\basind{\rho}}_{\;\;\basind{\alpha}}+
\diff{(\Lambda_R)^{\basind{\beta}}_{\;\;\basind{\rho}}}{\tau}(\Lambda^{-1}_R)^{\basind{\rho}}_{\;\;\basind{\alpha}}\,.		\label{eq:transport_matrix}
\end{equation}
We now introduce the tetrad components $\tilde a^{\basind{\alpha}}=(\Lambda_R)^{\basind{\alpha}}_{\;\;\basind{\beta}}\,a^{\basind{\beta}}$ of the four-acceleration with respect to the basis $\tilde e^\mu_{\;\basind{\alpha}}$ and the abbreviation 
\begin{equation}
\tilde\Theta^{\basind{\beta}}_{\;\;\basind{\alpha}}=\diff{(\Lambda_R)^{\basind{\beta}}_{\;\;\basind{\rho}}}{\tau}
(\Lambda^{-1}_R)^{\basind{\rho}}_{\;\;\basind{\alpha}}
\label{DefTheta}
\end{equation}
for the second term on the right hand side of eq.~(\ref{eq:transport_matrix}).

By substitution of eq.~(\ref{FermiWalkerTetradForm}) and usage of (\ref{InverseLorentz}), we are able to rewrite the coefficients ${\tilde \Omega_{\basind{\gamma\alpha}}=
\eta_{\basind{\gamma\beta}}\,\tilde \Omega^{\basind{\beta}}_{\;\;\,\basind{\alpha}}}$ according to
\begin{equation}
 \tilde \Omega_{\basind{\gamma\alpha}}=-\frac{1}{c}\left(\tilde a_{\basind{\gamma}}
\eta_{\basind{\alpha 0}}-
\tilde a_{\basind{\alpha}}\eta_{\basind{\gamma 0}}\right)+\tilde \Theta_{\basind{\gamma\alpha}}\,.
\label{ProperTransportMatrix}
\end{equation}
The first term on the right hand side represents the transport matrix of the Fermi-Walker transport, but with the  tetrad coefficients of the acceleration $\tilde a_{\basind{\alpha}}$ now corresponding to the new tetrad $\tilde e^\mu_{\;\basind{\alpha}}$. The second term $\tilde\Theta_{\basind{\gamma\alpha}}$ stems from the additional spatial rotation. As required by eq.~(\ref{TransportMatrixCondition}), it satisfies the antisymmetry relation
\begin{equation*}
 \tilde\Theta_{\basind{\gamma\alpha}}=-\tilde\Theta_{\basind{\alpha\gamma}}\,,
\end{equation*}
as can be verified by using
\begin{equation*}
 \diff{(\Lambda_R)^{\basind{\beta}}_{\;\;\basind{\rho}}}{\tau}
(\Lambda^{-1}_R)^{\basind{\rho}}_{\;\;\basind{\alpha}}=
-(\Lambda_R)^{\basind{\beta}}_{\;\;\basind{\rho}} \diff{(\Lambda^{-1}_R)^{\basind{\rho}}_{\;\;\basind{\alpha}}}{\tau}
\end{equation*}
and eq.~(\ref{InverseLorentz}) in the definition~(\ref{DefTheta}) of $\tilde\Theta_{\basind{\gamma\alpha}}$. 

Moreover, it follows directly from the definition~(\ref{LorentzRotation}) of the Lorentz matrix $\Lambda_R$ that 
\begin{equation*}
\tilde\Theta_{\basind{0\alpha}}=0\,.
\end{equation*}
Hence, the quantity $\tilde\Theta_{\basind{\gamma\alpha}}$ possesses only three independent, non-vanishing, spatial components $\tilde\Theta_{\basind{i k}}$. It is convenient to map them directly onto the spatial tetrad components $\tilde\omega^{\basind{k}}$ of a four-vector. This one-to-one mapping is accomplished by the tetrad components $\tilde \varepsilon_{\basind{\alpha\beta\gamma\delta}}$ of the totally antisymmetric tensor, eq.~(\ref{EpsilonTetradForm2}). The connection
\begin{equation}
\tilde\Theta_{\basind{\beta\gamma}}=\tilde\varepsilon_{\basind{0\alpha\beta\gamma}}\,
\tilde\omega^{\basind{\alpha}}    
\label{eq:antisymmetric_tensor_Omega}
\end{equation} 
ensures that the elements $\tilde\Theta_{\basind{0\alpha}}$ vanish and accounts for the antisymmetry of $\tilde\Theta_{\basind{\beta\gamma}}$. 
When we solve eq.~(\ref{eq:antisymmetric_tensor_Omega}) for $\tilde\omega^{\basind{\alpha}}$, we find that this one-to-one mapping\footnote{The definition~(\ref{eq:antisymmetric_tensor_Omega}) of the tetrad rotation vector $\tilde \omega^{\basind{\alpha}}$ guarantees the validity of the right-hand rule for the spatial components $\tilde \omega^{\basind{k}}$. We illustrate this statement by the set of time-dependent tetrad vectors $\tilde e^\mu_{\;\basind{0}}(\tau)=e^\mu_{\;\basind{0}}(\tau)$, 
$\tilde e^\mu_{\;\basind{1}}(\tau)=\cos(\omega\tau)\,e^\mu_{\;\basind{1}}(\tau)+ \sin(\omega\tau)\, e^\mu_{\;\basind{2}}(\tau)$,
$\tilde e^\mu_{\;\basind{2}}(\tau)=-\sin(\omega\tau)\, e^\mu_{\;\basind{1}}(\tau)+\cos(\omega\tau)\,
e^\mu_{\;\basind{2}}(\tau)$ and 
$\tilde e^\mu_{\;\basind{3}}(\tau)=e^\mu_{\;\basind{3}}(\tau)$, which describes a uniform, counter-clockwise rotation of the tetrad vectors $\tilde e^\mu_{\;\basind{1}}(\tau)$, $\tilde e^\mu_{\;\basind{2}}(\tau)$ relative to the Fermi-Walker transported basis around the common axis $e^\mu_{\;\basind{3}}(\tau)$. According to eqs.~(\ref{DefTheta})and~(\ref{eq:antisymmetric_tensor_Omega}), we obtain $\tilde\omega^{\basind{\alpha}}=(0,0,0,\omega)$.} implies 
\begin{equation}
\tilde\omega^{\basind{0}}=0\,.
\label{TetradRotationVector}
\end{equation} 
Hence, we can rewrite the transport eq.~(\ref{ProperTransportTetrad}) according to
\begin{equation}
\tilde e^\mu_{\;\basind{\alpha};\nu}\, u^\nu
=-\tilde \Omega_{\basind{\gamma\alpha}}\,\tilde e^{\mu\basind{\gamma}}\;\;\text{with}\;\;
\tilde \Omega_{\basind{\gamma\alpha}}=-\frac{1}{c}\left(\tilde a_{\basind{\gamma}}\eta_{\basind{\alpha 0}}-
\tilde a_{\basind{\alpha}}\eta_{\basind{\gamma 0}}\right)+\tilde\varepsilon_{\basind{0\beta\gamma\alpha}}\,
\tilde\omega^{\basind{\beta}}\, ,
\label{ProperTransportTetrad2}
\end{equation} 
which is the desired proper transport law formulated in tetrad language. 

In order to express eq.~(\ref{ProperTransportTetrad2}) in tensor form, we make use of the relations
${\tilde e^\mu_{\;\basind{0}}=u^\mu/c}$ and ${\tilde\omega_{\basind{\beta}}
=\tilde e^\sigma_{\;\basind{\beta}}\,\omega_\sigma}$ and arrive at
\begin{equation*}
\Omega^{\mu\nu}=\tilde e^{\mu\basind{\gamma}}\,\tilde e^{\nu\basind{\alpha}}\,\tilde \Omega_{\basind{\gamma\alpha}}=
-\frac{1}{c^2}\left(a^\mu u^\nu-u^\mu a^\nu \right)+\omega_\sigma\,\eta_{\basind{0\lambda}}\left(
\tilde\varepsilon^{\basind{\lambda\beta\gamma\alpha}}\,\tilde e^\sigma_{\;\basind{\beta}}\,\tilde e^{\mu}_{\;\basind{\gamma}}\,\tilde e^{\nu}_{\;\basind{\alpha}}
\right)\,.
\end{equation*} 
From eq.~(\ref{EpsilonTetradTrafo}) we obtain the identity
\begin{equation*}
\tilde\varepsilon^{\basind{\lambda\beta\gamma\alpha}}\,\tilde e^\sigma_{\;\basind{\beta}}\,\tilde e^{\mu}_{\;\basind{\gamma}}\,\tilde e^{\nu}_{\;\basind{\alpha}}=
\varepsilon^{\rho\sigma\mu\nu}\,\tilde e_{\rho}^{\;\,\basind{\lambda}}\,,
\end{equation*} 
which finally yields the proper tansport law 
\begin{equation}
\tilde e^\mu_{\;\basind{\alpha};\nu}\, u^\nu=-\Omega^{\mu\nu}\,\tilde e_{\nu\basind{\alpha}}\quad\text{with}\quad
\Omega^{\mu\nu}=-\frac{1}{c^2}\left(a^\mu u^\nu-a^\nu u^\mu \right)+\frac{1}{c}\,u_\rho\, \omega_\sigma \,
\varepsilon^{\rho\sigma\mu\nu}
\label{ProperTransportTensor}
\end{equation} 
in its tensorial form. Thus, we have established the proper transport which has been utilized by \cite{Ehlers73,MTW} to define the proper reference frame.


\section{Riemann normal coordinates and proper reference frames\label{AppendixSpecialCoordinates}}

In this appendix we give a rudimentary introduction into Riemann normal coordinates and proper reference frames within the framework of general relativity. For this purpose it is necessary to first examine the formal solution of the geodesic equation \cite{Eisenhart64}. In the subsequent discussion of Riemann normal coordinates, we concentrate on the expansion of the metric coefficients around a fixed point in spacetime, thereby following the approach given in \cite{AlvarezGaume81,Hatzinikitas00}. We then turn to the discussion of proper reference frame coordinates, which have been introduced in subsect.~\ref{ProperRef}. Finally, we also provide the expansion of the metric coefficients for this case \cite{Ehlers73,MTW,Nesterov76,Ni78,Li79a,Marzlin94,Nesterov99}.

\subsection{Formal solution of the geodesic equation\label{AppFormalSolution}}

In this subsection we examine the formal solution of the geodesic equation
\begin{equation}
 \frac{\D^2x^\mu}{\D\lambda^2}+\Gamma^\mu_{\;\alpha\beta}(x^\sigma)\diff{x^\alpha}{\lambda}
     \diff{x^\beta}{\lambda}=0 
\label{GeodesicEq2}
\end{equation}
discussed in appendix~\ref{WorldLineGeodesic}, for the initial conditions
\begin{align}
x^\mu(0)=p^\mu \quad\text{and}\quad \diff{x^\mu(0)}{\lambda} =v^\mu=v^{\basind{\alpha}}\,e^\mu_{\;\basind{\alpha}}\,.
\label{Initial}
 \end{align}
The corresponding solutions represent all geodesics, which start at the point $P$ with coordinates $p^\mu$, and which have arbitrarily directed tangent vectors $v^\mu$ at $P$. Without loss of generality we impose the restriction 
\begin{equation*}
 (v^{\basind{0}})^2+(v^{\basind{1}})^2+(v^{\basind{2}})^2+(v^{\basind{3}})^2=1
\end{equation*}
on the tetrad coefficients $v^{\basind{\alpha}}$ of our initial tangent vectors $v^\mu$. In this subsection we are interested in taking all kinds of geodesics into account: timelike, lightlike and spacelike geodesics. Hence, we denote the ``arclength" of the geodesics by $\lambda$.

With the initial conditions~(\ref{Initial}) the power-series expansion of the solution of the geodesic equation reads 
\begin{equation}
x^\mu(\lambda)=p^\mu +v^\mu \lambda + \sum_{n=2}^\infty  \frac{\D^n x^\mu(0)}{\D\lambda^n}\frac{\lambda^n}{n!}\,.
\label{Geoexpand}
\end{equation}
We now derive general expressions for the higher derivatives $\frac{{\small\D}^n x^\mu(0)}{{\small\D}\lambda^n}$ by making use of the geodesic equation. 
For this purpose we differentiate eq.~\eqref{GeodesicEq2} with respect to $\lambda$ and arrive at
\begin{equation}
 \diff{^3x^\mu}{\lambda^3}+\left(\diff{}{\lambda}\Gamma^\mu_{\;\alpha\beta}\right)\diff{x^\alpha}{\lambda}
     \diff{x^\beta}{\lambda} + \Gamma^\mu_{\;\alpha\beta}\diff{^2x^\alpha}{\lambda^2}
     \diff{x^\beta}{\lambda} + \Gamma^\mu_{\;\alpha\beta}\diff{x^\alpha}{\lambda}
     \diff{^2x^\beta}{\lambda^2}=0 \,.
\end{equation}
When we insert the geodesic equation into the terms containing the second derivatives, we find 
\begin{equation*}
 \diff{^3x^\mu}{\lambda^3}+\left( \Gamma^\mu_{\;\alpha\beta,\gamma} - \Gamma^\nu_{\;\alpha\gamma}
 \Gamma^\mu_{\;\nu\beta}- \Gamma^\nu_{\;\beta\gamma}\Gamma^\mu_{\;\alpha\nu}\right) \diff{x^\alpha}{\lambda}
 \diff{x^\beta}{\lambda}\diff{x^\gamma}{\lambda}=\diff{^3x^\mu}{\lambda^3}+ \Gamma^\mu_{\;\alpha\beta\gamma}
 \diff{x^\alpha}{\lambda} \diff{x^\beta}{\lambda}\diff{x^\gamma}{\lambda}=0\,,
\end{equation*}
where in the second step we have introduced the quantity
\begin{equation*}
\Gamma^\mu_{\;\alpha\beta\gamma} \equiv \Gamma^\mu_{\;\alpha\beta,\gamma} - \Gamma^\nu_{\;\alpha\gamma} \Gamma^\mu_{\;\nu\beta}- \Gamma^\nu_{\;\beta\gamma}\Gamma^\mu_{\;\alpha\nu}\,.
\end{equation*}
We now generalize the definition of the coefficients $\Gamma^\mu_{\;\alpha\beta\gamma}$ recursively by
\begin{equation}
\Gamma^\mu_{\;\alpha_1\ldots\alpha_n} \equiv \Gamma^\mu_{\;\alpha_1\ldots\alpha_{n-1},\alpha_n} -\sum_{i=1}^{n-1} \Gamma^\nu_{\;\alpha_i\alpha_n} \Gamma^\mu_{\;\alpha_1\ldots\alpha_{i-1}\; \nu \;\alpha_{i+1}\ldots\alpha_{n-1}}\,.
\label{Christoffeln}
\end{equation}
This definition looks formally like a covariant differentiation, which has been applied to the lower indices of the quantities $\Gamma^\mu_{\;\alpha_1\ldots\alpha_{n-1}}$ only. However, we emphasize that none of the $\Gamma$'s has tensorial character at all.

Using the recursive definition~(\ref{Christoffeln}), it follows by induction that all higher differentiations of the geodesic equation with respect to $\lambda$ can be expressed by the general formulas
\begin{equation}
\diff{^nx^\mu}{\lambda^n}+ \Gamma^\mu_{\;\alpha_1\ldots\alpha_n} \diff{x^{\alpha_1}}{\lambda}\cdot\ldots\cdot\diff{x^{\alpha_n}}{\lambda}=0\quad\text{for}\quad n \geq 2\,. 
\label{Geodesicn}
\end{equation}
We substitute eq.~(\ref{Geodesicn}) evaluated at $P$ into the power-series expansion of the geodesic, eq.~(\ref{Geoexpand}), to arrive finally at the formal solution  
\begin{equation}
x^\mu(\lambda)=p^\mu +v^\mu \lambda - \sum_{n=2}^\infty \Gamma^\mu_{\;\alpha_1\ldots\alpha_n}(p^\sigma)\, v^{\alpha_1} \cdot\ldots\cdot v^{\alpha_n}\frac{\lambda^n}{n!}
\label{Geoexpansion}
\end{equation}
of the geodesic equation. With this result at hand we now continue with the construction of Riemann normal coordinates.

\subsection{Riemann normal coordinates\label{AppRiemannNormal}}

The objective of the present subsection is to provide the definition of Riemann normal coordinates and to obtain the expansion of the metric around the new origin up to the third order.

\subsubsection{Definition and subtleties}

We start by recalling the situation considered in the preceding subsection. There, we were interested in all geodesics which emerge from the point $P$ with coordinates $p^\mu$ and with arbitrarily directed initial tangent vectors $v^\mu$, eq.~(\ref{Initial}). The formal solution of the geodesic equation~(\ref{GeodesicEq2}) was then given by eq.~(\ref{Geoexpansion}). 

Now,it is always possible to connect any point $X$ with coordinates $x^\mu$ by a unique geodesic with the initial point $P$, provided that $X$ lies within a sufficiently small region $\mathcal{D}$ around $P$. In this case, there exists a one-to-one correspondence between the coordinates $x^\mu$ of the point $X\in\mathcal{D}$ and one of the scaled initial tangent vectors $v^\mu\lambda$. This one-to-one correspondence allows us to establish Riemann normal coordinates $x^{\basind{\alpha}}$ in the following way: given a tetrad basis $e^\mu_{\basind{\alpha}}(p^\sigma)$ at $P$, we identify Riemann normal coordinates as the tetrad coefficients $x^{\basind{\alpha}}=v^{\basind{\alpha}}\lambda$ of the scaled initial four-velocity $v^\mu \lambda\,$, eq.~(\ref{Initial}), where $\lambda$ denotes the curve parameter of the geodesics at the point $X$. Taking eq.~(\ref{OrthoNice1}) into account, this statement reads explicitly
\begin{equation}
v^\mu\lambda= e^\mu_{\;\basind{\alpha}}(p^\sigma)\; x^{\basind{\alpha}}
\quad\text{or respectively}\quad 
x^{\basind{\alpha}}=e_\mu^{\;\,\basind{\alpha}}(p^\sigma)\; v^\mu\, \lambda
\,.
\label{DefRiemannNormalCoord}
\end{equation}
The connection between the original coordinates $x^\mu$ of a point $X\in\mathcal{D}$ and Riemann normal coordinates $x^{\basind{\alpha}}$ follows by substitution of eq.~(\ref{DefRiemannNormalCoord}) into the formal solution~(\ref{Geoexpansion}) of the geodesic equation leading to the expression
\begin{equation}
x^\mu(x^{\basind{\sigma}})=p^\mu + e^\mu_{\;\basind{\alpha}}\, x^{\basind{\alpha}} - \sum_{n=2}^\infty \frac{1}{n!}\, \Gamma^\mu_{\;\nu_1\ldots\nu_n}(p^\sigma)\, e^{\nu_1}_{\;\basind{\beta_1}}\cdot\ldots\cdot e^{\nu_n}_{\;\basind{\beta_n}}\, x^{\basind{\beta_1}} \cdot\ldots\cdot x^{\basind{\beta_n}}\,.
\label{RiemannNormalCoordTrafo}
\end{equation}
In this way, the origin of the Riemann normal coordinates is given by the spacetime point~$P$. Its new coordinates read ${p^{\basind{\mu}}=0}$ since they correspond to the starting point ${\lambda =0}$ of the geodesic in eq.~(\ref{DefRiemannNormalCoord}). Moreover, we conclude by inspection of eq.~(\ref{DefRiemannNormalCoord}), that all geodesics passing through $P$ become straight lines in Riemann normal coordinates $x^{\basind{\alpha}}$, in analogy to Cartesian coordinates in Euclidean geometry. Due to their definition as tetrad coefficients\footnote{We admit that the notation used in appendix~\ref{AppTetrad} might be misleading, since it suggests the relation ${x^\mu = e^\mu_{\;\basind{\alpha}} x^{\basind{\alpha}}}$ which does {\it not} hold true. This is due to the fact that the coordinates $x^\mu$ label points in the manifold, whereas the tetrad vectors $e^\mu_{\;\basind{\alpha}}$ are elements of the tangent space at the point $x^\mu$. Indeed, the superscript $\mu$ has two completely different meanings depending on the left or right hand side of the previous expression. The possible confusion in the notation originates from the desire to denote the new coordinates by the letter $x^{\basind{\alpha}}$, and not by ${(v\lambda)^{\basind{\alpha}}=e_\mu^{\;\,\basind{\alpha}}\,(v\lambda)^\mu}$ as would be self-evident according to appendix~\ref{AppTetrad}. Equation~(\ref{RiemannNormalCoordTrafo}) displays the correct relation between both sets of coordinates.} 
of the scaled tangent vector $v^\mu\lambda$, Riemann normal coordinates behave like invariants under general coordinate transformations.  Strictly speaking, Riemann normal coordinates are two-point invariants, since they depend on the two points $P$ and ${X\in\mathcal{D}}$ which are connected by a unique geodesic. 

\begin{figure}
\centering
\includegraphics[width=0.87\textwidth]{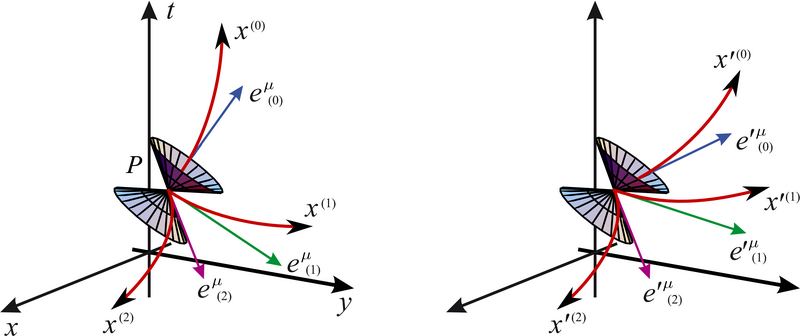}
\caption{Dependence of Riemann normal coordinates on the choice of the tetrad at the point $P$. The geodesics $x^{\basind{\alpha}}$ and $x'^{\basind{\alpha}}$ which emerge from $P$ for the two different initial tangent vectors $e^\mu_{\;\basind{\alpha}}$ (left) and $e'^\mu_{\;\basind{\alpha}}$ (right) define two different sets of ``coordinate axis'' for the corresponding Riemann normal coordinates.
In order to bring out the similarities and differences between the two pictures most clearly, we have marked in the right picture only those quantities which are different from the ones in the left picture.}
\label{FigRiemannNormalCoord}
\end{figure}
Riemann normal coordinates are not uniquely defined by eq.~(\ref{DefRiemannNormalCoord}). Indeed, they crucially depend on the choice of the particular tetrad at the origin $P$ as illustrated in fig.~\ref{FigRiemannNormalCoord}. The two pictures represent two examples of Riemann normal coordinates with the same origin $P$ resulting from two different tetrads. The tetrad $e^\mu_{\;\basind{\alpha}}$ in the left picture of fig.~\ref{FigRiemannNormalCoord} corresponds to the tetrad shown on the left of fig.~\ref{LocalLightcone}, and in analogy the tetrad in the right picture of fig.~\ref{FigRiemannNormalCoord} coincides with the tetrads $e'^\mu_{\;\basind{\alpha}}$ on the right of fig.~\ref{LocalLightcone}. In both pictures of fig.~\ref{FigRiemannNormalCoord} we have indicated some ``coordinate axes'' $x^{\basind{\alpha}}$ and $x'^{\basind{\alpha}}$ which are just the geodesics resulting from the initial point $P$ with initial tangent vectors $e^\mu_{\;\basind{\alpha}}$ and $e'^\mu_{\;\basind{\alpha}}$, respectively. Both sets of Riemann normal coordinates are connected to each other by a corresponding Lorentz transformation matrix.

Before we turn to the expansion of the metric coefficients in Riemann normal coordinates, we briefly address the range of validity of Riemann normal coordinates. Due to the fact, that according to eq.~(\ref{DetTetradMatrix}) the coordinate transformation~(\ref{RiemannNormalCoordTrafo}) possesses a non-vanishing Jacobian determinant  
\begin{equation*}
\text{det} \left(\pdiff{x^\mu}{x^{\basind{\alpha}}}\right)\bigg|_P =\text{det}\left(e^\mu_{\;\basind{\alpha}}\right)=\frac{1}{\sqrt{-g}}\neq 0\,,
\end{equation*} 
there exists a sufficiently small region around $P$ such, that the coordinate transformation~(\ref{RiemannNormalCoordTrafo}) can be inverted. As already pointed out in the beginning of this subsection, Riemann normal coordinates are thus defined only in a domain $\mathcal{D}$ around $P$, in which it is possible to connect any point $X$ in $\mathcal{D}$ by a unique geodesic with $P$, such that there exists a one-to-one correspondence between the scaled, initial tangent $v^\mu \lambda$ and $X$.

There exist several scenarios which are responsible for the restriction of Riemann normal coordinates to the domain $\mathcal{D}$. For example, it might happen that two geodesics, which emanate from $P$ with different initial tangent vectors, intersect in some point ${Y\notin\mathcal{D}}$ due to the curvature of the underlying spacetime. It could also happen, that a geodesic emerging from $P$ encounters a singularity ${S\notin\mathcal{D}}$ of the metric. Or the geodesic could be bent by the intrinsic rotation of the spacetime in such a way, that it propagates only within a finite spatial region around $P$. This situation occurs in e.\,g. G\"odel's Universe.

\subsubsection{Metric coefficients}

With the help of Riemann normal coordinates the laws of nature reduce to their special relativistic form in the neighborhood of any fixed event~$P$. It is our aim to substantiate this statement by considering the expansion 
\begin{equation}
g_{\basind{\mu\nu}}(x^{\basind{\sigma}})=g_{\basind{\mu\nu}}(0) +\sum_{n=1}^\infty \frac{1}{n!}\, g_{\basind{\mu\nu},\basind{\alpha_1},\ldots,\basind{\alpha_n}}(0)\, x^{\basind{\alpha_1}}\cdot\ldots\cdot x^{\basind{\alpha_n}}
\label{ExpansionMetric}
\end{equation}
of the metric coefficients $g_{\basind{\mu\nu}}(x^{\basind{\sigma}})$ expressed in Riemann normal coordinates around the origin ${p^{\basind{\mu}}=0}$.

\paragraph{Zeroth-order term}

The zeroth-order term $g_{\basind{\mu\nu}}(0)$ can easily be found using the transformation law of the metric coefficients for the coordinate transformation~(\ref{RiemannNormalCoordTrafo}) at~$P$ together with the orthonormality condition~(\ref{OrthonormalTetrad}), which yields
\begin{equation}
g_{\basind{\mu\nu}}(0)=\pdiff{x^\alpha}{x^{\basind{\mu}}}\bigg|_P \pdiff{x^\beta}{x^{\basind{\nu}}}\bigg|_P g_{\alpha\beta}(p^\sigma)=
e^\alpha_{\;\basind{\mu}}\, e^\beta_{\;\basind{\nu}}\, g_{\alpha\beta}(p^\sigma)=\eta_{\basind{\mu\nu}}\, .
\label{Riemanng0}
\end{equation}

\paragraph{First-order term}

In order to obtain the higher order terms of the expansion we consider the geodesic equation~(\ref{GeodesicEq2}) together with the corresponding initial conditions~(\ref{Initial}), but now expressed in Riemann normal coordinates, that is
\begin{equation}
 \diff{^2x^{\basind{\mu}}}{\lambda^2}+
\Gamma^{\basind{\mu}}_{\;\basind{\alpha}\basind{\beta}}(x^{\basind{\sigma}})
\diff{x^{\basind{\alpha}}}{\lambda} \diff{x^{\basind{\beta}}}{\lambda}=0 
\,,\quad
x^{\basind{\mu}}(0)=0 \;\;\text{and}\;\;  \diff{ x^{\basind{\mu}}(0)}{\lambda} = v^{\basind{\mu}}\,.
\label{RiemannInitial}
\end{equation}
In complete analogy to eq.~(\ref{Geoexpansion}) we can find the formal solution 
\begin{equation}
x^{\basind{\mu}}(\lambda)=v^{\basind{\mu}} \lambda -\sum_{n=2}^\infty \Gamma^{\basind{\mu}}_{\;\basind{\alpha_1}\ldots\basind{\alpha_n}}(0)\, v^{\basind{\alpha_1}}\cdot\ldots\cdot v^{\basind{\alpha_n}}\,\frac{\lambda^n}{n!} 
\label{RiemannGeoexpansion}
\end{equation}
of the geodesic equation in Riemann normal coordinates. Here the quantities $\Gamma^{\basind{\mu}}_{\;\basind{\alpha_1}\ldots\basind{\alpha_n}}$ are defined again in complete analogy to eq.~(\ref{Christoffeln}) with $\Gamma^{\basind{\mu}}_{\;\basind{\alpha}\basind{\beta}}$ denoting the Christoffel symbols in Riemann normal coordinates. 

However, it is now important to realize, that the formal solution was already given by eq.~(\ref{Geoexpansion}) in terms of the original coordinates $x^\mu$. This formal expression was then used in connection with eq.~(\ref{DefRiemannNormalCoord}) to establish the transformation law~(\ref{RiemannNormalCoordTrafo}) between the original and the Riemann normal coordinates. Therefore, the solution of the geodesic equation~(\ref{RiemannInitial}) reads
\begin{equation*}
x^{\basind{\mu}} (\lambda)=e^{\;\,\basind{\mu}}_{\alpha}\,v^\alpha\, \lambda= v^{\basind{\mu}} \,\lambda \,.
\end{equation*}
Since the formal solution~(\ref{RiemannGeoexpansion}) has to match the last equation for all possible values ${\lambda>0}$ and all possible tangent vectors $v^{\basind{\mu}}$, we conclude that the quantities $\Gamma^{\basind{\mu}}_{\;\basind{\alpha_1}\ldots\basind{\alpha_n}}$ have to satisfy the condition  
\begin{equation}
\Gamma^{\basind{\mu}}_{\;\{\basind{\alpha_1}\ldots\basind{\alpha_n}\}}(0)=0 \quad \forall\;\;n\geq 2  \label{ZeroGammas}
\end{equation}
at $P$. Here we used the symmetrization brackets~(\ref{SymmBrackets}). 

Since the Christoffel symbols 
$\Gamma^{\basind{\mu}}_{\;\basind{\alpha}\basind{\beta}}$ are symmetric in their lower indices, the condition~(\ref{ZeroGammas}) implies for $n=2$ at $P$
\begin{equation}
\Gamma^{\basind{\mu}}_{\;\basind{\alpha}\basind{\beta}}(0)=0 \,.    
\label{ZeroChristoffel}
\end{equation}
This relation brings out another interesting feature of Riemann normal coordinates: The covariant derivative of any tensor field reduces to the ordinary partial derivative at the point $P$. Moreover, since the covariant derivative of the metric always vanishes ${g_{\basind{\mu\nu};\basind{\alpha}}=0}$, we obtain with eq.~(\ref{ZeroChristoffel}) at $P$ the identity 
\begin{equation}
g_{\basind{\mu\nu},\basind{\alpha}}(0)=0 \,.
\label{Riemanng1}
\end{equation}
Hence, all first partial derivatives of the metric vanish at $P$ in Riemann normal coordinates.\\

\paragraph{Second- and third-order terms}

Equation~(\ref{Riemanng0}) together with eq.~\eqref{Riemanng1} provide the bedrock of local inertial coordinates: Coordinates in which the metric reduces to its Minkowski form ${g_{\basind{\mu\nu}}(0)=\eta_{\basind{\mu\nu}}}$ at a spacetime point~$P$, while the first partial derivatives $g_{\basind{\mu\nu},\basind{\alpha}}$ vanish at $P$, are called {\it local inertial coordinates}.

However, the second and the higher partial derivatives of the metric do not vanish at $P$, but they can be expressed in terms of the curvature tensor and its higher covariant derivatives. Here, we are solely interested in the second and third order terms. 

We start by drawing the connection between the symmetrized coefficients of the second and third partial derivatives $g_{\basind{\mu\nu}\{,\basind{\alpha_1},\basind{\alpha_2}\}}(0)$ and 
$g_{\basind{\mu\nu}\{,\basind{\alpha_1},\basind{\alpha_2},\basind{\alpha_3}\}}(0)$, 
which appear in eq.~(\ref{ExpansionMetric}), and the derivatives of the Christoffel symbols. For the second partial derivative we explicitly evaluate the vanishing second covariant derivative of the metric tensor ${g_{\basind{\mu\nu};\basind{\alpha_1};\basind{\alpha_2}}=0}$ at $P$. This procedure yields with eq.~(\ref{ZeroChristoffel}) the expression
\begin{align}
0&=g_{\basind{\mu\nu};\basind{\alpha_1};\basind{\alpha_2}}(0)=
\left(g_{\basind{\mu\nu};\basind{\alpha_1}}\right)_{,\basind{\alpha_2}} \big|_P 
\label{Zwischenrechnungg2}\\
&= g_{\basind{\mu\nu},\basind{\alpha_1},\basind{\alpha_2}}(0)-
\Gamma^{\basind{\rho}}_{\;\basind{\mu}\basind{\alpha_1},\basind{\alpha_2}}(0)\,g_{\basind{\rho\nu}}(0) -
\Gamma^{\basind{\rho}}_{\;\basind{\nu}\basind{\alpha_1},\basind{\alpha_2}}(0)\,g_{\basind{\mu\rho}}(0) 
\nonumber\,.
\end{align}
An analogous expansion of the vanishing third covariant derivative $g_{\basind{\mu\nu};\basind{\alpha_1};\basind{\alpha_2};\basind{\alpha_3}}=0$
leads after application of eqs.~(\ref{ZeroChristoffel}) and (\ref{Riemanng1}) and the fact, that all covariant derivatives of the metric components are zero, to the relation
\begin{equation*}
g_{\basind{\mu\nu},\basind{\alpha_1},\basind{\alpha_2},\basind{\alpha_3}}(0)-
\Gamma^{\basind{\rho}}_{\;\basind{\mu}\basind{\alpha_1},\basind{\alpha_2},\basind{\alpha_3}}(0)
\,g_{\basind{\rho\nu}}(0) -
\Gamma^{\basind{\rho}}_{\;\basind{\nu}\basind{\alpha_1},\basind{\alpha_2},\basind{\alpha_3}}(0)
\,g_{\basind{\mu\rho}}(0)=0\,.
\end{equation*}
Thus, the symmetrized second and third partial derivatives of the metric coefficients at $P$ can be expressed in terms of the Christoffel symbols according to 
\begin{equation}
g_{\basind{\mu\nu}\{,\basind{\alpha_1},\basind{\alpha_2}\}}(0)=g_{\basind{\rho\nu}}(0)\,
\Gamma^{\basind{\rho}}_{\;\basind{\mu}\{\basind{\alpha_1},\basind{\alpha_2}\}}(0)+
g_{\basind{\mu\rho}}(0)\,\Gamma^{\basind{\rho}}_{\;\basind{\nu}\{\basind{\alpha_1},\basind{\alpha_2}\}}(0)
\label{RiemannZwischen0}
\end{equation}
and
\begin{align}
&g_{\basind{\mu\nu}\{,\basind{\alpha_1},\basind{\alpha_2},\basind{\alpha_3}\}}(0)=\label{RiemannZwischen0b}\\
&\qquad g_{\basind{\mu\rho}}(0)\,
\Gamma^{\basind{\rho}}_{\;\basind{\nu}\{\basind{\alpha_1},\basind{\alpha_2},\basind{\alpha_3}\}}(0)  
+ g_{\basind{\rho\nu}}(0)\,
\Gamma^{\basind{\rho}}_{\;\basind{\mu}\{\basind{\alpha_1},\basind{\alpha_2},\basind{\alpha_3}\}}(0)
 \nonumber\,.
\end{align}
On the other hand, the curvature tensor~(\ref{DefCurvatureTensor}) and its first covariant derivative reduce in Riemann normal coordinates to the simple form 
\begin{equation}
 R^{\basind{\rho}}_{\;\,\basind{\alpha_1}\basind{\alpha_2}\basind{\alpha_3}}(0)=
\Gamma^{\basind{\rho}}_{\;\basind{\alpha_1}\basind{\alpha_3},\basind{\alpha_2}}(0)-
\Gamma^{\basind{\rho}}_{\;\basind{\alpha_1}\basind{\alpha_2},\basind{\alpha_3}}(0) 
\label{NormalRiemann1}
\end{equation}
and
\begin{equation}
 R^{\basind{\rho}}_{\;\,\basind{\alpha_1}\basind{\alpha_2}\basind{\alpha_3};\basind{\alpha_4}}(0)=
\Gamma^{\basind{\rho}}_{\;\basind{\alpha_1}\basind{\alpha_3},\basind{\alpha_2},\basind{\alpha_4}}(0)-
\Gamma^{\basind{\rho}}_{\;\basind{\alpha_1}\basind{\alpha_2},\basind{\alpha_3},\basind{\alpha_4}}(0)
\label{NormalRiemann2}
\end{equation}
at the spacetime point $P$.
The remaining task is to express the first and the second derivative of the Christoffel symbols which appear on the right hand side of eqs.~(\ref{RiemannZwischen0}) and~(\ref{RiemannZwischen0b}) in terms of the curvature tensor and its first covariant derivative at $P$. For this purpose, we first cast condition~(\ref{ZeroGammas}) for $n=3$ and $n=4$ in a more useful form and replace the quantities ${\Gamma^{\basind{\rho}}_{\;\basind{\alpha_1}\basind{\alpha_2}\basind{\alpha_3}}}$ and 
${\Gamma^{\basind{\rho}}_{\;\basind{\alpha_1}\basind{\alpha_2}\basind{\alpha_3}\basind{\alpha_4}}}$
in terms of their recursive definitions~(\ref{Christoffeln}). Together with eq.~(\ref{ZeroChristoffel}), we thus arrive at the relations
\begin{align}
\Gamma^{\basind{\rho}}_{\;\{\basind{\alpha_1}\basind{\alpha_2}\basind{\alpha_3}\}}(0)&= \Gamma^{\basind{\rho}}_{\;\{\basind{\alpha_1}\basind{\alpha_2},\basind{\alpha_3}\}}(0)=0
\label{ConditionGamma1}
\end{align}
and
\begin{align}
\Gamma^{\basind{\rho}}_{\;\{\basind{\alpha_1}\basind{\alpha_2}\basind{\alpha_3}\basind{\alpha_4}\}}(0)&= \Gamma^{\basind{\rho}}_{\;\{\basind{\alpha_1}\basind{\alpha_2},\basind{\alpha_3},\basind{\alpha_4}\}}(0)=0
\label{ConditionGamma2}  \,.
\end{align}
Using the symmetrization symbol, eq.~(\ref{SubsetSymmetrization}), it is not difficult to show, that eq.~(\ref{ConditionGamma1}) is equivalent to the expression
\begin{equation*}
\frac{1}{3}\,\symm{\alpha_1 \alpha_2}{
\Gamma^{\basind{\rho}}_{\;\basind{\alpha_1}\basind{\alpha_2},\basind{\alpha_3}}(0)+
\Gamma^{\basind{\rho}}_{\;\basind{\alpha_1}\basind{\alpha_3},\basind{\alpha_2}}(0)+
\Gamma^{\basind{\rho}}_{\;\basind{\alpha_3}\basind{\alpha_1},\basind{\alpha_2}}(0)}=0\,,
\end{equation*}
which reduces due to the symmetry of the Christoffel symbol in its lower indices to the identity
\begin{equation}
\Gamma^{\basind{\rho}}_{\;\{\basind{\alpha_1}\basind{\alpha_2}\},\basind{\alpha_3}}(0)
=-2\,\Gamma^{\basind{\rho}}_{\;\basind{\alpha_3}\{\basind{\alpha_1},\basind{\alpha_2}\}}(0)\,.
\label{ConditionGamma3}
\end{equation}
In complete analogy, we first rewrite eq.~(\ref{ConditionGamma2}) in terms of the symmetrization symbol
\begin{align*}
&\frac{1}{4}\,\symm{\alpha_1 \alpha_2 \alpha_4}{
\Gamma^{\basind{\rho}}_{\;\basind{\alpha_1}\basind{\alpha_2},\basind{\alpha_3},\basind{\alpha_4}}(0)+
\Gamma^{\basind{\rho}}_{\;\basind{\alpha_1}\basind{\alpha_2},\basind{\alpha_4},\basind{\alpha_3}}(0)}\\
&+\frac{1}{4}\,\symm{\alpha_1 \alpha_2 \alpha_4}{
\Gamma^{\basind{\rho}}_{\;\basind{\alpha_3}\basind{\alpha_1},\basind{\alpha_2},\basind{\alpha_4}}(0)+
\Gamma^{\basind{\rho}}_{\;\basind{\alpha_1}\basind{\alpha_3},\basind{\alpha_2},\basind{\alpha_4}}(0)
}=0\,,
\end{align*}
and use ${\Gamma^{\basind{\rho}}_{\;\basind{\alpha_1}\basind{\alpha_2},\basind{\alpha_3},\basind{\alpha_4}}=
\Gamma^{\basind{\rho}}_{\;\basind{\alpha_1}\basind{\alpha_2},\basind{\alpha_4},\basind{\alpha_3}}}$, as well as the symmetry in the lower indices of the Christoffel symbols, to finally arrive at
\begin{equation}
\Gamma^{\basind{\rho}}_{\;\{\basind{\alpha_1}\basind{\alpha_2},\basind{\alpha_4}\},\basind{\alpha_3}}(0)=
-\Gamma^{\basind{\rho}}_{\;\basind{\alpha_3}\{\basind{\alpha_1},\basind{\alpha_2},\basind{\alpha_4}\}}(0)
\,.
\label{ConditionGamma4}
\end{equation}
Hence, applying the symmetrization symbol with respect to the indices $(\alpha_1,\alpha_2)$ to eq.~(\ref{NormalRiemann1}) and with respect to the indices $(\alpha_1,\alpha_2,\alpha_4)$ to eq.~(\ref{NormalRiemann2}), we find with the help of eqs.~(\ref{ConditionGamma3}) and~(\ref{ConditionGamma4}) the identities
\begin{align}
\Gamma^{\basind{\rho}}_{\;\basind{\alpha_3}\{\basind{\alpha_1},\basind{\alpha_2}\}}(0)=&
\frac{1}{3}\,\symm{\alpha_1 \alpha_2}{ R^{\basind{\rho}}_{\;\,\basind{\alpha_1}\basind{\alpha_2}\basind{\alpha_3}}(0)}
\label{ConditionGamma5}
\end{align}
and
\begin{align}
\Gamma^{\basind{\rho}}_{\;\basind{\alpha_3}\{\basind{\alpha_1},\basind{\alpha_2},\basind{\alpha_4}\}}(0)=&
\frac{1}{2}\,\symm{\alpha_1 \alpha_2 \alpha_4}{ R^{\basind{\rho}}_{\;\,\basind{\alpha_1}\basind{\alpha_2}\basind{\alpha_3};\basind{\alpha_4}}(0)}
\label{ConditionGamma6}\,.
\end{align} 
Thus, we have accomplished our task to express the symmetrized first and second derivative of the Christoffel symbols in terms of the curvature tensor and its first covariant derivative at $P$. We now insert the last two expressions into eqs.~(\ref{RiemannZwischen0}) and~\eqref{RiemannZwischen0b} and obtain
\begin{align}
g_{\basind{\mu\nu}\{,\basind{\alpha_1},\basind{\alpha_2}\}}(0)=&
\frac{2}{3}\,\symm{\alpha_1 \alpha_2}{
R_{\basind{\mu}\basind{\alpha_1}\basind{\alpha_2}\basind{\nu}}(0)}
\label{Riemanng2}
\end{align}
and
\begin{align}
g_{\basind{\mu\nu}\{,\basind{\alpha_1},\basind{\alpha_2},\basind{\alpha_3}\}}(0)=&
\symm{\alpha_1 \alpha_2 \alpha_3}{ R_{\basind{\mu}\basind{\alpha_1}\basind{\alpha_2}\basind{\nu};\basind{\alpha_3}}(0)}\label{Riemanng3}\,,
\end{align} 
where we have used the symmetries of the curvature tensor (\ref{RiemannTensorCovSymm}).

\subsubsection{Summary}

In conclusion, we insert the results~(\ref{Riemanng0}), (\ref{Riemanng1}), (\ref{Riemanng2}) and (\ref{Riemanng3}) into the expansion~(\ref{ExpansionMetric}) of the metric in Riemann normal coordinates and obtain up to third order 
\begin{align}
g_{\basind{\mu\nu}}(x^{\basind{\sigma}})=&\;\eta_{\basind{\mu\nu}}+\frac{1}{3}\,
R_{\basind{\mu}\basind{\alpha_1}\basind{\alpha_2}\basind{\nu}}(0)\,x^{\basind{\alpha_1}} x^{\basind{\alpha_2}}
\label{RiemannMetricExpansion}\\
&+\frac{1}{6}\,R_{\basind{\mu}\basind{\alpha_1}\basind{\alpha_2}\basind{\nu};\basind{\alpha_3}}(0)\,
x^{\basind{\alpha_1}} x^{\basind{\alpha_2}} x^{\basind{\alpha_3}} 
+\mathcal{O}((x^{\basind{\alpha}})^4)\,.\nonumber  
\end{align}
We emphasize that the preceding derivation is based on the recursive determination of the expansion coefficients of arbitrary tensors in Riemann normal coordinates given in \cite{AlvarezGaume81,Hatzinikitas00}. Moreover, some of the higher order contributions to the expansion of the metric can be found there.

In summary, Riemann normal coordinates provide local inertial coordinates around a fixed spacetime point $P$, where the curvature of the underlying spacetime appears in the second and higher orders of the expansion of the metric coefficients. However, Riemann normal coordinates neither represent appropriate coordinates for the description of a freely falling inertial observer, nor for the more general case of an arbitrarily rotating observer which moves along any world line. Nevertheless, our inspection of Riemann normal coordinates provides a solid basis for the discussion of the so-called proper reference frames, which represent appropriate local coordinates for such arbitrarily moving observers.

\subsection{Local coordinates of a proper reference frame\label{AppProperRef}}

The basic ingredients and the geometrical concepts behind the definition of proper reference frame coordinates  were already given in subsect.~\ref{ProperRef}. Nevertheless, we want to recall their definition in a more explicit and formal manner in the beginning of this subsection, thereby bringing forward the analogies and differences to Riemann normal coordinates. We then proceed with the analysis of the metric expansion around the world line of an accelerating and rotating observer and utilize the same techniques which were previously used in the derivation of eq.~(\ref{RiemannMetricExpansion}). We close by giving an approximate solution of the geodesic equation in proper reference frame coordinates which is valid in the local neighborhood of the observer's world line.

\subsubsection{Definition}

The construction of proper reference frame coordinates is primarily based on the world line $p^\mu(\tau)$ of the accelerating and rotating observer with four-velocity $u^\mu(\tau)$ and four-acceleration $a^\mu(\tau)$. Furthermore, the observer carries with him a tetrad $e^\mu_{\;\basind{\alpha}}(\tau)$ in order to distinguish different spatial directions in his rest frame. This tetrad satisfies the relativistic orthonormality condition (\ref{RelOrthogonality}) and the proper transport eq.~(\ref{ProperTransport}) along $p^\mu(\tau)$. The local coordinates of the proper reference frame are then established as follows: draw a spacelike geodesic $x^\mu=x^\mu(\tau,v^\sigma s)$ from a point $p^\mu(\tau)$ on the world line to a point $x^\mu$ in its neighborhood in such a way, that the initial tangent vector $v^\mu(\tau)$ is orthogonal to the four-velocity $u^{\mu}(\tau)$ at the initial point $p^{\mu}(\tau)$. This construction is indicated in fig.~\ref{FigProperReferenceFrame} for the initial point $P_1$ with coordinate values $p^\mu(\tau_1)$. Within a sufficiently small region around the world line $p^\mu(\tau)$, this definition guarantees a one-to-one correspondence between the original coordinate values $x^\mu$ and the local coordinates represented by the arclength $s$ and the initial direction of the spacelike geodesic together with the proper time $\tau$, that corresponds to the initial point $p^\mu(\tau)$.

In order to substantiate the connection between the original and the local coordinates, we once more take advantage of the formal solution of the geodesic equation presented in appendix~\ref{AppFormalSolution}. However, this time we consider the spacelike geodesic $x^\mu=x^\mu(\tau,v^\sigma s)$. 

We start from the geodesic equation
\begin{equation}
 \diff{^2x^\mu}{ s^2}+\Gamma^\mu_{\;\alpha\beta}(x^\sigma)\diff{x^\alpha}{ s}
     \diff{x^\beta}{s}=0 \,.
\label{DefGeoProper}
\end{equation}
In contrast to Riemann normal coordinates, the initial conditions 
\begin{equation}
x^\mu(\tau,0)=p^\mu(\tau) \quad\text{and}\quad \diff{}{s} x^\mu(\tau,0) =v^\mu(\tau)=v^{\basind{i}}(\tau)\,e^\mu_{\;\basind{i}}(\tau)
\label{PropInitialCond}
 \end{equation}
for the spacelike geodesic now depend explicitly on the proper time $\tau$ of the observer.

The orthogonality relation
\begin{equation}
v^\mu(\tau)\,u_\mu(\tau)=0\,,
\label{OrthoTangent}
\end{equation} 
between the initial tangent vector $v^\mu(\tau)$ and the four-velocity ${u^\mu(\tau)=c\,e^\mu_{\;\basind{0}}(\tau)}$ of the observer is automatically satisfied by its tetrad decomposition ${v^\mu(\tau)=v^{\basind{i}}(\tau)\,e^\mu_{\;\basind{i}}(\tau)}$. Moreover, we assume without loss of generality that the tetrad components of the initial tangent vector are normalized, that is
\begin{equation*}
 {(v^{\basind{1}}(\tau))^2+(v^{\basind{2}}(\tau))^2+(v^{\basind{3}}(\tau))^2=1}\,.
\end{equation*} 
The quantities $v^{\basind{i}}$ just characterize the initial direction in which the spacelike geodesic emanates from  $p^\mu(\tau)$.

Hence, the formal solution~(\ref{Geoexpansion}) of the geodesic equation~(\ref{DefGeoProper}) reads with eq.~(\ref{PropInitialCond})
\begin{equation}
 x^\mu(\tau,v^\sigma s)=p^\mu(\tau) +v^\mu(\tau) \,s - \sum_{n=2}^\infty
 \Gamma^\mu_{\;\alpha_1\ldots\alpha_n}(p^\sigma(\tau))\, v^{\alpha_1}(\tau) \cdot\ldots\cdot
 v^{\alpha_n}(\tau)\,\frac{s^n}{n!}\,.
\label{PropFormalSolution}
\end{equation}
According to subsect.~\ref{ProperRef}, the local coordinates\footnote{We characterize both proper reference frame coordinates and Riemann normal coordinates by superscripts in parenthesis. In this way we want to highlight the analogies between both sets of coordinates. However, throughout this article the symbol $x^{\basind{\alpha}}$ denotes proper reference frame coordinates. The only exception is appendix~\ref{AppRiemannNormal}.}
of the proper reference frame $x^{\basind{\alpha}}$ are given by the proper time $\tau$ of the observer located at $p^\mu(\tau)$, the spatial direction $v^{\basind{i}}$ of the outgoing spacelike geodesic, and the value of the arclength $s$ corresponding to the spacetime point $x^\mu$. In particular, they are established by the relations
\begin{equation}
\begin{array}{lll}
\tau=x^{\basind{0}}/c & \text{or} & \quad x^{\basind{0}}=c\,\tau \\
 v^\mu(\tau)\,s= e^\mu_{\;\basind{i}}(\tau)\; x^{\basind{i}} \quad & & \quad
 x^{\basind{i}}=e_\mu^{\;\,\basind{i}}(\tau)\; v^\mu(\tau)\,s=v^{\basind{i}}\,s\,.
\end{array}
\label{DefProperRef}
\end{equation}
Our notation might suggest that the time coordinate ${x^{\basind{0}}=c\,\tau}$ is a tetrad component. However, this is not the case -- it is simply a scalar that is proportional to the proper time of the observer. Nevertheless, our notation proves to be useful in pointing out the analogies to Riemann normal coordinates. On the other hand the spatial coordinates $x^{\basind{i}}$ are real tetrad indices coming from the scaled initial tangent vector $v^\mu(\tau)\,s$, which itself is an element of the three-dimensional spacelike hypersurface of the tangent space that is orthogonal to the four-velocity $u^\mu(\tau)$ at $p^\mu(\tau)$.

With this definition at hand, the connection between both sets of coordinates is simply established by inserting (\ref{DefProperRef}) into the formal solution~(\ref{PropFormalSolution}) of the geodesic equation, that is
\begin{align}
 x^\mu(x^{\basind{\sigma}})&=p^\mu\big(\textstyle{\frac{x^{\basind{0}}}{c}}\big) 
+ e^\mu_{\;\basind{i}}\big(\textstyle{\frac{x^{\basind{0}}}{c}}\big)\, x^{\basind{i}}
\label{PropRefFrameTrafo}\\
 &- \sum_{n=2}^\infty \frac{1}{n!}\, \Gamma^\mu_{\;\nu_1\ldots\nu_n}
\big(p^\sigma\big(\textstyle{\frac{x^{\basind{0}}}{c}}\big)\big)\, 
e^{\nu_1}_{\;\basind{i_1}}\big(\textstyle{\frac{x^{\basind{0}}}{c}}\big)
\cdot\ldots\cdot 
e^{\nu_n}_{\;\basind{i_n}}\big(\textstyle{\frac{x^{\basind{0}}}{c}}\big)\, x^{\basind{i_1}} \cdot\ldots\cdot x^{\basind{i_n}}\,.\nonumber
\end{align}
As a consequence of eq.~(\ref{DefProperRef}), the world line of the observer reads ${p^{\basind{\mu}}(\tau)=(c \tau,0,0,0)}$ when expressed in proper reference frame coordinates, with the origin ${p^{\basind{i}}=0}$ in the spatial subspace simply corresponding to the arclength ${s=0}$. 

That proper reference frame coordinates are indeed valid within a sufficiently small region around the world line $p^{\basind{\sigma}}(\tau)$ can be verified with the Jacobi determinant. Making use of (\ref{PropRefFrameTrafo}) we find for the transformation matrix along the world line $p^{\basind{\sigma}}(\tau)$
\begin{equation}
 \pdiff{x^\mu\phantom{^x}}{x^{\basind{\alpha}}}\Big|_{p^{\basind{\sigma}}(\tau)}=e^\mu_{\;\basind{\alpha}}(\tau)\,.
\label{PropTrafoMatrix}
\end{equation}
We have already discussed the determinant of a given tetrad in the context of Riemann normal coordinates. In analogy to the result given there, we conclude with the help of eq.~(\ref{DetTetradMatrix}) that
\begin{equation}
\det\left( \pdiff{x^\mu\phantom{^x}}{x^{\basind{\alpha}}}\right)\Big|_{p^{\basind{\sigma}}(\tau)}=
\det\left(e^\mu_{\;\basind{\alpha}}(\tau)\right)
=\frac{1}{\sqrt{-g}}\Big|_{p^{\sigma}(\tau)}\neq 0\,,
\end{equation} 
which proves the validity of the proper reference frame coordinates within a sufficiently small region around the world line of the observer.

\subsubsection{Metric \label{MetricExpandPropRef}}

We now come to the determination of the power-series expansion of the metric coefficients in proper reference frame coordinates up to second order around the world line $p^{\basind{\sigma}}(\tau)$ of the observer. Before we begin, we mention, that the second- and third-order contributions were already determined by \cite{Nesterov76,Ni78} and \cite{Li79a,Nesterov99}, respectively\footnote{The third-order contributions in \cite{Li79a} and \cite{Nesterov99} differ from each other and should therefore be considered with caution.\label{FootnoteDifference}}. In the case of a weak gravitational field, all orders of the power-series were investigated by~\cite{Marzlin94}.

The metric expansion enables us, to estimate the main physical effects which arise from the acceleration and rotation of the observer, as well as from the curvature of spacetime in the spatial neighborhood of his world line. In order to shorten the notation, we omit the explicit time dependence of ${p^{\basind{\sigma}}(\tau)}$ and simply write $p^{\basind{\sigma}}$ henceforth. However, it should be kept in mind that $p^{\basind{\sigma}}$ represents the world line of the observer. 

We start the derivation by providing the spatial power-series expansion
\begin{equation}
g_{\basind{\mu \nu}}(x^{\basind{\sigma}})=g_{\basind{\mu\nu}}(p^{{\basind{\sigma}}}) +\sum_{n=1}^\infty \frac{1}{n!}\, g_{\basind{\mu \nu},\basind{i_1},\ldots,\basind{i_n}}(p^{{\basind{\sigma}}})\, x^{\basind{i_1}}\cdot\ldots\cdot x^{\basind{i_n}} 
\label{ExpansionMetricProper}
\end{equation} 
of the metric around the observer's world line $p^{\basind{\sigma}}$. Since we use the spatial tetrad indices $x^{\basind{i}}$ of the scaled initial tangent vector $v^\mu(\tau)\,s$ as expansion parameters, the power-series~(\ref{ExpansionMetricProper}) solely applies to the spatial subspace of the tangent space at $p^{\basind{\sigma}}(\tau)$, that is orthogonal to the four-velocity $u^{\basind{\mu}}(\tau)$. This ansatz is in contrast to the corresponding expansion~(\ref{ExpansionMetric}) for Riemann normal coordinates.

\paragraph{Zeroth-order term}

According to the transformation matrix~(\ref{PropTrafoMatrix}), the zeroth-order coefficients of the metric expansion read
\begin{equation}
g_{\basind{\mu\nu}}(p^{\basind{\sigma}})=\pdiff{x^\alpha}{x^{\basind{\mu}}}\bigg|_{p^{\basind{\sigma}}}
 \pdiff{x^\beta}{x^{\basind{\nu}}}\bigg|_{p^{\basind{\sigma}}}
g_{\alpha\beta}(p^\sigma)=
e^\alpha_{\;\basind{\mu}}(\tau)\, e^\beta_{\;\basind{\nu}}(\tau)\, g_{\alpha\beta}(p^\sigma)=\eta_{\basind{\mu\nu}}\,,
\label{Properg0}
\end{equation}
and they are given by the metric of flat Minkowski spacetime, in complete analogy to Riemann normal coordinates.

\paragraph{Determination of higher-order terms}

However, the first- and second-order contributions will differ from the corresponding expressions in Riemann normal coordinates. They can be derived most conveniently in two steps: 

(i) The purely spatial components ${g_{\basind{j k},\basind{i_1}}(p^{{\basind{\sigma}}})}$ and 
${g_{\basind{j k}\{,\basind{i_1},\basind{i_2}\}}(p^{{\basind{\sigma}}})}$ follow in analogy to the results~(\ref{Riemanng1}) and~(\ref{Riemanng2}) for Riemann normal coordinates. 

(ii) The remaining components ${g_{\basind{0 \nu},\basind{i_1}}(p^{{\basind{\sigma}}})}$ and 
${g_{\basind{0 \nu}\{,\basind{i_1},\basind{i_2}\}}(p^{{\basind{\sigma}}})}$ are found by making use of the proper transport, eq.~(\ref{ProperTransport}), expressed in proper reference frame coordinates.\\

We start by reformulating the geodesic eq.~(\ref{DefGeoProper}) 
\begin{equation}
 \diff{^2x^{\basind{\mu}}}{s^2}
 +\Gamma^{\basind{\mu}}_{\;\basind{\alpha}\basind{\beta}}(x^{\basind{\sigma}})\diff{x^{\basind{\alpha}}}{s}
 \diff{x^{\basind{\beta}}}{s}=0 \,,
\label{DefGeoProper2}
\end{equation}
as well as the corresponding initial conditions~(\ref{PropInitialCond})
\begin{equation*}
x^{\basind{\mu}}(\tau,0)=p^{\basind{\mu}}(\tau)=(c\tau,0,0,0) \quad\text{and}\quad \diff{}{s} x^{\basind{\mu}}(\tau,0) =v^{\basind{\mu}}(\tau)
 \end{equation*}
in proper reference frame coordinates. With the help of eq.~(\ref{PropTrafoMatrix}), the vector components $v^{\basind{\mu}}$ follow from the transformation law of vectors and are thus in agreement with the tetrad components of $v^\mu(\tau)$, eq.~(\ref{PropInitialCond}). In particular, we find from eq.~(\ref{OrthoTangent}) for the zeroth component
\begin{equation}
{v^{\basind{0}}=e_\mu^{\;\,\basind{0}}\,v^\mu=0}\,.
\label{OrthoTangentProper}
\end{equation} 

These considerations enable us to rewrite the formal solution~(\ref{Geoexpansion}) of the geodesic equation in terms of proper reference frame coordinates
\begin{equation}
x^{\basind{\mu}}(\tau,v^{\basind{r}} s)=p^{\basind{\mu}}(\tau) +v^{\basind{\mu}}(\tau) \,s - \sum_{n=2}^\infty \Gamma^{\basind{\mu}}_{\;\basind{i_1}\ldots\basind{i_n}}(p^{\basind{\sigma}})\, v^{\basind{i_1}}(\tau) \cdot\ldots\cdot v^{\basind{i_n}}(\tau)\,\frac{s^n}{n!}\,.
\label{PropFormalSol}
\end{equation} 
As for Riemann normal coordinates, the exact solution of the geodesic eq.~(\ref{DefGeoProper2}) is already known by construction. Indeed, according to the definition~(\ref{DefProperRef}) of proper reference frame coordinates, it is given by the straight line
\begin{align}
x^{\basind{0}}(\tau,v^{\basind{r}} s)&=c\,\tau\,,\label{PropStraightLines}\\
x^{\basind{i}}(\tau,v^{\basind{r}} s)&=v^{\basind{i}}(\tau)\,s  \,.
\nonumber
\end{align} 
However, the formal solution~(\ref{PropFormalSol}) and the exact expression~(\ref{PropStraightLines}) coincide for all values of the curve parameter $s$ and all initial tangent vectors $v^{\basind{i}}$ only if
\begin{equation}
\Gamma^{\basind{\mu}}_{\;\{\basind{i_1}\ldots\basind{i_n}\}}(p^{\basind{\sigma}})=0 \quad \forall\;\;\tau\in\mathds{R}\quad \text{and}\quad \forall\;\;n\geq 2\,.
\label{PropZeroGammas}
\end{equation} 
Unlike the constraints~(\ref{ZeroGammas}) for Riemann normal coordinates, eq.~(\ref{PropZeroGammas}) provides just a necessary condition for the quantities ${\Gamma^{\basind{\mu}}_{\;\basind{i_1}\ldots\basind{i_n}}(p^{\basind{\sigma}})}$ with spatial lower indices. Moreover, expressions with zero indices, such as ${\Gamma^{\basind{\mu}}_{\;\basind{i_1}\basind{0}}(p^{\basind{\sigma}})}$, are not covered by eq.~(\ref{PropZeroGammas}) at all. Therefore, we cannot expect to establish {\it all}\;metric coefficients with the help of~(\ref{PropZeroGammas}), but we obtain at least the spatial coefficients 
${g_{\basind{j k},\basind{i_1}}(p^{{\basind{\sigma}}})}$ and 
${g_{\basind{j k},\{\basind{i_1},\basind{i_2}\}}(p^{{\basind{\sigma}}})}$.

\paragraph{First-order terms}

The first-order contributions can be derived without much effort by taking advantage of eq.~(\ref{PropZeroGammas}) for the special case $n=2$, which yields
\begin{equation}
\Gamma^{\basind{\mu}}_{\;\basind{i_1}\basind{i_2}}(p^{\basind{\sigma}})=0\,.
\label{ZeroChristoffelSpatial}
\end{equation} 
By using ${g_{\basind{j k};\basind{i_1}}=0}$ once more, we realize that these spatial contributions vanish,
\begin{equation}
g_{\basind{j k},\basind{i_1}}(p^{\basind{\sigma}})=
\Gamma^{\basind{\rho}}_{\;\basind{j}\basind{i_1}}(p^{\basind{\sigma}})\,
g_{\basind{\rho k}}(p^{\basind{\sigma}}) +
 \Gamma^{\basind{\rho}}_{\;\basind{k}\basind{i_1}}(p^{\basind{\sigma}})\,
g_{\basind{j \rho}}(p^{\basind{\sigma}})=0 \,,
\label{Properg1spatial}
\end{equation} 
as they did for Riemann normal coordinates.

\paragraph{Second-order terms}

The similarity between (\ref{ZeroGammas}) and (\ref{PropZeroGammas}) also plays a crucial role in the determination of the second-order coefficients 
${g_{\basind{j k},\{\basind{i_1},\basind{i_2}\}}(p^{\basind{\sigma}})}$. Indeed, when we express them in terms of the first partial derivatives of the Christoffel symbols -- in analogy to eq.~(\ref{RiemannZwischen0}) -- we find with eq.~(\ref{ZeroChristoffelSpatial}) the relation
\begin{equation*}
g_{\basind{j k}\{,\basind{i_1},\basind{i_2}\}}(p^{\basind{\sigma}})=
g_{\basind{\rho k}}(p^{\basind{\sigma}})\,
\Gamma^{\basind{\rho}}_{\;\basind{j}\{\basind{i_1},\basind{i_2}\}}(p^{\basind{\sigma}})+
g_{\basind{j \rho}}(p^{\basind{\sigma}})\,
\Gamma^{\basind{\rho}}_{\;\basind{k}\{\basind{i_1},\basind{i_2}\}}(p^{\basind{\rho}})\,.
\end{equation*} 
It is interesting to note, that we can formally establish this result by replacing the indices in the identity~(\ref{RiemannZwischen0}) according to ${(\mu)\rightarrow(j)}$, ${(\nu)\rightarrow(k)}$, ${(\alpha_1)\rightarrow(i_1)}$ and  ${(\alpha_2)\rightarrow(i_2)}$.
Moreover, we can apply this substitution law also to eqs.~(\ref{NormalRiemann1}), (\ref{ConditionGamma1}) and (\ref{ConditionGamma3}), since the underlying calculations are similar to the Riemann normal coordinate case. In particular, we obtain from eq.~(\ref{ConditionGamma5}) the expression
\begin{equation}
\Gamma^{\basind{\rho}}_{\;\basind{i_3}\{\basind{i_1},\basind{i_2}\}}(p^{\basind{\sigma}})=
\frac{1}{3}\,\symm{i_1 i_2}{ R^{\basind{\rho}}_{\;\,\basind{i_1}\basind{i_2}\basind{i_3}}(p^{\basind{\sigma}})}\,.
 \label{PropConditionGamma1}
\end{equation}
These considerations directly aim at the index substitution of eq.~(\ref{Riemanng2}), which results in the expression
\begin{equation}
g_{\basind{j k}\{,\basind{i_1},\basind{i_2}\}}(p^{\basind{\sigma}})=
\frac{2}{3}\,\symm{i_1 i_2}{
R_{\basind{j}\basind{i_1}\basind{i_2}\basind{k}}(p^{\basind{\sigma}})}\,.
\label{Properg2spatial}
\end{equation}
 for the spatial components of the second-order contribution.

\paragraph{Proper transport in local coordinates}

So far, we have provided all spatial components of the metric coefficients up to second order in proper reference frame coordinates. The starting point of their derivation was the formal solution of the geodesic equation~(\ref{PropFormalSol}) in combination with the exact solution~(\ref{PropStraightLines}). From a comparison of both expressions, we have obtained eq.~(\ref{PropZeroGammas}), which made the spatial components of the metric expansion available to us. Motivated by this idea, we start our search for the remaining coefficients by rewriting the proper transport, eq.~(\ref{ProperTransport}), in the local coordinates of the proper reference frame which yields
\begin{equation}
e^{\basind{\mu}}_{\;\basind{\underline{\alpha}};\basind{\nu}}\, u^{\basind{\nu}}=
-\Omega^{\basind{\mu}}_{\;\;\basind{\nu}}\,e^{\basind{\nu}}_{\;\,\basind{\underline{\alpha}}}
\label{ProperTransportProper}
\end{equation}
with
\begin{equation}
\Omega_{\basind{\mu \nu}}=
-\frac{1}{c^2}\left(a_{\basind{\mu}} u_{\basind{\nu}}
- a_{\basind{\nu}} u_{\basind{\mu}} \right)
+\frac{1}{c}\,u^{\basind{\rho}}\, \omega^{\basind{\sigma}} \,\varepsilon_{\basind{\rho\sigma\mu\nu}}\,.
\label{ProperTransportProperb}
\end{equation} 
At this point, we unfortunately encounter conceptional problems in our notation which reveal themselves in the additionally underlined lower tetrad indices. The reason for this problem are the different interpretations associated with the superscript and subscript of the tetrad $e^{\basind{\mu}}_{\;\basind{\underline{\alpha}}}$. It is important to bear in mind that the superscript $(\mu)$ denotes a tensor index, whereas the subscript $(\underline{\alpha})$ just labels the individual tetrad vectors, as discussed in appendix~\ref{AppTetrad}. This distinction is especially important for the covariant derivative
\begin{equation*}
e^{\basind{\mu}}_{\;\basind{\underline{\alpha}};\basind{\nu}}
=e^{\basind{\mu}}_{\;\basind{\underline{\alpha}},\basind{\nu}}
+\Gamma^{\basind{\mu}}_{\;\,\basind{\rho}\basind{\nu}}\,
e^{\basind{\rho}}_{\;\basind{\underline{\alpha}}}\,,
\end{equation*}
where the only relevant index is $(\mu)$, and not the subscript~$(\underline{\alpha})$. For this purpose, we underline the indices which should be ignored when taking covariant derivatives. However, we will silently drop this notation as soon as the distinction between tensor and tetrad indices is of no further importance in the calculations.

Coming back to eq.~(\ref{ProperTransportProper}), we observe that with the inverse of the transformation matrix (\ref{PropTrafoMatrix}), the tetrad components read in local coordinates along $p^{\basind{\sigma}}(\tau)$
\begin{equation}
e^{\basind{\mu}}_{\;\,\basind{\underline{\alpha}}}(\tau)= \pdiff{x^{\basind{\mu}}}{x^\nu\phantom{^x}}\Big|_{p^{\basind{\sigma}}(\tau)}\; 
e^{\nu}_{\;\,\basind{\underline{\alpha}}}(\tau) 
= e^{\;\,\basind{\mu}}_{\nu}(\tau)\; e^\nu_{\;\basind{\underline{\alpha}}}(\tau)
=\delta^{\basind{\mu}}_{\basind{\alpha}}\,.
\label{ProperTetrad}
\end{equation} 
This result demonstrates that the orthogonal tetrad vectors 
$e^{\basind{\mu}}_{\;\,\basind{\underline{\alpha}}}(\tau)$ remain fixed along their directions, when expressed in the adapted coordinates of the proper reference frame. 

When we insert the last equation into the left hand side of the proper transport law (\ref{ProperTransportProper}), we find with ${u^{\basind{\nu}}=c\, e^{\basind{\nu}}_{\;\,\basind{\underline{0}}}=c\, \delta^{\basind{\nu}}_{\basind{0}}}$ the relation
\begin{equation*}
e^{\basind{\mu}}_{\;\,\basind{\underline{\alpha}};\basind{\nu}}\, u^{\basind{\nu}}\big|_{p^{\basind{\sigma}}}=
\diff{}{\tau}e^{\basind{\mu}}_{\;\,\basind{\underline{\alpha}}}+
\Gamma^{\basind{\mu}}_{\;\basind{\rho}\basind{\nu}}(p^{\basind{\sigma}})\, e^{\basind{\rho}}_{\;\,\basind{\underline{\alpha}}}(\tau)\,u^{\basind{\nu}}(\tau)=
c\,\Gamma^{\basind{\mu}}_{\;\basind{\alpha}\basind{0}}(p^{\basind{\sigma}})\,.
\end{equation*} 
Substitution of eq.~(\ref{ProperTetrad}) into eq.~(\ref{ProperTransportProper}) thus provides the expression 
\begin{equation}
\Gamma^{\basind{\mu}}_{\;\basind{\alpha}\basind{0}}(p^{\basind{\sigma}})
=-\frac{1}{c}\,\Omega^{\basind{\mu}}_{\;\;\basind{\alpha}}(\tau)
\label{GammaOmega}
\end{equation}
for the missing components of the Christoffel symbols along $p^{\basind{\sigma}}(\tau)$. 
It is possible to combine the results~(\ref{ZeroChristoffelSpatial}) and (\ref{GammaOmega}) in a single equation by taking advantage of the identity
\begin{equation*}
\Gamma^{\basind{\rho}}_{\;\basind{\mu}\basind{\nu}}(p^{\basind{\sigma}})=
\eta_{\basind{0\mu}}\,\Gamma^{\basind{\rho}}_{\;\basind{0}\basind{\nu}}(p^{\basind{\sigma}})+
\eta_{\basind{0\nu}}\,\Gamma^{\basind{\rho}}_{\;\basind{\mu}\basind{0}}(p^{\basind{\sigma}})-
\eta_{\basind{0\mu}}\,\eta_{\basind{0\nu}}\,\Gamma^{\basind{\rho}}_{\;\basind{0}\basind{0}}(p^{\basind{\sigma}})\,.
\end{equation*} 
Substitution of eq.~(\ref{GammaOmega}) into the last equation thus yields
\begin{equation}
\Gamma^{\basind{\rho}}_{\;\basind{\mu}\basind{\nu}}(p^{\basind{\sigma}})=-\frac{1}{c}\left(
\eta_{\basind{0\mu}}\,\Omega^{\basind{\rho}}_{\;\;\basind{\nu}}+
\eta_{\basind{0\nu}}\,\Omega^{\basind{\rho}}_{\;\;\basind{\mu}}-
\eta_{\basind{0\mu}}\,\eta_{\basind{0\nu}}\,\Omega^{\basind{\rho}}_{\;\;\basind{0}}
\right)\,.
\label{GammaAtP}
\end{equation}

\paragraph{Remaining first-order terms}

Being equipped with this identity, it is not difficult to find the remaining first-order contributions to the metric expansion. Once more we make use of the vanishing covariant derivative of the metric, $g_{\basind{0\nu};\basind{i_1}}=0$, which then leads us to the analog of eq.~(\ref{Properg1spatial}), that is
\begin{equation*}
g_{\basind{0\nu},\basind{i_1}}(p^{\basind{\sigma}})=
\Gamma^{\basind{\rho}}_{\;\basind{0}\basind{i_1}}(p^{\basind{\sigma}})\,
g_{\basind{\rho \nu}}(p^{\basind{\sigma}}) +
 \Gamma^{\basind{\rho}}_{\;\basind{\nu}\basind{i_1}}(p^{\basind{\sigma}})\,
g_{\basind{0 \rho}}(p^{\basind{\sigma}})\,.
\end{equation*} 
By inserting the zeroth-order term (\ref{Properg0}) and the Christoffel symbols~(\ref{GammaAtP}) along the world line, we arrive at the remaining first-order terms
\begin{equation}
g_{\basind{0\nu},\basind{i_1}}(p^{\basind{\sigma}})=
-\frac{1}{c}\left(\eta_{\basind{\nu\rho}}+\eta_{\basind{0\nu}}\eta_{\basind{0\rho}}\right)
\Omega^{\basind{\rho}}_{\;\;\basind{i_1}}\,.
\label{g1SpaceAndTime}
\end{equation} 
Again, it is possible to combine this expression with the spatial components of the first-order term, eq.~(\ref{Properg1spatial}). In analogy to the treatment of the Christoffel symbols given above, we make use of the relation
\begin{equation*}
g_{\basind{\mu\nu},\basind{i_1}}(p^{\basind{\sigma}})=
\eta_{\basind{0\mu}}\,g_{\basind{0\nu},\basind{i_1}}(p^{\basind{\sigma}})+
\eta_{\basind{0\nu}}\,g_{\basind{\mu 0},\basind{i_1}}(p^{\basind{\sigma}})-
\eta_{\basind{0\mu}}\,\eta_{\basind{0\nu}}\,g_{\basind{0 0},\basind{i_1}}(p^{\basind{\sigma}})
\end{equation*} 
in order to establish the general expression
\begin{equation}
g_{\basind{\mu\nu},\basind{i_1}}(p^{\basind{\sigma}})=
-\frac{1}{c}\left(\eta_{\basind{0\mu}}\,\Omega_{\basind{\nu i_1}}+\eta_{\basind{0\nu}}\,
\Omega_{\basind{\mu i_1}}\right)
\label{Properg1}
\end{equation} 
for the first-order contribution by making use of eq.~(\ref{g1SpaceAndTime}).

\paragraph{Remaining second-order terms}

Hence, we are left with the determination of the remaining second-order contributions 
${g_{\basind{0 \nu}\,\{,\basind{i_1},\basind{i_2}\}}(p^{{\basind{\sigma}}})}$, which we now express in terms of the first derivatives of the Christoffel symbols. For this purpose, we translate the arguments associated with eq.~(\ref{Zwischenrechnungg2}) to the present case, and obtain
\begin{align}
\label{ZwischenPropg2} 
g_{\basind{0 \nu}\{,\basind{i_1},\basind{i_2}\}}(p^{\basind{\sigma}})=&\;
\Gamma^{\basind{\rho}}_{\;\,\basind{0}\{\basind{i_1},\basind{i_2}\}}(p^{\basind{\sigma}})\,
\eta_{\basind{\rho\nu}} +
\Gamma^{\basind{\rho}}_{\;\,\basind{\nu}\{\basind{i_1},\basind{i_2}\}}(p^{\basind{\sigma}})\,
\eta_{\basind{0\rho}}  \\ \nonumber 
&+ S_{\basind{\nu}\basind{i_1 i_2}}  
\end{align}
where we have introduced the abbreviation 
\begin{align}
S_{\basind{\nu}\basind{i_1 i_2}}\equiv\symm{i_1 i_2}{
\Gamma^{\basind{\rho}}_{\;\,\basind{0}\basind{i_1}}(p^{\basind{\sigma}})\,
g_{\basind{\rho\nu},\basind{i_2}}(p^{\basind{\sigma}})+
\Gamma^{\basind{\rho}}_{\;\,\basind{\nu}\basind{i_1}}(p^{\basind{\sigma}})\,
g_{\basind{0\rho},\basind{i_2}}(p^{\basind{\sigma}})
}
\,.
\nonumber
\end{align} 
Now it is most convenient to treat the different terms which appear on the right hand side separately. We start with the term $S_{\basind{\nu}\basind{i_1 i_2}}$ since it contains only quantities that have been worked out so far. Inserting eqs.~(\ref{GammaAtP}) and~(\ref{Properg1}) into it, a straight forward calculation shows that 
\begin{align}
S_{\basind{\nu}\basind{i_1 i_2}} =\frac{1}{c^2}\,\symm{i_1 i_2}{\Omega_{\basind{0 i_1}}\Omega_{\basind{\nu i_2}}
+\eta_{\basind{0 \nu}}\left(\Omega_{\basind{0 i_1}}\Omega_{\basind{0 i_2}}
+2\,\Omega^{\basind{\rho}}_{\;\;\basind{i_1}}\Omega_{\basind{\rho\, i_2}}\right)}\,.
\label{Zwischen2Propg2}
\end{align} 

Next, we consider the first term on the right hand side of eq.~(\ref{ZwischenPropg2}), which can be found by the following argument: since the spatial components of the Christoffel symbols vanish for all $p^{\basind{\sigma}}(\tau)$, eq.~(\ref{ZeroChristoffelSpatial}), their first partial derivative with respect to $x^{\basind{0}}=c\,\tau$ does also vanish, that is
\begin{equation}
 \Gamma^{\basind{\rho}}_{\;\basind{i_1}\basind{i_2},\basind{0}}(p^{\basind{\sigma}})=0\,.
\label{Zwischen3Propg2}
\end{equation}
Moreover, recalling the definition~(\ref{DefCurvatureTensor}) of the curvature tensor, we obtain
\begin{equation*}
 R^{\basind{\rho}}_{\;\,\basind{i_1}\basind{i_2}\basind{0}}(p^{\basind{\sigma}})=\left[
 \Gamma^{\basind{\rho}}_{\;\basind{i_1}\basind{0},\basind{i_2}}-
 \Gamma^{\basind{\rho}}_{\;\basind{i_1}\basind{i_2},\basind{0}}+
 \Gamma^{\basind{\rho}}_{\;\basind{\mu}\basind{i_2}}
 \Gamma^{\basind{\mu}}_{\;\basind{i_1}\basind{0}}-
 \Gamma^{\basind{\rho}}_{\;\basind{\mu}\basind{0}}
 \Gamma^{\basind{\mu}}_{\;\basind{i_1}\basind{i_2}}
\right]\left|_{p^{\basind{\sigma}}}\right.
\,.
\end{equation*}
We solve this last equation for the unknown term $\Gamma^{\basind{\rho}}_{\;\basind{i_1}\basind{0},\basind{i_2}}$ and arrive with the help of eqs.~(\ref{GammaAtP}) and~(\ref{Zwischen3Propg2}) after some algebra at
\begin{equation*}
\Gamma^{\basind{\rho}}_{\;\basind{i_1}\basind{0},\basind{i_2}}(p^{\basind{\sigma}})=
 R^{\basind{\rho}}_{\;\,\basind{i_1}\basind{i_2}\basind{0}}(p^{\basind{\sigma}})-\frac{1}{c^2}\,
 \Omega_{\basind{0 i_1}}\Omega^{\basind{\rho}}_{\;\;\basind{i_2}}\,.
\end{equation*}
Symmetrization with respect to the indices $(i_1)$ and $(i_2)$ of the last expression thus leads us to the first term on the right hand side of eq.~(\ref{ZwischenPropg2})
\begin{equation}
 \Gamma^{\basind{\rho}}_{\;\,\basind{0}\{\basind{i_1},\basind{i_2}\}}(p^{\basind{\sigma}})\,
 \eta_{\basind{\rho\nu}}=
 \symm{i_1 i_2}{R_{\basind{\nu}\basind{i_1}\basind{i_2}\basind{0}}(p^{\basind{\sigma}})-\frac{1}{c^2}\,
\Omega_{\basind{0 i_1}}\Omega_{\basind{\nu i_2}}}\,.
\label{Zwischen4Propg2}
\end{equation}

Finally, there is only one term left to be found, namely the second contribution on the right hand side of eq.~(\ref{ZwischenPropg2}). It is most suitable here to consider the cases ${(\nu)=(0)}$ and 
${(\nu)=(k)}$ separately. For ${(\nu)=(0)}$ we encounter the pleasant situation, that the missing term follows as a special case from eq.~(\ref{Zwischen4Propg2}) leading us to the expression
\begin{equation}
\Gamma^{\basind{\rho}}_{\;\,\basind{0}\{\basind{i_1},\basind{i_2}\}}(p^{\basind{\sigma}})\,
\eta_{\basind{0\rho}}=
\symm{i_1 i_2}{R_{\basind{0}\basind{i_1}\basind{i_2}\basind{0}}(p^{\basind{\sigma}})-\frac{1}{c^2}\,
\Omega_{\basind{0 i_1}}\Omega_{\basind{0 i_2}}}\,.
\label{Zwischen5PropgNu0}
\end{equation} 
For $(\nu)=(k)$ we can take advantage of the identity~(\ref{PropConditionGamma1}) to arrive at
\begin{equation}
\Gamma^{\basind{\rho}}_{\;\,\basind{k}\{\basind{i_1},\basind{i_2}\}}(p^{\basind{\sigma}})\,
\eta_{\basind{0\rho}}=\frac{1}{3}
\,\symm{i_1 i_2}{ R_{\basind{0}\basind{i_1}\basind{i_2}\basind{k}}(p^{\basind{\sigma}})}\,.
\label{Zwischen5PropgNuk}
\end{equation}

With these results at hand, we are finally able to obtain the remaining second-order contributions. Combining eqs.~(\ref{Zwischen2Propg2}), (\ref{Zwischen4Propg2}), and (\ref{Zwischen5PropgNu0}) for ${(\nu)=(0)}$ yields
\begin{equation}
g_{\basind{0 0}\{,\basind{i_1},\basind{i_2}\}}(p^{\basind{\sigma}})=2\,\symm{i_1 i_2}{
R_{\basind{0}\basind{i_1}\basind{i_2}\basind{0}}(p^{\basind{\sigma}})+\frac{1}{c^2}\,
\Omega^{\basind{\rho}}_{\;\;\basind{i_1}}\Omega_{\basind{\rho\, i_2}}
}\,.
\label{Properg2time}
\end{equation} 
In accordance, we add up eqs.~(\ref{Zwischen2Propg2}), (\ref{Zwischen4Propg2}), and (\ref{Zwischen5PropgNuk}) for ${(\nu)=(k)}$, and use the symmetry properties of the curvature tensor, eq.~(\ref{RiemannTensorCovSymm}), to obtain
\begin{equation}
g_{\basind{0 k}\{,\basind{i_1},\basind{i_2}\}}(p^{\basind{\sigma}})=\frac{4}{3}\,\symm{i_1 i_2}{
R_{\basind{0}\basind{i_1}\basind{i_2}\basind{k}}(p^{\basind{\sigma}})
}\,.
\label{Properg2mixed}
\end{equation}

\paragraph{Summary}

Thus, we have found all coefficients of the metric expansion up to second \nobreak{order}, and again, it is most convenient to summarize them separately for the metric coefficients 
$g_{\basind{0 0}}(x^{\basind{\sigma}})$, $g_{\basind{0 k}}(x^{\basind{\sigma}})$ and 
$g_{\basind{j k}}(x^{\basind{\sigma}})$. Inserting the results~(\ref{Properg0}), \eqref{Properg1spatial}, (\ref{Properg2spatial}), (\ref{Properg1}), (\ref{Properg2time}) and (\ref{Properg2mixed}) into the power-series~(\ref{ExpansionMetricProper}), we obtain the metric coefficients 
\begin{align}
\label{ProperMetricExpansionOmega}
g_{\basind{0 0}}(x^{\basind{\sigma}})=&\;1
-\frac{2}{c}\,\Omega_{\basind{0 i_1}} x^{\basind{i_1}}
+\bigg(R_{\basind{0}\basind{i_1}\basind{i_2}\basind{0}}(p^{\basind{\sigma}})\bigg.\\  \nonumber
&\left.+\frac{1}{c^2}\,
\Omega^{\basind{\rho}}_{\;\;\basind{i_1}}\Omega_{\basind{\rho\, i_2}}\right) x^{\basind{i_1}} x^{\basind{i_2}}  
+\mathcal{O}(x^3)\,,\\
g_{\basind{0 k}}(x^{\basind{\sigma}})=&
-\frac{1}{c}\,\Omega_{\basind{k i_1}} x^{\basind{i_1}} + \frac{2}{3}\,
R_{\basind{0}\basind{i_1}\basind{i_2}\basind{k}}(p^{\basind{\sigma}})\, x^{\basind{i_1}} x^{\basind{i_2}}
+\mathcal{O}(x^3)\,,
\nonumber\\
g_{\basind{j k}}(x^{\basind{\sigma}})=&-\delta_{\basind{j k}}
+\frac{1}{3}\,R_{\basind{j}\basind{i_1}\basind{i_2}\basind{k}}(p^{\basind{\sigma}})\, x^{\basind{i_1}} x^{\basind{i_2}}
+\mathcal{O}(x^3)
\nonumber
\end{align} 
in proper reference frame coordinates.

We close this appendix with an explicit expression for the transport matrix $\Omega_{\basind{\mu \nu}}(\tau)$, eq.~(\ref{ProperTransportProperb}), in proper reference frame coordinates. According to the transformation matrix~(\ref{PropTrafoMatrix}), we obtain for the components of the four-velocity $u_{\basind{\mu}}(\tau)$ and the four-acceleration $a_{\basind{\mu}}(\tau)$ the following constraints 
\begin{equation*}
u^{\basind{\mu}}=c\,e^{\basind{\mu}}_{\;\,\basind{\underline{0}}}=c\,\delta^{\basind{\mu}}_{\basind{0}}
\quad\text{and}\quad 
a_{\basind{0}}=e^\mu_{\;\basind{0}}\,a_\mu=\frac{1}{c}\,u^\mu a_\mu=0\,,
\end{equation*}
where we made use of eq.~(\ref{constraintonaccel}) in the last step of $a_{\basind{0}}$.

Moreover, we know from appendix~\ref{AppTetradTransport} and in particular from eq.~(\ref{TetradRotationVector}) that the four-vector $\omega^{\basind{\mu}}(\tau)$, which characterizes the rotation of the tetrad along the world line, has a vanishing zeroth component in proper reference frame coordinates, that is $\omega^{\basind{0}}=0$.
Thus, we end up with the matrix
\begin{equation*}
 \Omega_{\basind{\mu \nu}}=
-\frac{1}{c}\left(a_{\basind{\mu}}\eta_{\basind{\nu 0}}-
a_{\basind{\nu}}\eta_{\basind{\mu 0}}\right)+\varepsilon_{\basind{0 p \mu \nu}}\,
\omega^{\basind{p}}\,,
\end{equation*}
in analogy to the tetrad form of the transport matrix, eq.~(\ref{ProperTransportTetrad2}). With this explicit expression at hand, we can now show without much effort that
\begin{equation*}
 \Omega^{\basind{\rho}}_{\;\;\basind{i_1}}\Omega_{\basind{\rho\, i_2}}=\frac{1}{c^2}\,
 a_{\basind{i_1}} a_{\basind{i_2}}+\omega_{\basind{i_1}} \omega_{\basind{i_2}}
 -\omega^{\basind{l}}\omega_{\basind{l}}\,\eta_{\basind{i_1 i_2}}\,.
\end{equation*}
Insertion of these last two results in the power-series~(\ref{ProperMetricExpansionOmega}) finally yields the individual coefficients, eq.~(\ref{ProperMetricExpansion}), for the metric expansion in proper reference frame coordinates.

\subsubsection{Approximate solution of the geodesic equation\label{AppRadarDistance}}

We now briefly sketch how to arrive at an approximate formula for arbitrary geodesics in the spatial neighborhood of the observer's world line $p^{\basind{\mu}}(\tau)=(c\tau,0,0,0)$. We thereby take advantage of the derivation of the metric expansion up to second order presented in the preceding subsect.~\ref{MetricExpandPropRef}. Our construction of the approximate geodesics will have the same level of accuracy as the metric expansion, eq.~(\ref{ProperMetricExpansionOmega}).

We start by recalling the geodesic equation
\begin{equation}
\diff{^2x^{\basind{\mu}}}{ \lambda^2}
 +\Gamma^{\basind{\mu}}_{\;\basind{\alpha}\basind{\beta}}(x^{\basind{\sigma}})\diff{x^{\basind{\alpha}}}{\lambda} \diff{x^{\basind{\beta}}}{\lambda}=0
\label{DefGeoProper3}
\end{equation}
in proper reference frame coordinates.
Again, we suppose that the geodesics start from the world line of the observer, but now with an arbitrarily directed initial tangent vector, giving rise to the initial conditions
\begin{equation*}
x^{\basind{\mu}}(\tau,0)=p^{\basind{\mu}}(\tau)=(c\tau,0,0,0) \quad\text{and}\quad 
\diff{}{\lambda} x^{\basind{\mu}}(\tau,0) =v^{\basind{\mu}}(\tau)\,.
\end{equation*}
In contrast to the spacelike geodesics used in subsect.~\ref{MetricExpandPropRef} which satisfy the condition~(\ref{OrthoTangentProper}), the geodesics considered in the present discussion do not have any restrictions concerning the directions of their initial tangents. Here, we simply require
\begin{equation*} {(v^{\basind{0}}(\tau))^2+(v^{\basind{1}}(\tau))^2+(v^{\basind{2}}(\tau))^2+(v^{\basind{3}}(\tau))^2=1}\,.
\end{equation*}
Hence, we allow for timelike, spacelike and null geodesics.

Next, we recall that the derivation of the metric coefficients up to second order, eq.~\eqref{ProperMetricExpansionOmega}, was based on the formal solution of the geodesic equation~(\ref{PropFormalSol}) and the resulting condition~(\ref{PropZeroGammas}) up to third order $n=3$. In this spirit, we use as ansatz for the approximate solution of the geodesic equation~(\ref{DefGeoProper3}) a truncated expression of the formal solution, eq.~\eqref{Geoexpansion}. Thus, we start from the approximate formula
\begin{equation}
x^{\basind{\mu}}(\tau,v^{\basind{\sigma}} \lambda)=p^{\basind{\mu}} +v^{\basind{\mu}} \,\lambda - \sum_{n=2}^3 \Gamma^{\basind{\mu}}_{\;\{\basind{\alpha_1}\ldots\basind{\alpha_n}\}}(p^{\basind{\sigma}})\; v^{\basind{\alpha_1}} \cdot\ldots\cdot v^{\basind{\alpha_n}}\,\frac{\lambda^n}{n!}+\mathcal{O}(\lambda^4)
\label{PropFormalSol2}
\end{equation} 
for the geodesics in the spatial neighborhood around the world line $p^{\basind{\mu}}(\tau)$ and express the coefficients $\Gamma^{\basind{\mu}}_{\;\basind{\alpha_1}\basind{\alpha_2}}(p^{\basind{\sigma}})$ and $\Gamma^{\basind{\mu}}_{\;\{\basind{\alpha_1}\basind{\alpha_2}\basind{\alpha_3}\}}(p^{\basind{\sigma}})$ in terms of the transport matrix $\Omega_{\basind{\alpha\beta}}$ and of the curvature tensor, in analogy to the metric expansion~(\ref{ProperMetricExpansionOmega}).
Fortunately, we have already found some of the coefficients in subsect.~\ref{MetricExpandPropRef}. According to eq.~\eqref{GammaAtP}, the second-order terms are completely determined by
\begin{equation}
\Gamma^{\basind{\mu}}_{\;\basind{\alpha_1}\basind{\alpha_2}}(p^{\basind{\sigma}})=
-\frac{1}{c}\left(\eta_{\basind{0\alpha_1}}\,\Omega^{\basind{\mu}}_{\;\;\basind{\alpha_2}}+
\eta_{\basind{0\alpha_2}}\,\Omega^{\basind{\mu}}_{\;\;\basind{\alpha_1}}-
\eta_{\basind{0\alpha_1}}\,\eta_{\basind{0\alpha_2}}\,\Omega^{\basind{\mu}}_{\;\;\basind{0}}\right)\,.
\label{GammaAtP2} 
\end{equation}
Moreover, the purely spatial components of the third-order coefficients $\Gamma^{\basind{\mu}}_{\;\{\basind{i_1}\basind{i_2}\basind{i_3}\}}(p^{\basind{\sigma}})$ 
vanish due to eq.~(\ref{PropZeroGammas}). All third-order coefficients which at least possess one zero-index, namely $\Gamma^{\basind{\mu}}_{\;\{\basind{0}\basind{i_1}\basind{i_2}\}}(p^{\basind{\sigma}})$, $\Gamma^{\basind{\mu}}_{\;\{\basind{0}\basind{0}\basind{i_1}\}}(p^{\basind{\sigma}})$ and $\Gamma^{\basind{\mu}}_{\;\basind{0}\basind{0}\basind{0}}(p^{\basind{\sigma}})$ can be obtained by applying the same techniques that were used in subsect.~\ref{MetricExpandPropRef}. 

With the help of the identity
\begin{equation*}
 \Gamma^{\basind{\mu}}_{\;\basind{\alpha_1}\basind{\alpha_2}\basind{\alpha_3}}=
 \Gamma^{\basind{\mu}}_{\;\basind{\alpha_1}\basind{\alpha_2},\basind{\alpha_3}}-
 \Gamma^{\basind{\nu}}_{\;\basind{\alpha_1}\basind{\alpha_3}}
 \Gamma^{\basind{\mu}}_{\;\basind{\nu}\basind{\alpha_2}}-
\Gamma^{\basind{\nu}}_{\;\basind{\alpha_2}\basind{\alpha_3}}
 \Gamma^{\basind{\mu}}_{\;\basind{\alpha_1}\basind{\nu}}
\,,
\end{equation*}
which follows from the definition~(\ref{Christoffeln}) for $n=3$, we can express the third-order coefficients $\Gamma^{\basind{\mu}}_{\;\{\basind{\alpha_1}\basind{\alpha_2}\basind{\alpha_3}\}}$ in terms of the first partial derivatives and products of Christoffel symbols. In particular, eqs.~\eqref{PropZeroGammas}, \eqref{PropConditionGamma1}, \eqref{GammaAtP}, \eqref{Zwischen3Propg2} and~\eqref{Zwischen4Propg2}, allow for the substitution of the Christoffel symbols as well as their first derivatives by the transport matrix and by the curvature tensor. We refrain from presenting the details of this calculation and just provide the final expressions
\begin{align}
\Gamma^{\basind{\mu}}_{\;\{\basind{i_1}\basind{i_2}\basind{i_3}\}}(p^{\basind{\sigma}})&=\;0\,,
\label{GammaAtP3} \\
\Gamma^{\basind{\mu}}_{\;\{\basind{0}\basind{i_1}\basind{i_2}\}}(p^{\basind{\sigma}})&=\;
\frac{2}{3}\;R^{\basind{\mu}}_{\;\,\lbrace\basind{i_1}\basind{i_2}\rbrace\basind{0}}(p^{\basind{\sigma}}) -\frac{2}{c^2}\;\symm{i_1 i_2}{\Omega_{\basind{0 i_1}}(\tau)\,
\Omega^{\basind{\mu}}_{\;\;\basind{i_2}}(\tau)} \,,\nonumber\\
\Gamma^{\basind{\mu}}_{\;\{\basind{0}\basind{0}\basind{i_1}\}}(p^{\basind{\sigma}})&=\;
\frac{1}{3}\;R^{\basind{\mu}}_{\;\,\basind{0}\basind{i_1}\basind{0}}(p^{\basind{\sigma}}) -\frac{1}{c}\;\Omega^{\basind{\mu}}_{\;\;\basind{i_1},\basind{0}}(\tau) 
-\frac{1}{c^2}\;\Omega^{\basind{\mu}}_{\;\;\basind{\rho}}(\tau)\,\Omega^{\basind{\rho}}_{\;\;\basind{i_1}}(\tau)
,\nonumber\\
\Gamma^{\basind{\mu}}_{\;\{\basind{0}\basind{0}\basind{0}\}}(p^{\basind{\sigma}})&=
-\frac{1}{c}\;\Omega^{\basind{\mu}}_{\;\;\basind{0},\basind{0}}(\tau) 
-\frac{2}{c^2}\;\Omega^{\basind{\mu}}_{\;\;\basind{\rho}}(\tau)\,\Omega^{\basind{\rho}}_{\;\;\basind{0}}(\tau)
\nonumber
\end{align}
for the individual third-order coefficients.

The formal solution of the geodesic equation up to third order in the curve parameter is now simply established by substitution of the second- and third-order coefficients, eqs.~(\ref{GammaAtP2}) and ~(\ref{GammaAtP3}) into eq.~(\ref{PropFormalSol2}).

We emphasize, that this formal solution offers the possibility for an operational definition of proper reference frame coordinates, at least up to a certain level of accuracy. The basic idea is to take advantage of the radar method for light rays in order to explore the spatial neighborhood around the world line of the observer. For this purpose, the formal solution, eq.~\eqref{PropFormalSol2}, can be utilized to provide the connection between proper reference frame and {\it radar coordinates} \cite{Perlick01}.

Moreover, when we take advantage of the approximate solution, eq.~\eqref{PropFormalSol2}, for the local description of null geodesics, it should be possible to independently verify Pirani's method \cite{Laemmerzahl01,Synge,Pirani65}, which uses
the ``bouncing photon'' as a measure for the rotation of a reference frame relative to the inertial compass.


\section{Expansion of the Sagnac time delay in proper reference frame coordinates \label{CalcInPropRef}}

In the present appendix we derive a series expansion of the Sagnac time delay. The expansion is performed in orders of the moments of ``unit fluxes'' through the area $\mathcal{A}$, which encloses the spatial curve $\mathcal{S}$. We close this appendix by providing the two dominant contributions in the Sagnac time delay.

\subsection{General expansion}

As mentioned in sect.~\ref{ProperRefSagnac}, the proper reference frame coordinates are established by a slowly accelerating and rotating observer who moves along the world line ${p^{\basind{\sigma}}(x^{\basind{0}})=(x^{\basind{0}},0,0,0)}$ in such a way, that the metric coefficients can be considered as quasi-stationary for the time it takes to perform the Sagnac interferometer experiment. A second observer, who measures the proper time difference $\Delta\tau_{Sq}$, is supposed to be located at the fixed spatial point $q^{\basind{i}}$ on the closed curve $\mathcal{S}$ in proper reference frame coordinates. His world line reads ${q^{\basind{\sigma}}(x^{\basind{0}})=(x^{\basind{0}},q^{\basind{1}},q^{\basind{2}},q^{\basind{3}})}$.

We start by recalling the expression
\begin{equation}
\Delta\tau_S=-\frac{2}{c}\,\sqrt{g_{\basind{00}}(q^{\basind{r}})}
\oint\limits_\mathcal{S}\frac{g_{\basind{0k}}}{g_{\basind{00}}}\,\D s^{\basind{k}}
\label{GenSagnacTimeDelay2}
\end{equation} 
for the Sagnac time delay derived in \ref{SecDerivationTimeDelay}. 

Next we represent the integrand of the integral in eq.~(\ref{GenSagnacTimeDelay2}) by the power-series expansion
\begin{equation}
\frac{g_{\basind{0 k}}(x^{\basind{r}})}{g_{\basind{00}}(x^{\basind{r}})}
=\sum_{n=0}^\infty \,C_{\basind{k}\,\basind{i_1\ldots i_n}}\, x^{\basind{i_1}}\cdot\ldots\cdot x^{\basind{i_n}}
\label{DefCExpansion}
\end{equation}
valid in a sufficiently small spatial neighborhood around the world line $p^{\basind{\sigma}}(x^{\basind{0}})$. Here we have introduced the constant coefficients
\begin{equation}
C_{\basind{k}\,\basind{i_1\ldots i_n}}\equiv\frac{1}{n!}\,\pdiff{^n}{x^{\basind{i_1}}
\ldots \dd x^{\basind{i_n}}} \left(\frac{g_{\basind{0 k}}}{g_{\basind{00}}}\right)
\bigg|_{p^{\basind{r}}} \,.
\label{DefCCoefficients}
\end{equation}
We emphasize that in this expression the case $n=0$ does not involve any derivatives. Moreover, we note that the quantities ${C_{\basind{k}\,\basind{i_1\ldots i_n}}}$ are by construction totally symmetric in their indices 
${(i_1 \ldots i_n)}$.

When we insert the power-series expansion~(\ref{DefCExpansion}) into the Sagnac time delay~(\ref{GenSagnacTimeDelay2}), we obtain 
\begin{equation}
\Delta\tau_{Sq}=-\frac{2}{c}\,\sqrt{g_{\basind{00}}(q^{\basind{r}})}
\sum_{n=0}^\infty \oint\limits_\mathcal{S} C_{\basind{k}\,\basind{i_1\ldots i_n}} \,
s^{\basind{i_1}}\cdot\ldots\cdot s^{\basind{i_n}}\,\D s^{\basind{k}} 
\,.
\label{SagnacTimeDelayCalc1}
\end{equation} 
As shown in fig.~\ref{SagnacTubeRestFrame}, the line integral is evaluated along the closed spatial curve $\mathcal{S}$ with positive orientation according to the coordinate representation $s^{\basind{i}}(\phi)$.

\subsection{Stokes' theorem} 

\begin{figure}
\centering
\includegraphics[height=0.2\textwidth]{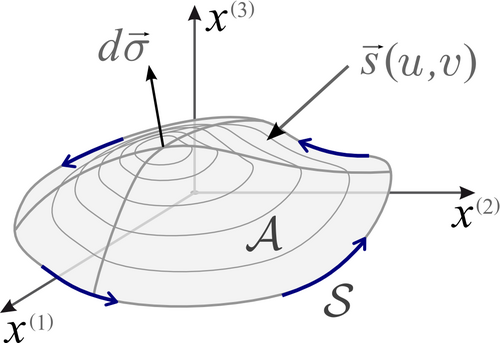}
\caption{Stokes' theorem in Cartesian coordinates. The parameterization $\mathbf{s}(u,v)$ of the surface $\mathcal{A}$ with positively oriented boundary $\mathcal{S}$ has to satisfy the right-hand  rule in the definition of the infinitesimal surface normal $d\boldsymbol{\sigma}$.}
\label{FigStokes}
\end{figure}

Since the spatial coordinates $x^{\basind{i}}$ of the proper reference frame coordinates are defined in the three-dimensional, spacelike subspace of the tangent space at $p^{\basind{\sigma}}(\tau)$ which is orthogonal to the four-velocity $u^{\basind{\sigma}}(\tau)$, we can take advantage of Stokes' theorem in its familiar formulation in Cartesian coordinates\footnote{In the literature one frequently finds a general formulation of Stokes' theorem in terms of differential forms. However, for our purpose the three-dimensional formulation in terms of Cartesian coordinates is sufficient. Throughout these lecture notes, Stokes' theorem refers to the familiar three-dimensional version.}. For a three-dimensional, continuously differentiable vector field $\mathbf{f}(\mathbf{s})$ it reads
\begin{equation*}
\oint\limits_\mathcal{S} \mathbf{f}\,\D \mathbf{s}
=\iint\limits_{\mathcal{A}} \op{rot}\,\mathbf{f}\; \D \boldsymbol{\sigma}\,,\quad\text{with}\quad 
\D \boldsymbol{\sigma}=\left(\pdiff{\mathbf{s}}{u}\times
\pdiff{\mathbf{s}}{v}\right)\,\D u\,\D v\,.
\end{equation*} 
The positively oriented curve $\mathcal{S}$ with parameterization $\mathbf{s}(\phi)$ encloses the surface $\mathcal{A}$ whose parameterization $\mathbf{s}(u,v)$ defines the infinitesimal surface normal $d\boldsymbol{\sigma}$ according to the right-hand rule as illustrated in fig.~\ref{FigStokes}.

Stokes' theorem can be translated into a tensorial form based on proper reference frame coordinates by identifying the parameterization $\mathbf{s}(u,v)$ with the proper reference frame coordinates $s^{\basind{i}}(u,v)$ and the vectors $\mathbf{f}(\mathbf{s})$ and $d\boldsymbol{\sigma}$ with the contravariant spatial vector components $f^{\basind{i}}(s^{\basind{r}})$
and $d\sigma^{\basind{i}}$. When we recall eqs.~(\ref{EpsilonTetradForm1}) and~(\ref{EpsilonTetradForm2}), we find the equivalent tensorial version 
\begin{equation}
\oint\limits_\mathcal{S} f_{\basind{k}}\,\D s^{\basind{k}}
=-\iint\limits_{\mathcal{A}} \varepsilon^{\basind{0 a j k}} \pdiff{f_{\basind{k}}}{s^{\basind{j}}}  
\,\D \sigma_{\basind{a}},\quad\text{with}\quad
\D \sigma_{\basind{a}}=\varepsilon_{\basind{0 a m n}}\pdiff{s^{\basind{m}}}{u}
\pdiff{s^{\basind{n}}}{v}\,\D u\,\D v
\label{StokesTensor}
\end{equation} 
of Stokes' theorem.

\subsection{Complete expansion of the Sagnac time delay}

We now apply Stokes' theorem in the form of eq.~(\ref{StokesTensor}) to each of the line integrals in the sum of eq.~(\ref{SagnacTimeDelayCalc1}). After minor algebra, we obtain the relation 
\begin{equation*}
\oint\limits_\mathcal{S} C_{\basind{k}\,\basind{i_1 \ldots i_n}} \,
s^{\basind{i_1}}\cdot\ldots\cdot s^{\basind{i_n}}\,\D s^{\basind{k}} 
=-n \;\varepsilon^{\basind{0ajk}}\,C_{\basind{k}\,\basind{j\, i_1 \ldots i_{n-1}}}
 \iint\limits_{\mathcal{A}}\, s^{\basind{i_1}}\cdot\ldots\cdot s^{\basind{i_{n-1}}}\,\D \sigma_{\basind{a}}\,,
\end{equation*} 
which after substitution into eq.~(\ref{SagnacTimeDelayCalc1}) leads to formula
\begin{equation}
\Delta \tau_{Sq}=\frac{2}{c}\,\sqrt{g_{\basind{00}}(q^{\basind{r}})}\,
\sum_{n=0}^\infty (n+1)\,\varepsilon^{\basind{0ajk}}\,C_{\basind{k}\,\basind{j\,i_1\ldots i_n}} 
\iint\limits_{\mathcal{A}}\, s^{\basind{i_1}}\cdot\ldots\cdot s^{\basind{i_n}}\,\D \sigma_{\basind{a}}
\label{SagnacCalc1}
\end{equation}
for the Sagnac time delay.

In order to highlight the structure of this expression for $\Delta\tau_{Sq}$, it is convenient to introduce the constant tensorial quantity
\begin{align}
\label{DefWQuantities}
W^{\basind{a}}_{\;\;\basind{i_1\ldots i_{n}}}=&\;\frac{c}{2} \;(n+1) \;\varepsilon^{\basind{0ajk}}\,C_{\basind{k}\,\basind{j\, i_1 \ldots i_n}} \\ \nonumber
=&\;\frac{c}{2 n!}\;\varepsilon^{\basind{0ajk}} 
\pdiff{^{n+1}}{x^{\basind{j}}  \dd x^{\basind{i_1}}
\ldots \dd x^{\basind{i_n}}} \left(\frac{g_{\basind{0 k}}}{g_{\basind{00}}}\right)
\bigg|_{p^{\basind{r}}} \,,
\end{align}
and the covariant components of the $n$-th moment of the unit fluxes
\begin{equation}
A_{\basind{a}}^{\;\;\basind{i_1 \ldots i_{n}}}= \iint\limits_{\mathcal{A}}\, s^{\basind{i_1}}\cdot\ldots\cdot s^{\basind{i_{n}}}  \D \sigma_{\basind{a}} 
 \,.
\label{MomentsArea}
\end{equation}
We briefly note, that we call the $A_{\basind{a}}^{\;\;\basind{i_1 \ldots i_{n}}}$ the $n$-th moment of the unit fluxes for the following reason: suppose we are given a unit vector field ${e^{\basind{i}}(x^{\basind{r}})=\delta^{\basind{i}}_{\basind{1}}}$ within the three-dimensional, spatial subspace of our proper reference frame coordinates. Then, the quantity
\begin{equation*}
A_{\basind{1}}= \iint\limits_{\mathcal{A}} \delta^{\basind{i}}_{\basind{1}}\,
\D \sigma_{\basind{i}}=
\iint\limits_{\mathcal{A}} \D \sigma_{\basind{1}}
\end{equation*}
just corresponds to the ordinary flux of $\delta^{\basind{i}}_{\basind{1}}$ through the surface $\mathcal{A}$. In analogy, we obtain all the covariant components of the zeroth moment $A_{\basind{a}}$ by considering the flux with respect to the constant vector fields $\delta^{\basind{i}}_{\basind{a}}$. The same argumentation applies also to the higher moments.

Coming back to the Sagnac time delay, we rewrite eq.~(\ref{SagnacCalc1}) with the help of eq.~(\ref{DefWQuantities}) and~(\ref{MomentsArea}) according to
\begin{equation}
 \Delta\tau_{Sq}=\frac{4}{c^2}\,\sqrt{g_{\basind{00}}(q^{\basind{r}})}\,
\sum_{n=0}^\infty W^{\basind{a}}_{\;\;\basind{i_1\ldots i_{n}}}\,
A_{\basind{a}}^{\;\;\basind{i_1 \ldots i_{n}}}\,.
\label{SagnacTimeDelayCalc2}
\end{equation}
The moments $A_{\basind{a}}^{\;\;\basind{i_1 \ldots i_{n}}}$ of the unit fluxes constitute the order parameters of this series expansion and they are completely determined by the choice of the closed spatial curve $\mathcal{S}$ and the shape of the enclosed surface $\mathcal{A}$. In contrast to this, the quantities $W^{\basind{a}}_{\;\;\basind{i_1\ldots i_{n}}}$ follow directly from the metric coefficients and their partial derivatives evaluated along the world line $p^{\basind{\sigma}}(\tau)$.

In this context, we would like to point out that although the moments of the unit fluxes depend on the shape of the surface $\mathcal{A}$, the Sagnac time delay itself does not. The latter is completely determined by the choice of the closed spatial curve $\mathcal{S}$. It is the splitting of the individual contributions in the Sagnac time delay into surface dependent and metric dependent expressions, $A_{\basind{a}}^{\;\;\basind{i_1 \ldots i_{n}}}$ and $W^{\basind{a}}_{\;\;\basind{i_1\ldots i_{n}}}$, which unfortunately obfuscates this fact after the utilization of Stokes' theorem.

\subsection{Leading-order Contributions}

We now focus on the lowest-order contributions in this series expansion. With the zeroth- and first-order terms
\begin{align*}
g_{\basind{00}}(p^{\basind{r}})&=1\,,\quad
g_{\basind{00},\basind{i_1}}(p^{\basind{r}})=-\frac{2}{c^2}\, a_{\basind{i_1}}\,,\\
g_{\basind{0k}}(p^{\basind{r}})&=0\,,\quad
g_{\basind{0k},\basind{i_1}}(p^{\basind{r}})=\frac{1}{c}\,
\varepsilon_{\basind{0 k l i_1}}\omega^{\basind{l}}
\end{align*} 
of the metric expansion, eq.~(\ref{ProperMetricExpansion}), we find from the definition~(\ref{DefWQuantities}) the expression
\begin{equation}
W^{\basind{a}}=-\omega^{\basind{a}}.
\label{W0}
\end{equation} 
Taking advantage of the second-order term, eq.~\eqref{Properg2mixed},
\begin{equation*}
g_{\basind{0k},\basind{i_1},\basind{i_2}}(p^{\basind{r}})=
\frac{4}{3}\,R_{\basind{0}\{\basind{i_1}\basind{i_2}\}\basind{k}}(p^{\basind{r}})\,,
\end{equation*} 
we arrive after some algebra at
\begin{equation*}
 W^{\basind{a}}_{\;\;\basind{i_1}}=\frac{2 c}{3}\,\varepsilon^{\basind{0 a j k}}\, 
R_{\basind{0}\{\basind{i_1}\basind{j}\}\basind{k}}(p^{\basind{r}})+
\frac{1}{c^2}\left(\omega^{\basind{l}} a_{\basind{l}}\,\delta^{\basind{a}}_{\basind{i_1}}-
3\, \omega^{\basind{a}} a_{\basind{i_1}}\right)\,.
\end{equation*}
When we finally substitute these expressions into eq.~(\ref{SagnacTimeDelayCalc2}), we obtain the two leading-order contributions of the Sagnac time delay
\begin{equation*}
\begin{split}
 \Delta\tau_{Sq}=\frac{4}{c^2}\,\sqrt{g_{\basind{00}}(q^{\basind{r}})}\,
&\left[-\omega^{\basind{a}} A_{\basind{a}}+
\frac{2 c}{3}\,\varepsilon^{\basind{0 a j k}}\, 
R_{\basind{0}\{\basind{i_1}\basind{j}\}\basind{k}}(p^{\basind{r}})\,A_{\basind{a}}^{\;\;\basind{i_1}}
\right.\\
&+\left.
\frac{1}{c^2}\left(\omega^{\basind{l}} a_{\basind{l}}\,\delta^{\basind{a}}_{\basind{i_1}}-
3\,\omega^{\basind{a}} a_{\basind{i_1}}\right) A_{\basind{a}}^{\;\;\basind{i_1}}
+\mathcal{O}\Big(A_{\basind{a}}^{\;\;\basind{i_1}\basind{i_2}}\Big)\right]	
\,.
\end{split}
\end{equation*}
It is this result, which allows for an operational definition of the compass of inertia within general relativity.


\section{Integration of null geodesics in G\"odel's Universe\label{Intnullgeo}}

In this appendix we solve the geodesic equation for light rays which emerge from the origin and possess the initial conditions 
\begin{equation}
r(0)=0,\quad z(0)=0,\quad t(0)=0\,,
\label{GeoIntInitialCond}
\end{equation} 
used in subsect.~\ref{LightPropagationGoe}. As suggested in \cite{Kundt56}, we can take advantage of the intrinsic symmetries of G\"odel's Universe to simplify the calculations. We refer to \cite{Schuecking03,Chandrasekhar61,Novello83,Novello93} for a more detailed discussion of the geodesics in G\"odel's spacetime.

\subsection{Killing vectors\label{AppKillings}}

We start by summarizing the five Killing vectors for the G\"odel metric. Three of them follow immediately from the invariance condition eq.~(\ref{iso}), since G\"odel's metric does not depend explicitly on the coordinates $t$, $\phi$, and $z$. We note without prove, that the complete solution of the Killing equations~(\ref{Killinggl}) reads 
\begin{equation}
\xi^\alpha(t,r,\phi,z)=A\delta^\alpha_0+B\delta^\alpha_2+C\delta^\alpha_3+D\zeta^\alpha(r,\phi)
+E\zeta^\alpha\textstyle{\left(r,\phi-\frac{\pi}{2}\right)},
\label{killigoe}
\end{equation}
with the integration constants $A$, $B$, $C$, $D$, $E$ and the vector field $\zeta^\alpha(r,\phi)$ defined by
\begin{equation}
\left( \begin{array}{c} \zeta^0\\ \zeta^1 \\ \zeta^2 \\ \zeta^3 \end{array} \right)\equiv
\frac{1}{\sqrt{1+\left(\frac{r}{2a}\right)^2}}\left( 
\begin{array}{c} \frac{r}{\sqrt{2}c}\cos\phi \\ a\left(1+\left(\frac{r}{2a}\right)^2\right)\sin\phi \\ 
\frac{a}{r}\left(1+2\left(\frac{r}{2a}\right)^2\right)\cos\phi\ \\ 0 \end{array} \right) \,.
\end{equation}

\subsection{Constants of motion}

As described in \ref{WorldLineGeodesic}, each symmetry represented by a Killing vector corresponds to a constant of motion of the geodesic equation. For simplicity, we focus on the first three Killing vectors in eq.~(\ref{killigoe}) and determine with eq.~(\ref{ConstMotion}) the three constants of motion
\begin{equation}
\A=u_0({\lambda})\,, \qquad\qquad \B=u_2({\lambda})\,, \qquad\qquad \C=u_3({\lambda}) \,.
\label{const_Clambda}
\end{equation}
When we take into account the identity $u_{\mu}(\lambda)=g_{\mu\nu}(x^\sigma(\lambda))u^{\nu}(\lambda)$ together with the metric coefficients
\begin{equation*}
(g_{\mu\nu})=\left(%
\begin{array}{c c c c}
c^2&0&r^2\,\Omega_G&\phantom{-}0\vspace{2mm}\\
0&-\frac{1}{1+\left(\frac{r}{2a}\right)^2}&0&\phantom{-}0\\
r^2\,\Omega_G&0&-r^2\left(1-\left(\frac{r}{2a}\right)^2\right)&\phantom{-}0\vspace{2mm}\\
0&0&0&-1
\end{array}%
\right) \,,
\end{equation*}
following from the line element~(\ref{metric}) and eq.~(\ref{rotskalar}), we arrive at
\begin{eqnarray}
\A&=u_0(0)&=c^2 u^0(0)+r^2(0)\,\Omega_G\, u^2(0)=c^2 u^0(0)\,,   \label{const_C0}\\
\B&=u_2(0)&=r^2(0)\,\Omega_G \, u^0(0)-r^2(0)
\left(1-\left(\frac{r(0)}{2a}\right)^2\right) u^2(0) =0\,,  \label{const_C2}\\
\C&=u_3(0)&=-u^3(0)\, \label{const_C3}.
\end{eqnarray}
The simplicity of these expressions is due to the initial conditions~(\ref{GeoIntInitialCond}). 

When we furthermore insert the latter expressions into the condition
\begin{equation}
g_{\mu\nu}u^{\mu}u^{\nu}=0  \label{const_of_four_velocity}
\end{equation}
we obtain
\begin{equation}
c^2(u^0(0))^2=(u^1(0))^2+(u^3(0))^2\,.  
\label{condition_u1}
\end{equation}
On the other hand, using the equivalent form 
$g^{\mu\nu}(x^{\alpha}(\lambda))\,u_{\mu}(\lambda)u_{\nu}(\lambda)=0$ of eq.~(\ref{const_of_four_velocity}) 
and the corresponding contravariant components 
\begin{equation}
(g^{\mu\nu})=\left(%
\begin{array}{c c c c}
\frac{1}{c^2}\,\frac{1-\left(\frac{r}{2a}\right)^2}{1+\left(\frac{r}{2a}\right)^2}&0&\frac{\Omega_G}{c^2}\,\frac{1}{1+\left(\frac{r}{2a}\right)^2}&\phantom{-}0\vspace{1mm}\\
0&-\left(1+\left(\frac{r}{2a}\right)^2\right)&0&\phantom{-}0\vspace{1mm}\\
\frac{\Omega_G}{c^2}\,\frac{1}{1+\left(\frac{r}{2a}\right)^2}&0&-\frac{1}{r^2}\,\frac{1}{1+\left(\frac{r}{2a}\right)^2}&\phantom{-}0\vspace{2mm}\\
0&0&0&-1
\end{array}%
\right)
\label{contrametriccoeff}
\end{equation}
of the metric tensor together with the constants of motion (\ref{const_Clambda}-\ref{const_C3}), we find the additional condition
\begin{eqnarray}
c^2(u^0(0))^2\;\frac{1-\left(\frac{r(\lambda)}{2a}\right)^2}{1+\left(\frac{r(\lambda)}{2a}\right)^2}
-(u_1(\lambda))^2\,\left(1+\left(\frac{r(\lambda)}{2a}\right)^2\right)
-(u^3(0))^2=0\,.       
\label{radeqn}
\end{eqnarray}

\subsection{Explicit expressions for the coordinates}

We now present the differential equation for the radial coordinate and its solution. This result enables us to deduce expressions for the other coordinates.

\subsubsection{Integration of the radial coordinate\label{AppIntegrationOfR}}

When we substitute $u_1=g_{11}u^1$ into eq.~(\ref{radeqn}), we finally arrive at
\begin{equation}
(u^{1}(\lambda))^2=\left(\diff{r}{\lambda}\right)^2=
\Big( c^2(u^0(0))^2 - (u^3(0))^2 \Big) -
\Big( c^2(u^0(0))^2 + (u^3(0))^2 \Big) \left(\frac{r}{2a}\right)^2 
\,.    \label{eq:u1_sq}
\end{equation}
It is useful to introduce the constants
\begin{equation*}
\omega = \frac{1}{2a}\sqrt{c^2 (u^0(0))^2+(u^3(0))^2} = \frac{u^1(0)}{2a}\sqrt{1+2\left(\frac{u^3(0)}{u^1(0)}\right)^2}\,
\end{equation*}
and
\begin{equation}
A =\sqrt{\frac{c^2 (u^0(0))^2-(u^3(0))^2}{c^2 (u^0(0))^2+(u^3(0))^2}} = \frac{u^1(0)}{2a\omega}\,,
\label{constant_A2}
\end{equation}
which cast eq.~(\ref{eq:u1_sq}) into the form 
\begin{equation}
 (u^{1}(\lambda))^2=\left(\diff{r}{\lambda}\right)^2=(2aA\omega)^2
\left(1-\left(\frac{r(\lambda)}{2aA}\right)^2\right)\,. \label{condition_for_u1}
\end{equation}
We take the derivative of eq.~(\ref{condition_for_u1}) with respect to $\lambda$ and arrive at a second-order differential equation 
\begin{equation}
\diff{^2 r}{\lambda^2}+\omega^2\,r(\lambda)=0\, \label{HarmonicOsciDGL}
\end{equation}
for the radial coordinate $r(\lambda)$, which is identical to the equation of motion of a classical harmonic oscillator.

Together with the definition~(\ref{constant_A2}), the solution of~(\ref{HarmonicOsciDGL}) subjected to our initial conditions reads
\begin{equation}
r(\lambda)=2aA\sin\left(\omega\lambda\right) \,.
\label{rad_simple_version}
\end{equation}
Here we have restricted ourselves to the parameter values $0\leq\lambda\leq\frac{\pi}{\omega}$. 

At $\lambda=\frac{\pi}{\omega}$ the radial coordinate returns to its initial value $r(\pi/\omega)=r(0)=0$. Moreover, due to the non-uniqueness of polar coordinates at the origin $r=0$ we encounter a jump in the radial velocity $u^1(\lambda)$, eq.~(\ref{condition_for_u1}), from $-|u^1(0)|$ to $+|u^1(0)|$. Hence, the periodicity in the time evolution of the radial coordinate can be expressed concisely by the formula
\begin{equation}
r(\lambda) = 2aA\left|\sin\left(\omega\lambda\right)\right|\,.  \label{r_lambda2}
\end{equation}
The maximum value $r_{\text{max}}$ of $r(\lambda)$ is found to be $2aA$. In particular, we obtain for vanishing $u^3(0)$ the value $A=1$ and thus $r_{\text{max}}=2a$. However, with increasing $u^3(0)$, the value of $A$ gets smaller and $r_{\text{max}}$ falls below the critical G\"odel radius $2a$.

\subsubsection{Time coordinate}

The knowledge of the radial coordinate enables us to calculate the remaining functions 
$t(\lambda)$ and $\phi(\lambda)$. For the time coordinate we start with the relation ${u^0=g^{0\mu}u_{\mu}}$ which yields after substitution of our constants of motion (\ref{const_Clambda}-\ref{const_C3})
\begin{equation}
u^0(\lambda)=\diff{t}{\lambda} =  u^0(0)\; \frac{1-\left(\frac{r(\lambda)}{2a}\right)^2}
{1+\left(\frac{r(\lambda)}{2a}\right)^2}\,. \label{diff_gl_t}
\end{equation} 
Separation of variables and substitution of eq.~(\ref{r_lambda2}) in eq.~(\ref{diff_gl_t}) leads to
\begin{eqnarray}
\D t = u^0(0) \left(-1 + \frac{2}{1+A^2\sin^2(\omega\lambda)}\right)  \D \lambda\,.  \label{rad_eq_before_int}
\end{eqnarray}
The right-hand side of the above equation is a periodic function with period $\lambda=\pi/\omega$, and it can be integrated piecewise in the separate intervals $\lambda\in[n\pi/(2\omega),(n+1)\pi/(2\omega)]$ with $n\in\mathds{N}_0$.
Taking the different contributions of the integration carefully into account, we finally arrive at the continuous function 
\begin{equation}
t(\lambda)=-u^0(0)\,\lambda + 
\frac{2}{\Omega_G}
\left( \arctan\left( \frac{\Omega_G}{\omega}\,u^0(0)\, \tan(\omega\lambda)\right) + 
\pi \left\lfloor\frac{\omega}{\pi}\lambda+\frac{1}{2}\right\rfloor \right)\,,
\label{t_lambda2} 
\end{equation}
for the time coordinate. Here we have made use of the floor function $\lfloor x\rfloor$, which is defined according to
\begin{equation*}
\lfloor x\rfloor=\max\{n\in\mathbb{Z}\;|\;n\leq x\}\,.
\end{equation*}

\subsubsection{Polar coordinate}

In analogy to the coordinate time $t$,  we now recall the relation $u^2=g^{2\mu}u_{\mu}$ to derive the functional dependence of the polar angle $\phi(\lambda)$. Insertion of the radial function (\ref{r_lambda2}) yields
\begin{equation}
u^2(\lambda)=\diff{\phi}{\lambda}=\frac{\Omega_G \,u^0(0)}{1+A^2\sin^2(\omega\lambda)}\,.
\end{equation}
Hence, we encounter the same periodic function for the integration of $\phi(\lambda)$ as in the previous case for the time coordinate $t(\lambda)$. In addition another subtlety occurs due to the non-uniqueness of polar coordinates at the origin. Every continuously differentiable curve which passes through the origin suffers a jump in its angular coordinate $\phi(\lambda)$ by $\pi$. Therefore, we have to substract an extra term $\pi\lfloor \omega \lambda/\pi\rfloor$ in comparison to the integration of the periodic function given above.
Taking the initial polar angle $\phi(0)$ of the geodesic into account, the final expression for our angular coordinate reads
\begin{equation}
\phi(\lambda) = 
  \arctan\left( \frac{\Omega_G}{\omega}\,u^0(0)\, \tan(\omega\lambda)\right) + 
\pi\left\lfloor\frac{\omega}{\pi}\lambda + \frac12 \right\rfloor - \pi\bigg\lfloor\frac{\omega}{\pi}\lambda \bigg\rfloor + \phi(0)\,.  \label{phi_lambda2}
\end{equation}

\subsubsection{Vertical coordinate}

The remaining function $z(\lambda)$ is easily found by recalling the constant of motion (\ref{const_C3}) in combination with (\ref{const_Clambda}). An integration provides the final result
\begin{equation}
z(\lambda)=u^{(3)}(0)\,\lambda \,.
\label{z_lambda2}
\end{equation}
Thus, we have established the general solution of the geodesic equation for light rays which emerge from the origin.


\end{document}